%% file: main.tex
\documentclass[runningheads]{llncs}
\usepackage{amsfonts}
\usepackage{graphicx}
\usepackage{balance}  
\usepackage[caption=false,font=footnotesize]{subfig}
\usepackage{amsmath}
\usepackage{float}
\usepackage{amsopn}
\usepackage{comment}
\usepackage{multirow}
\usepackage{enumitem}
\usepackage{multirow} 
\usepackage{subfloat}

\usepackage{CJK}
\makeatletter
\newif\if@restonecol
\makeatother

\usepackage[linesnumbered,ruled,vlined]{algorithm2e}
\usepackage{algpseudocode}
\usepackage{amsmath}
\usepackage{array}
\usepackage{color}

\newcommand{\Paragraph} [1] {\smallskip\noindent{\bf #1.}}
\begin{document}
\title{What Have We Learned from OpenReview?}
%

%

\author{Gang Wang \and
Qi Peng \and Yanfeng Zhang \and
Mingyang Zhang}
\authorrunning{G.Wang et al.}

\institute{Northeastern University, China \\
\email{1910636@stu.neu.edu.cn}\\
\email{ffpengqi@stumail.neu.edu.cn}\\
\email{zhangyf@mail.neu.edu.cn}\\
\email{theremay@outlook.com}}

\maketitle              
\begin{abstract}
Anonymous peer review is used by the great majority of computer science conferences. OpenReview is such a platform that aims to promote openness in peer review process. The paper, (meta) reviews, rebuttals, and final decisions are all released to public. We collect 5,527 submissions and their 16,853 reviews from the OpenReview platform. We also collect these submissions' citation data from Google Scholar and their non-peer-reviewed versions from arXiv.org. By acquiring deep insights into these data, we have several interesting findings that could help understand the effectiveness of the public-accessible double-blind peer review process. Our results can potentially help writing a paper, reviewing it, and deciding on its acceptance.

\end{abstract}
\input{intro}

\input{dataset}
\input{proreview}

\input{weakpoint}

\input{cluster}
\input{citation}
\input{arxiv}
\input{related}

%
%
%

%
%
%
\bibliographystyle{splncs04}
\bibliography{ref}
\end{document}


\title{Supplementary Material for Submission Entitled ``What Have We Learned from OpenReview?''}
\titlerunning{Supplementary Material for Submission 209}

%

\maketitle              

\section{Word Cloud Of ICLR Submissions}

\begin{figure}
\centering
\subfigure{
\begin{minipage}[b]{0.2\textwidth}
\centerline{
	\subfloat[2017]{\includegraphics[width=3.0in]{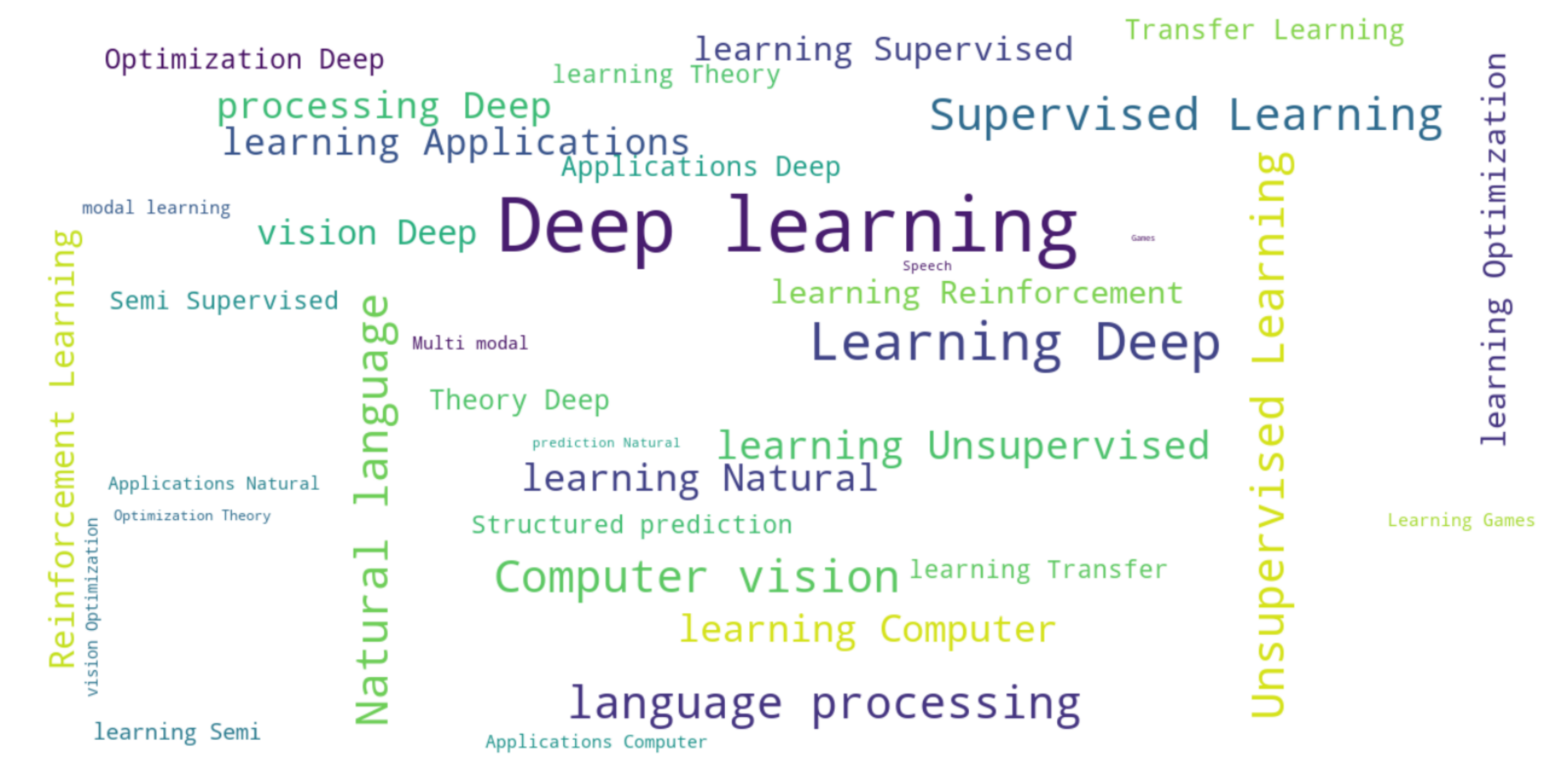}
    \label{fig:reviewer:2017}}
    \subfloat[2018]{\includegraphics[width=3.0in]{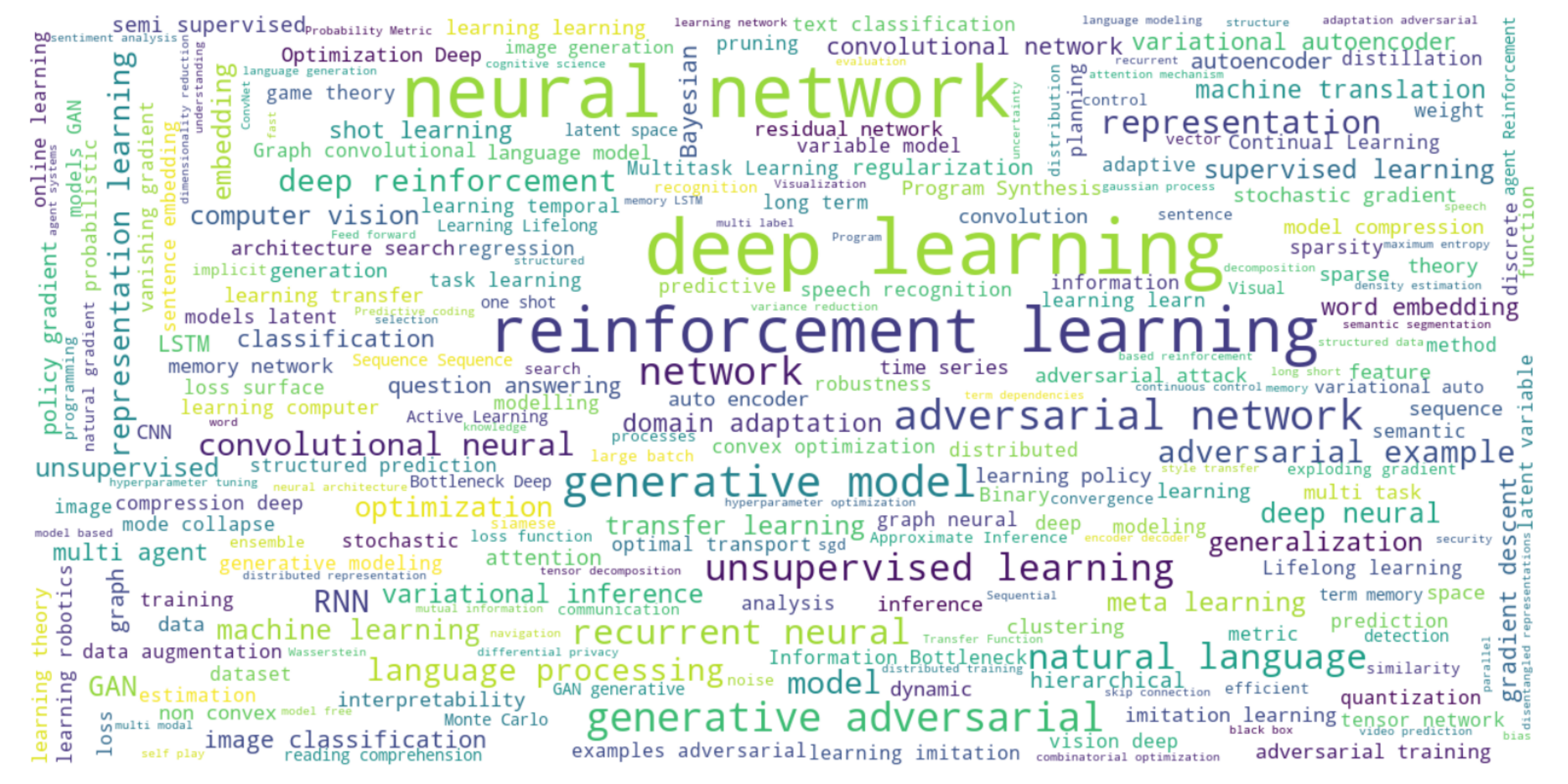}
    \label{fig:reviewer:2018}}
    }
\end{minipage}
}

\subfigure{
\begin{minipage}[b]{0.2\textwidth}
\centerline{
	\subfloat[2017]{\includegraphics[width=3.0in]{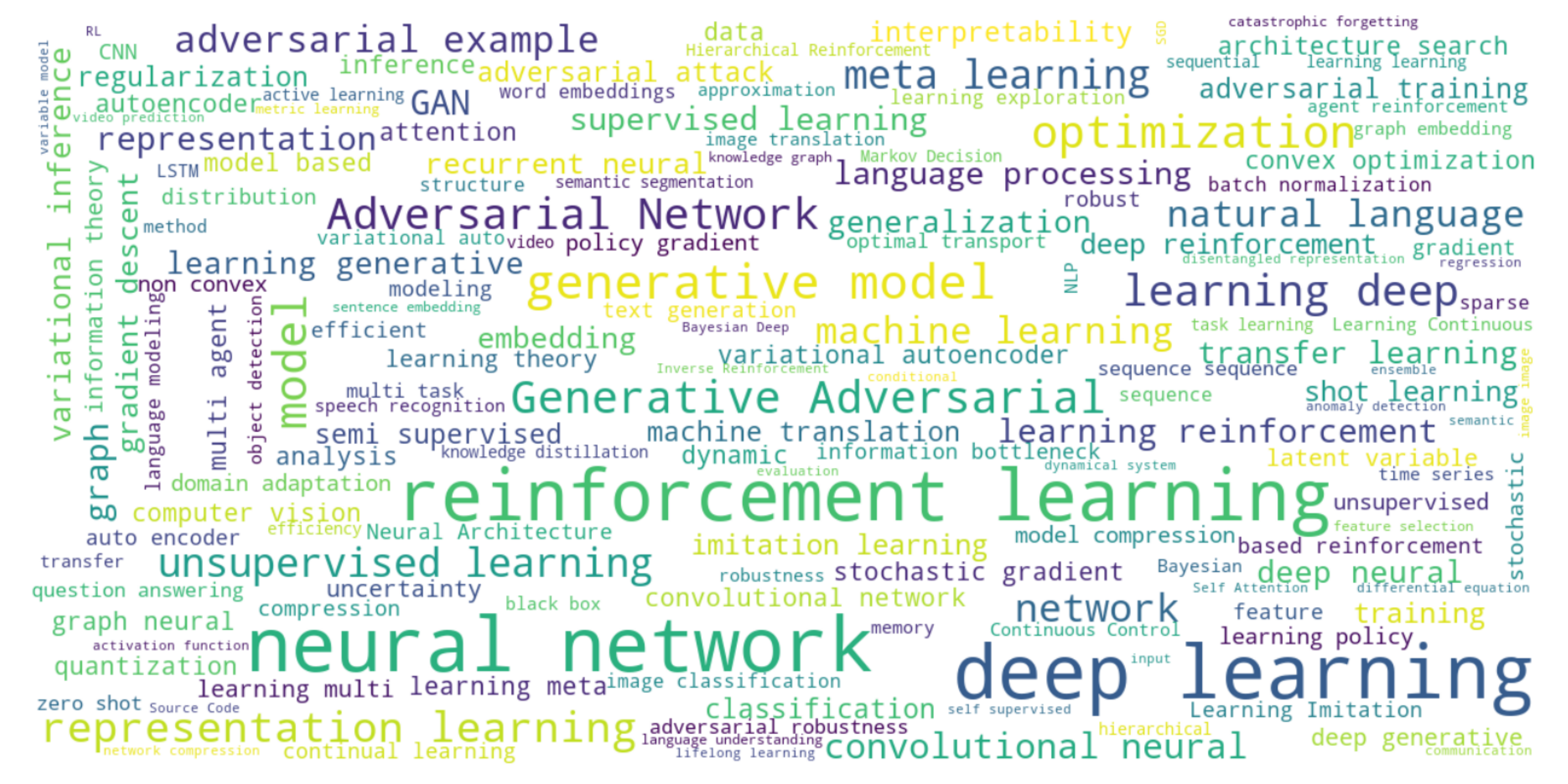}
    \label{fig:reviewer:2017}}
    \subfloat[2018]{\includegraphics[width=3.0in]{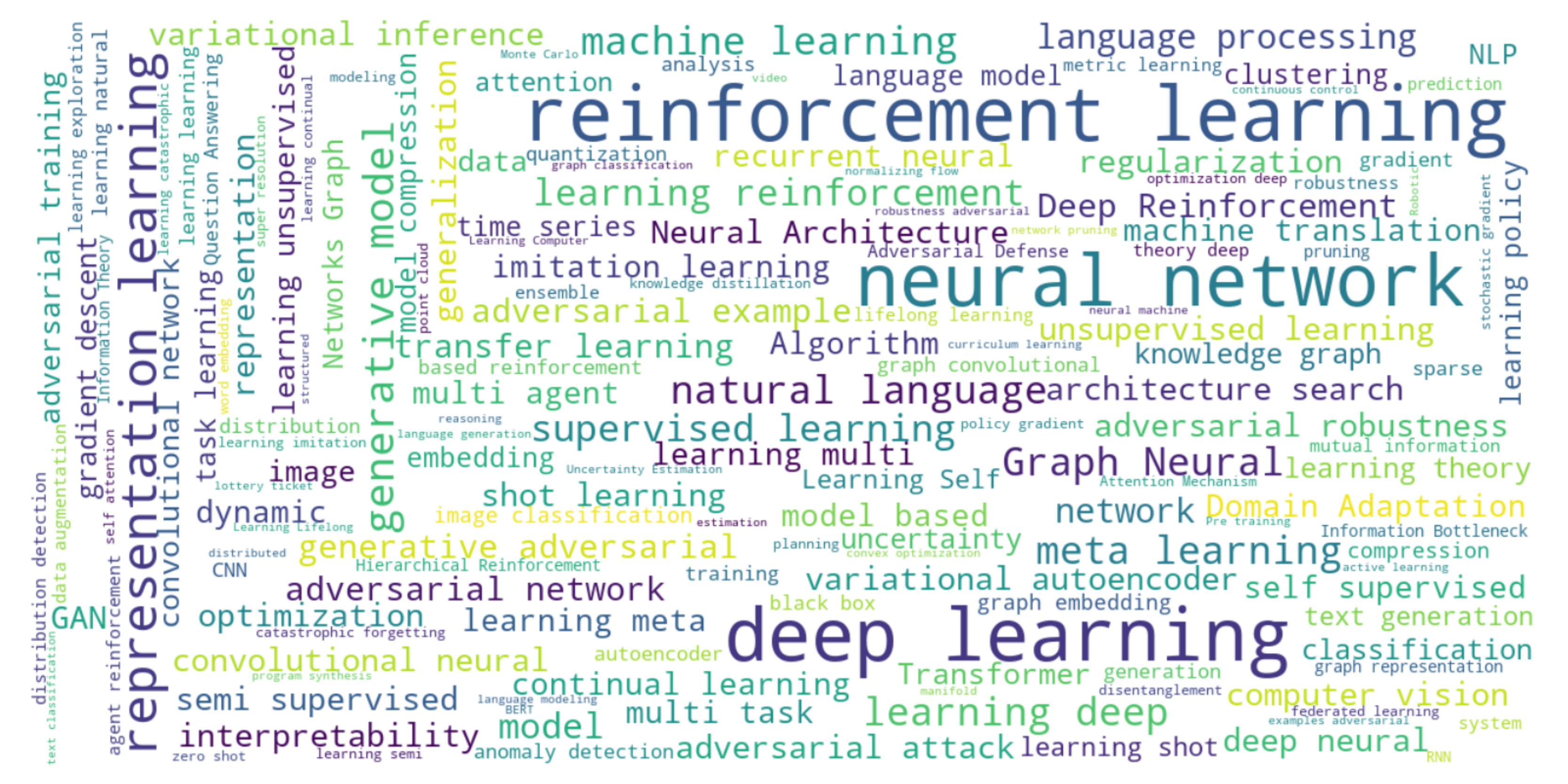}
    \label{fig:reviewer:2018}}
    }
\end{minipage}
}
\caption{The distribution of different reviewer scores of different confidence level reviews of 2017-2020. }
\label{fig:word_cloud_2017_2020}
\end{figure}

We draw the word cloud according to the word frequency appeared in the title and abstract submissions. We can see the research hotspots from Fig.\ref{fig:word_cloud_2017_2020}.


\section{More Results on the Impact of Non-Expert reviewers}

We show the additional results to Section 3.1 of the main manuscript in the following figures.

\begin{figure}[h]
     \centering 
     \includegraphics[width=3in]{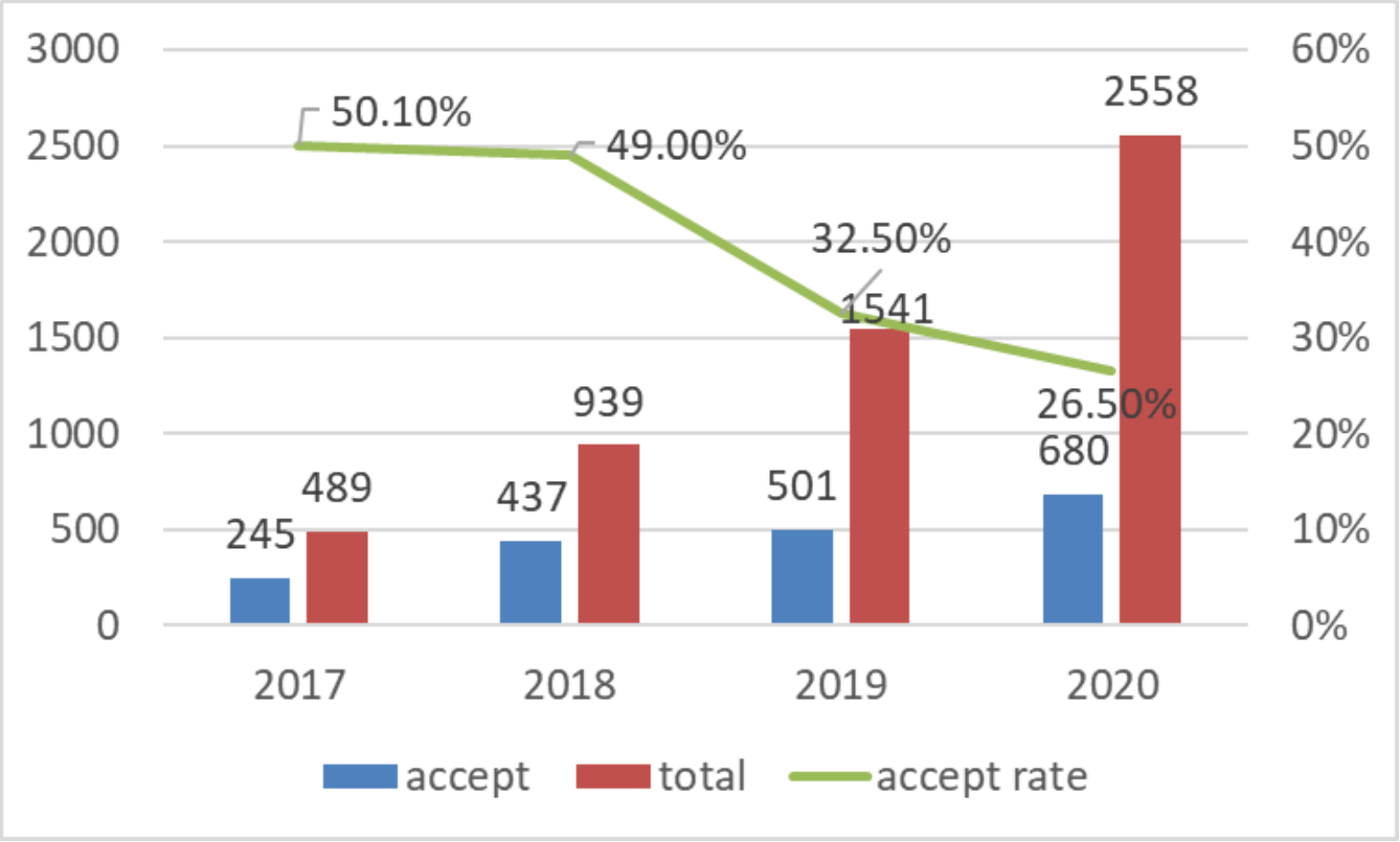}
     \caption{The number of submissions is increasing extensively from 2017 to 2020. This accordingly leads to heavy demand for reviewer volunteers, and at the same time leads to large number of low confidence reviews which will be shown in Fig. \ref{fig:pdfresizer.com-pdf-crop}.} 
     \label{fig:tendency_2017-2020} 
 \end{figure}
 
 \begin{figure}[t]
     \centering 
     \includegraphics[width=3in]{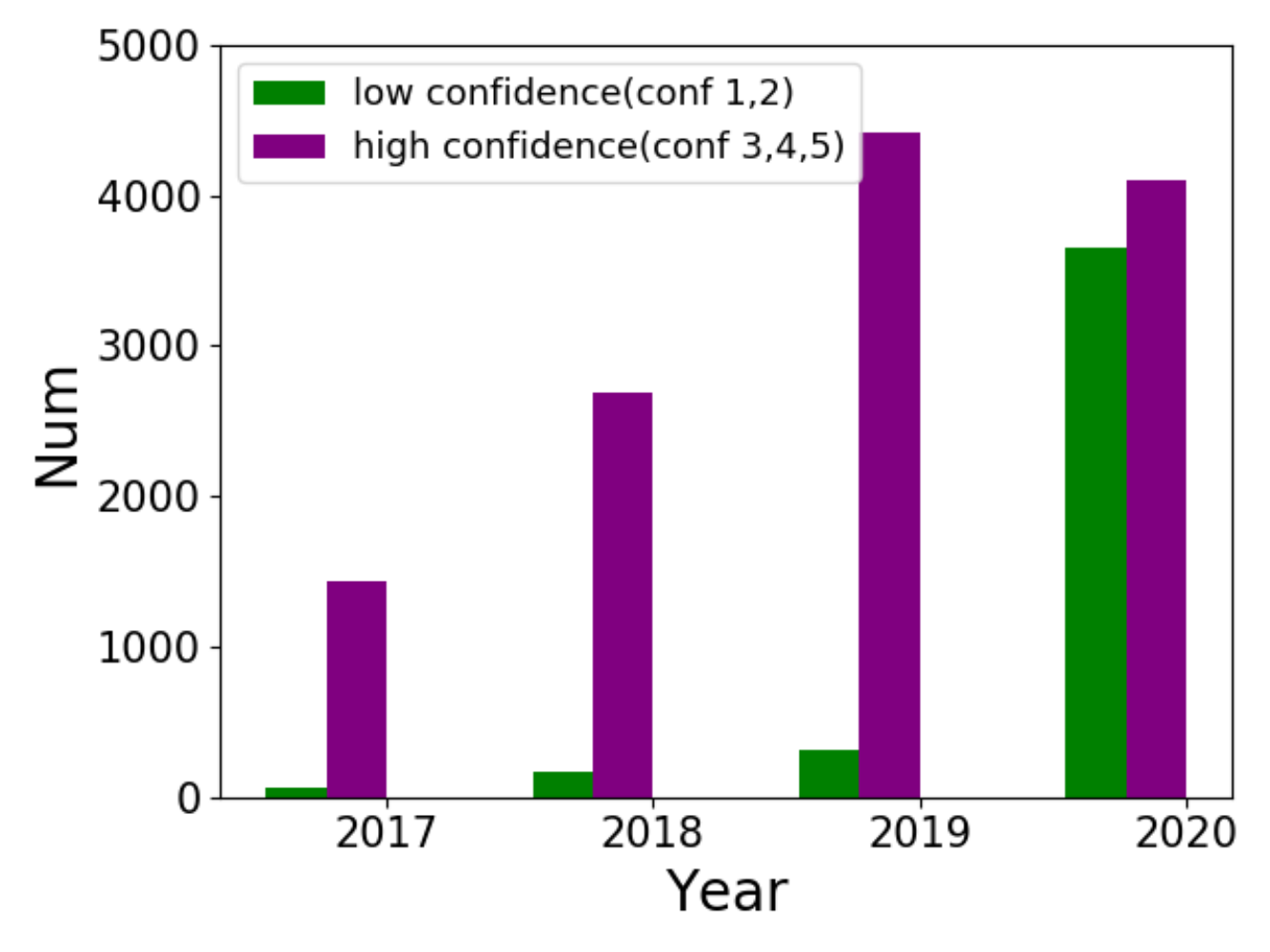}
     \caption{The numbers of high-confidence reviews and low-confidence reviews. For 2017-2019, reviews with confidence level 3, 4, and 5 are considered as high-confidence reviews, and reviews with confidence level 1 and 2 are considered as low-confidence reviews. For 2020, reviews with confidence level 3 and 4 are considered as high-confidence reviews, and reviews with confidence level 1 and 2 are considered as low-confidence reviews.} 
     \label{fig:pdfresizer.com-pdf-crop} 
 \end{figure}
 
  \begin{figure}[t]
     \centering 
     \includegraphics[width=4in]{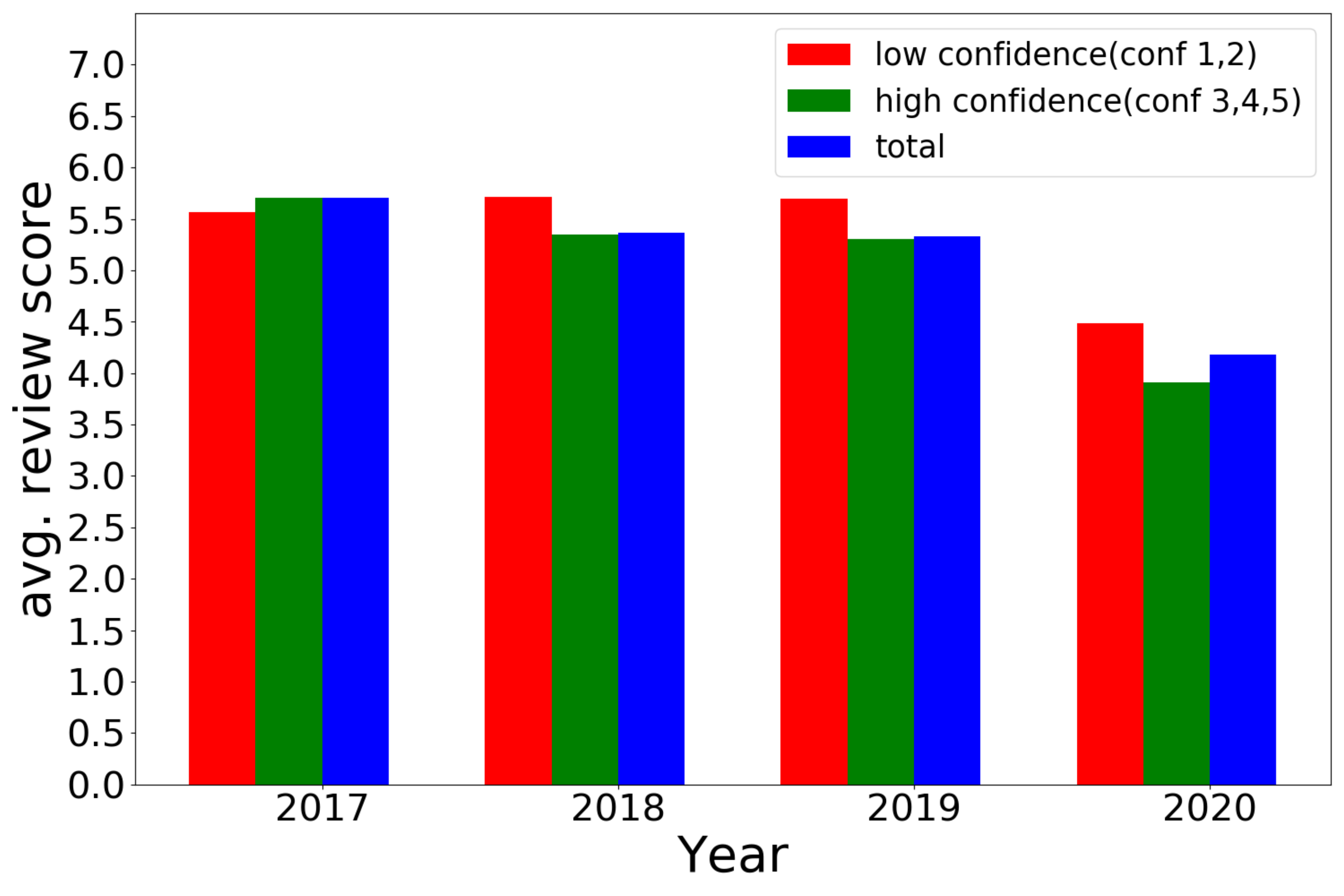}
     \caption{The average review score of high-confidence reviews,  low-confidence reviews, and mixed total. The`low-confidence reviewer tend to be more tolerant because they are not confident about their decision, while high-confidence reviewer tend to be more tough and rigorous because they are confident in the weakness they identified.} 
     \label{fig:pdfavg} 
 \end{figure}
 


\begin{figure}
\centering
\subfigure{
\begin{minipage}[b]{0.2\textwidth}
\centerline{
	\subfloat[2017]{\includegraphics[width=3.0in]{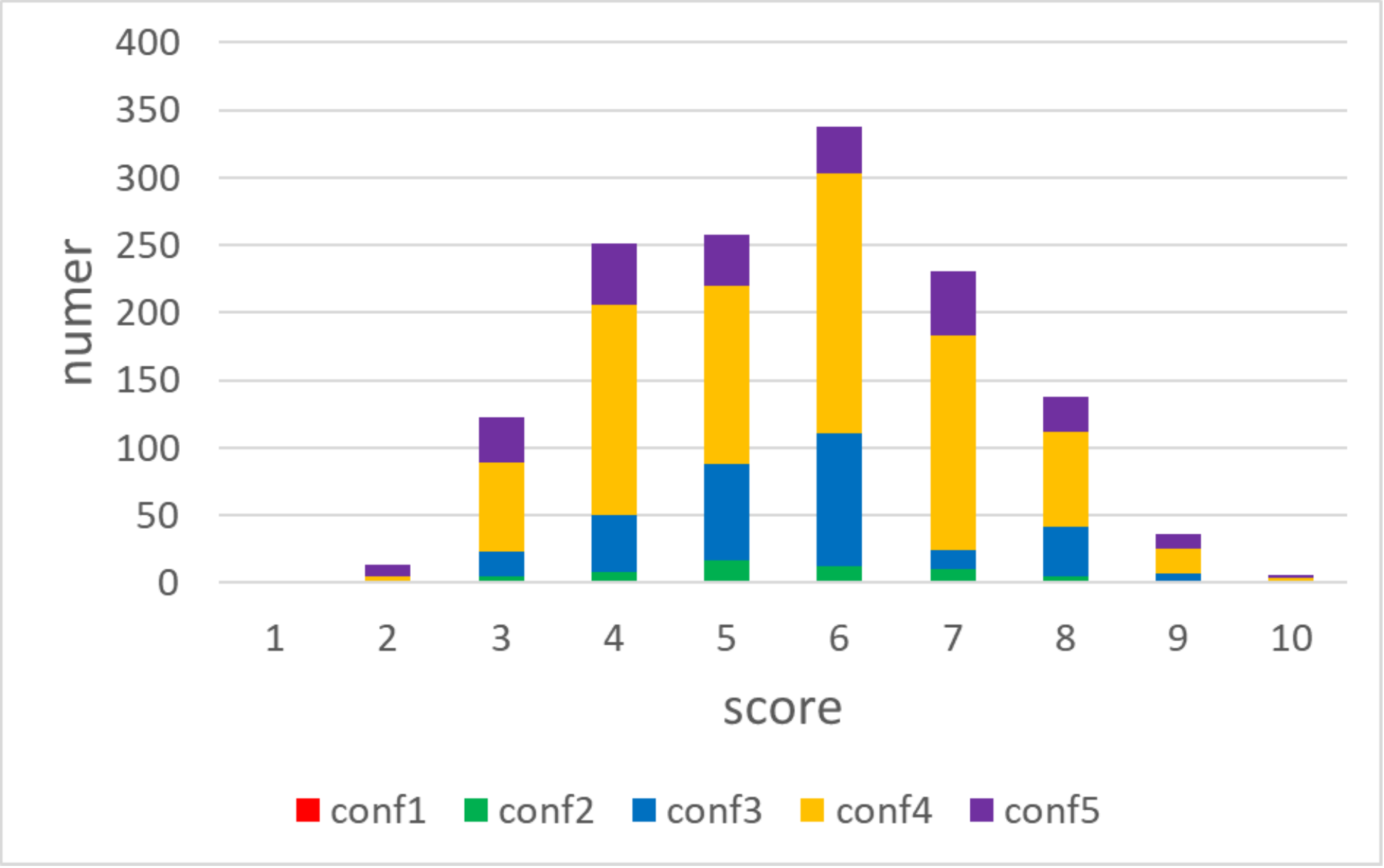}
    \label{fig:reviewer:2017}}
    \subfloat[2018]{\includegraphics[width=3.0in]{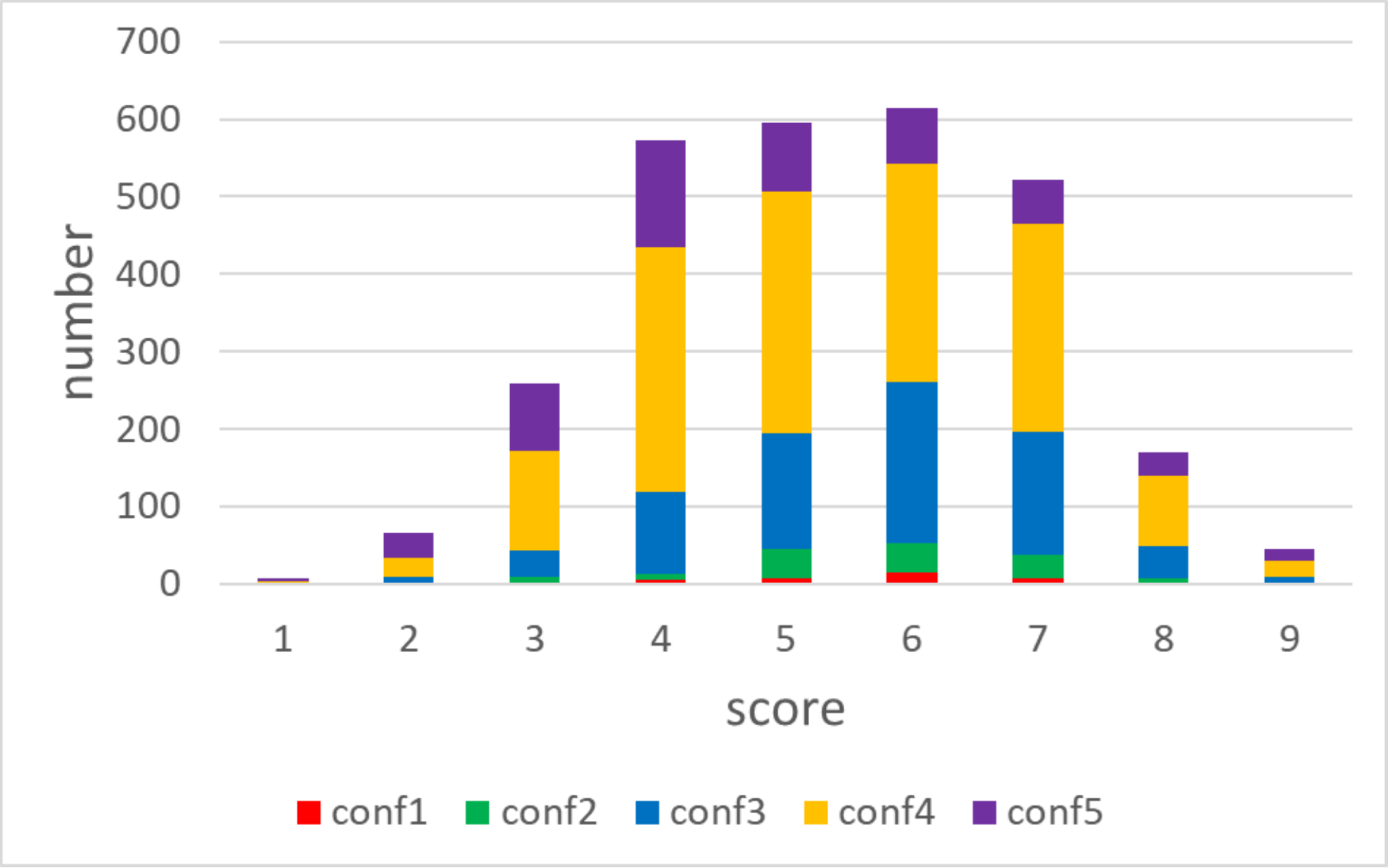}
    \label{fig:reviewer:2018}}
    }
\end{minipage}
}

\subfigure{
\begin{minipage}[b]{0.2\textwidth}
\centerline{
	\subfloat[2017]{\includegraphics[width=3.0in]{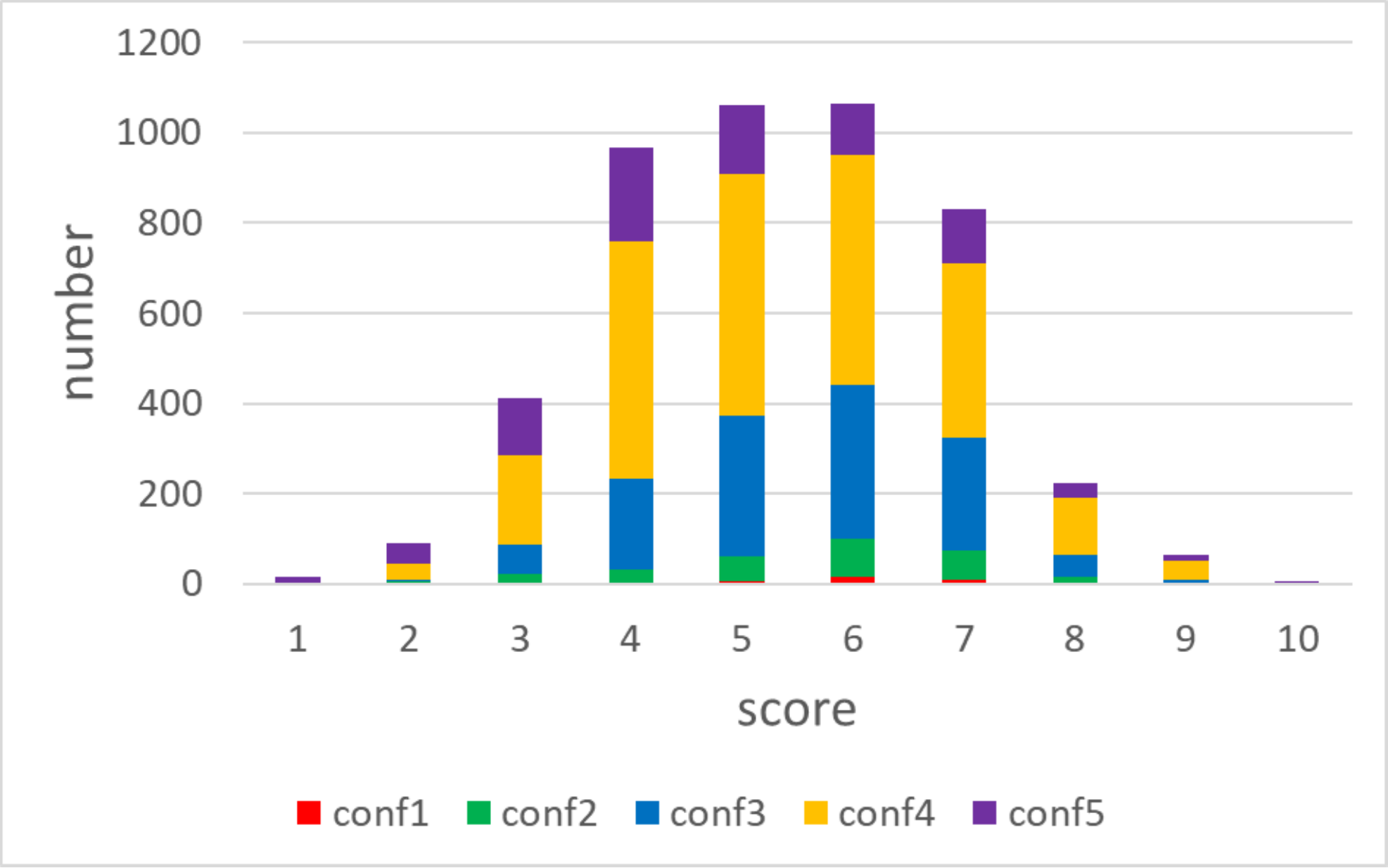}
    \label{fig:reviewer:2017}}
    \subfloat[2018]{\includegraphics[width=3.0in]{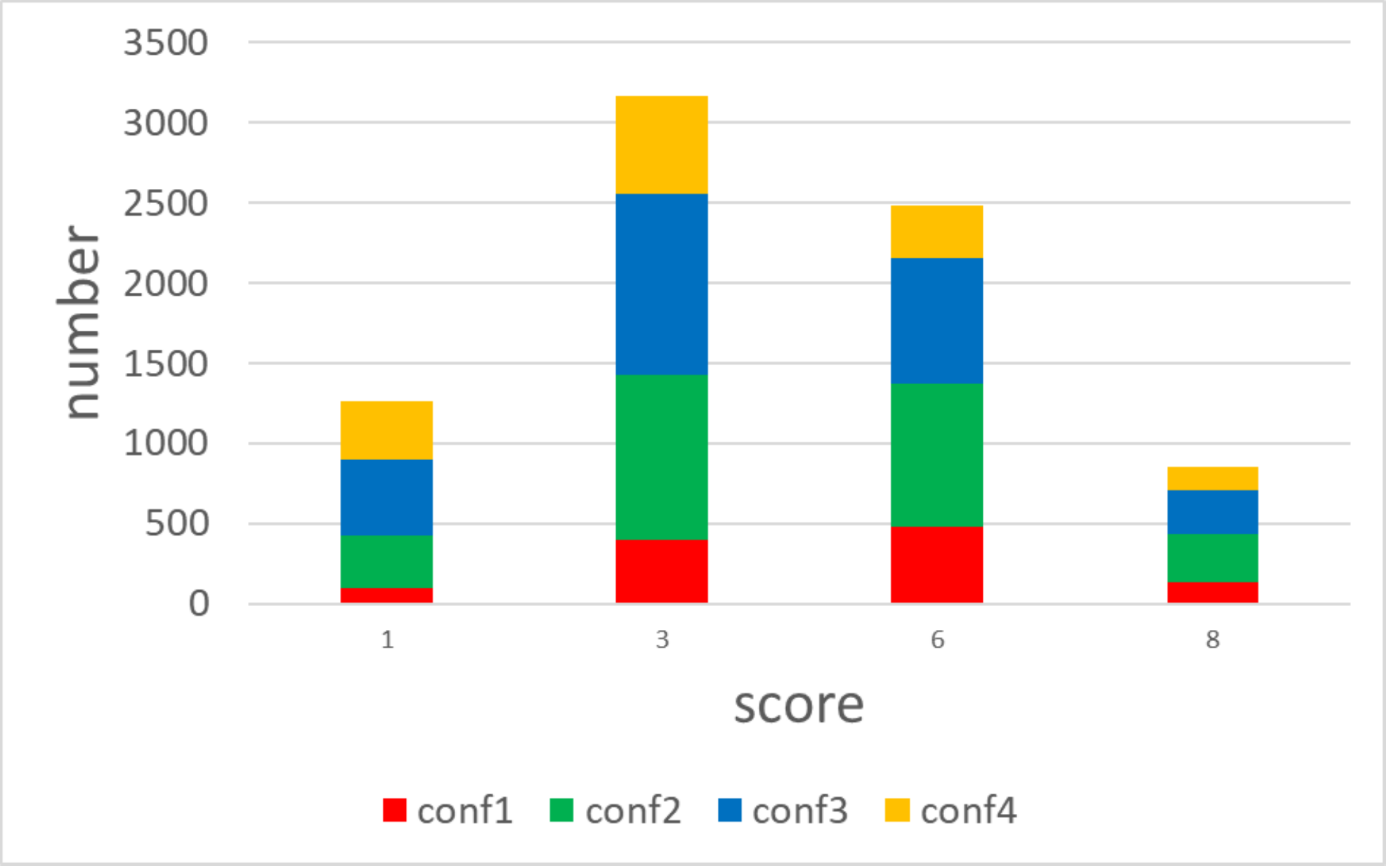}
    \label{fig:reviewer:2018}}
    }
\end{minipage}
}
\caption{The distribution of different reviewer scores of different confidence level reviews of 2017-2020. }
\end{figure}

\begin{figure}
\vspace{-0.2in}
	\centerline{
	\subfloat[2017-2019]{\includegraphics[width=2.5in]{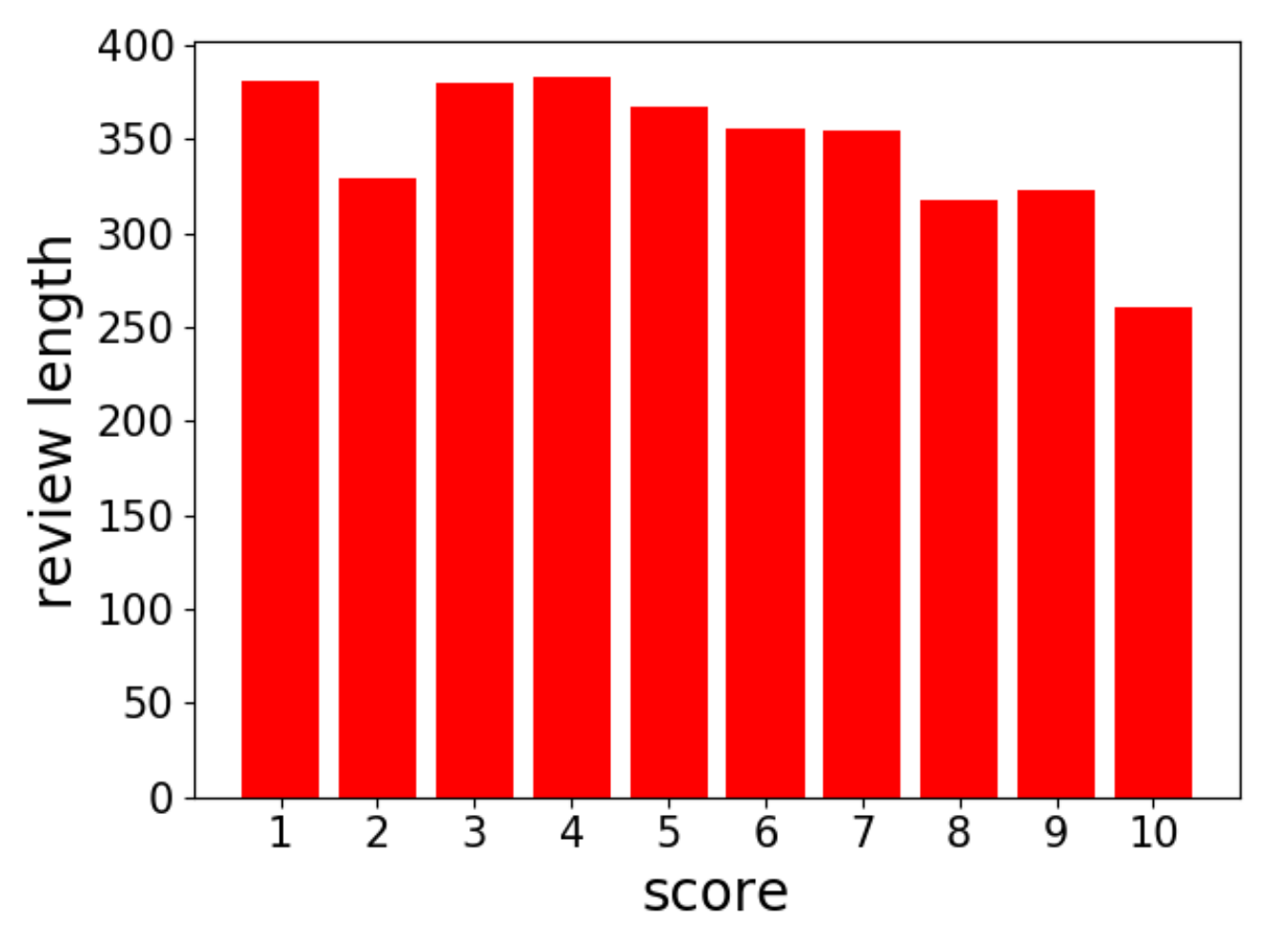}
    \label{fig:relation_between_review_length_score:2017-2019}}
    \subfloat[2020]{\includegraphics[width=2.5in]{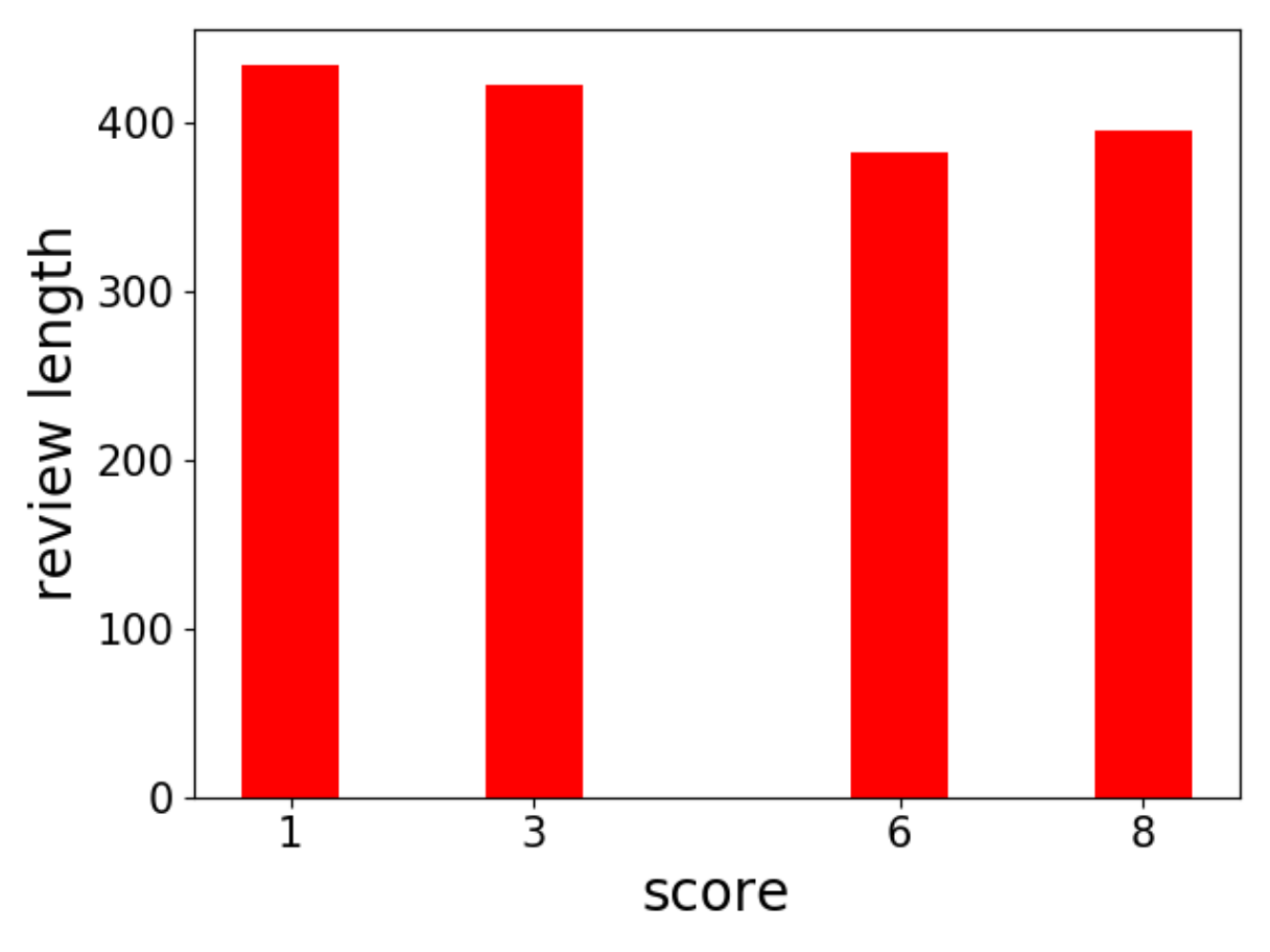}
    \label{fig:relation_between_review_length_score:2020}}
   
    }
    \vspace{-0.05in}
	\caption{Review score vs. length of review (i.e., number of words). The reviews with the same review score are grouped together to compute average review length. The reviews with higher score are likely to be short. Great paper is so good that it is not necessary to give more details about the reason for it.}
	\label{fig:relation_between_review_length_score}
\vspace{-0.2in}
\end{figure}



\begin{figure}
\centering
\subfigure{
\begin{minipage}[b]{0.2\textwidth}
\centerline{
	\subfloat[2017]{\includegraphics[width=3.0in]{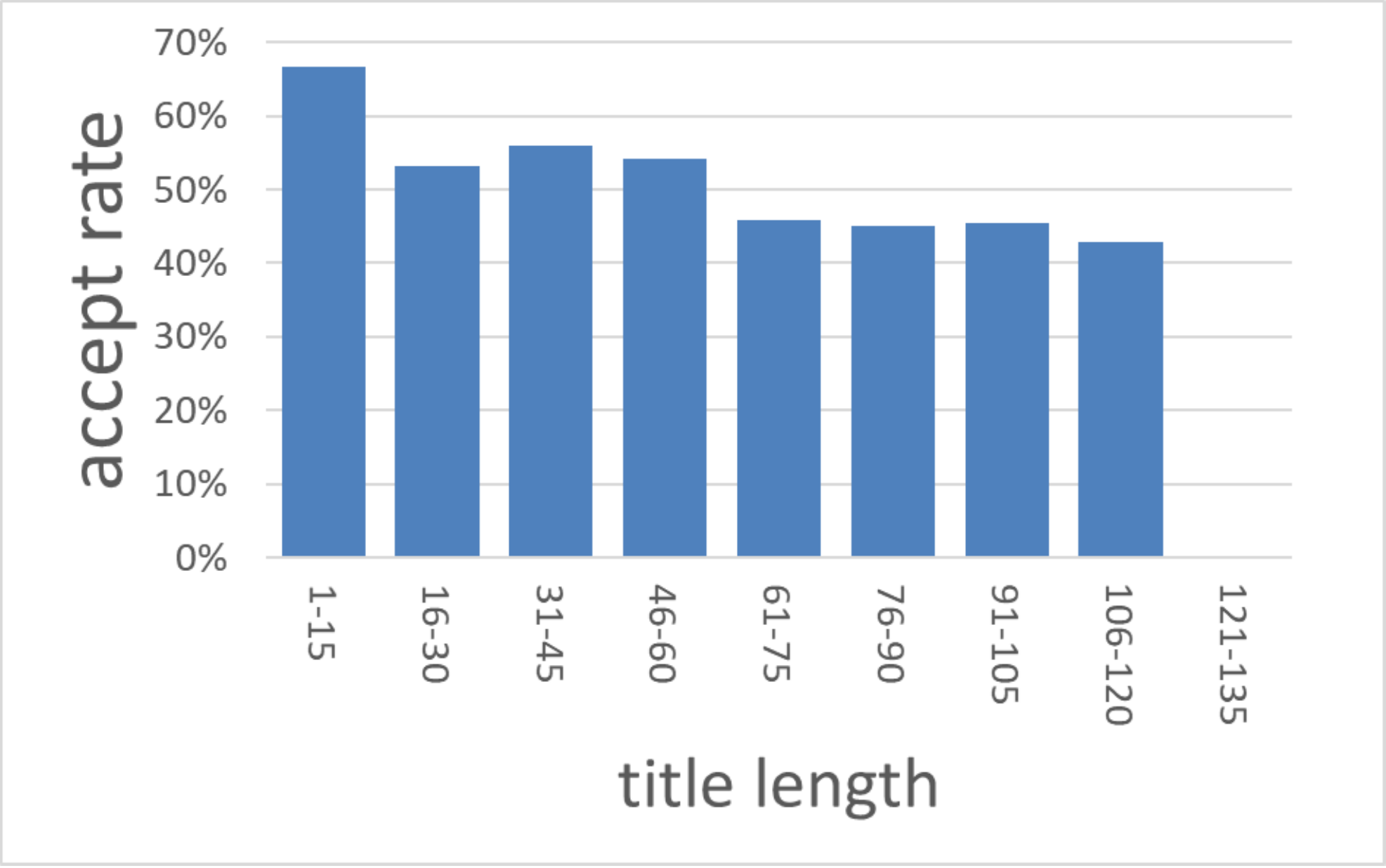}
    \label{fig:reviewer:2017}}
    \subfloat[2018]{\includegraphics[width=3.0in]{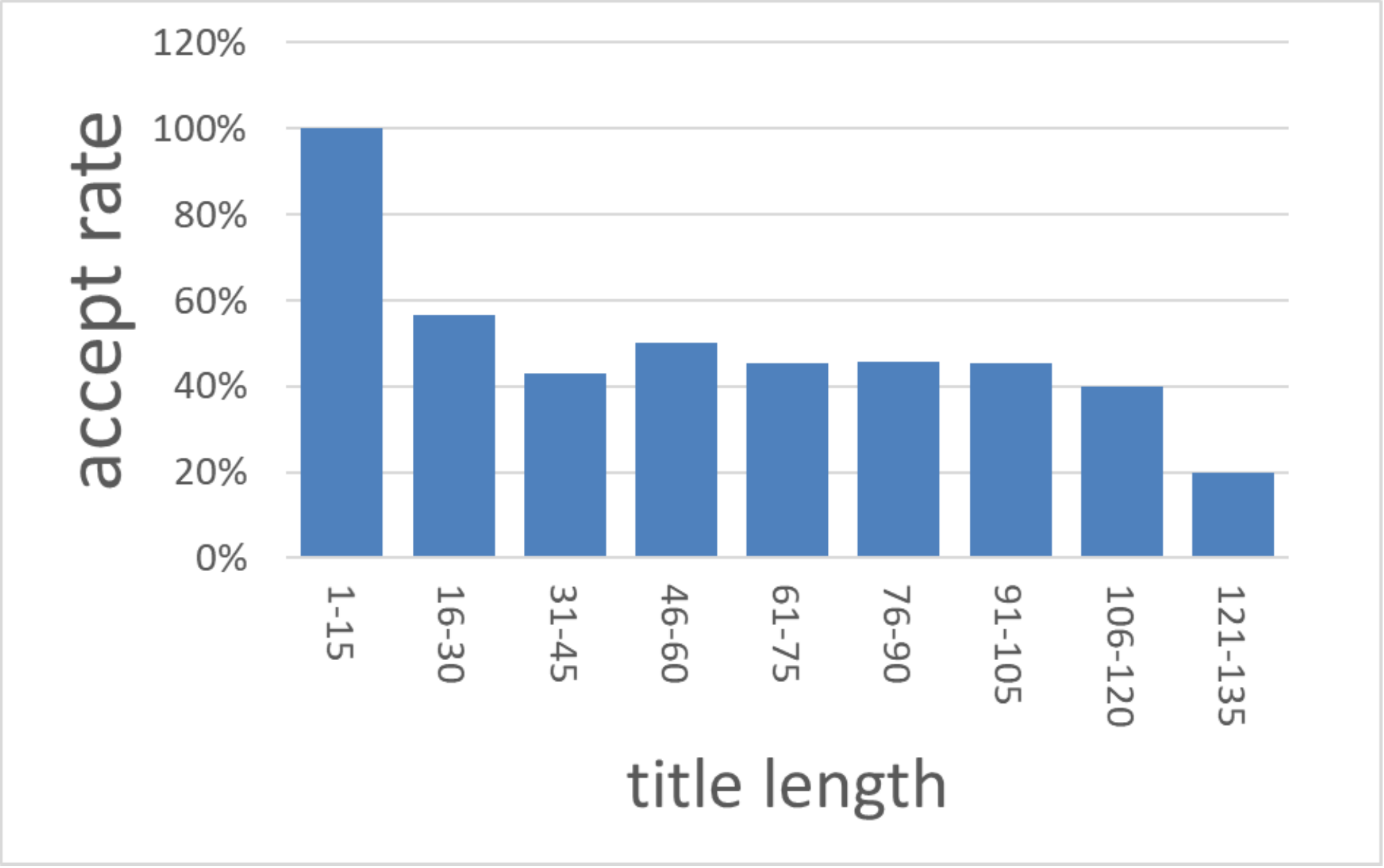}
    \label{fig:reviewer:2018}}
    }
\end{minipage}
}

\subfigure{
\begin{minipage}[b]{0.2\textwidth}
\centerline{
	\subfloat[2017]{\includegraphics[width=3.0in]{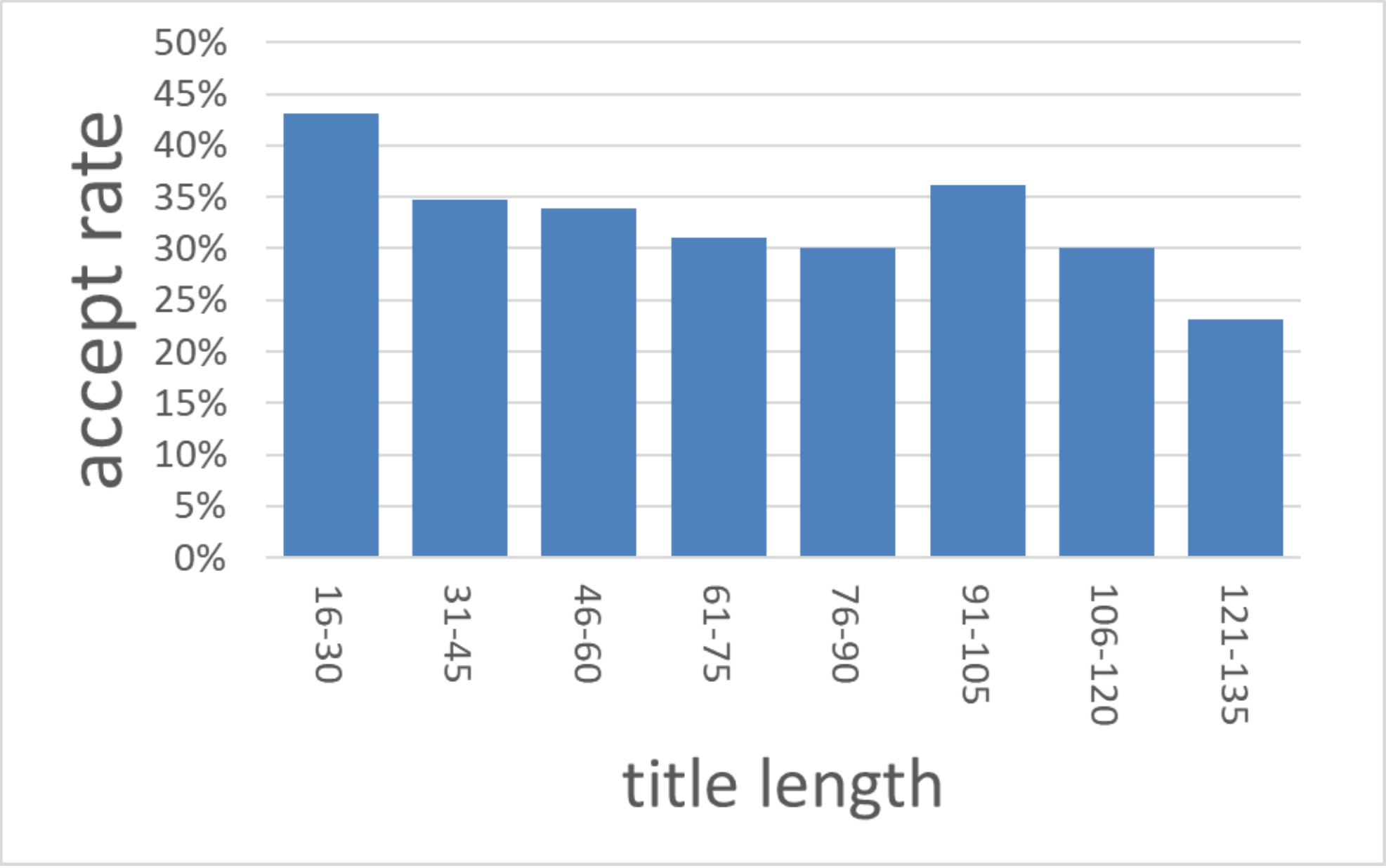}
    \label{fig:reviewer:2017}}
    \subfloat[2018]{\includegraphics[width=3.0in]{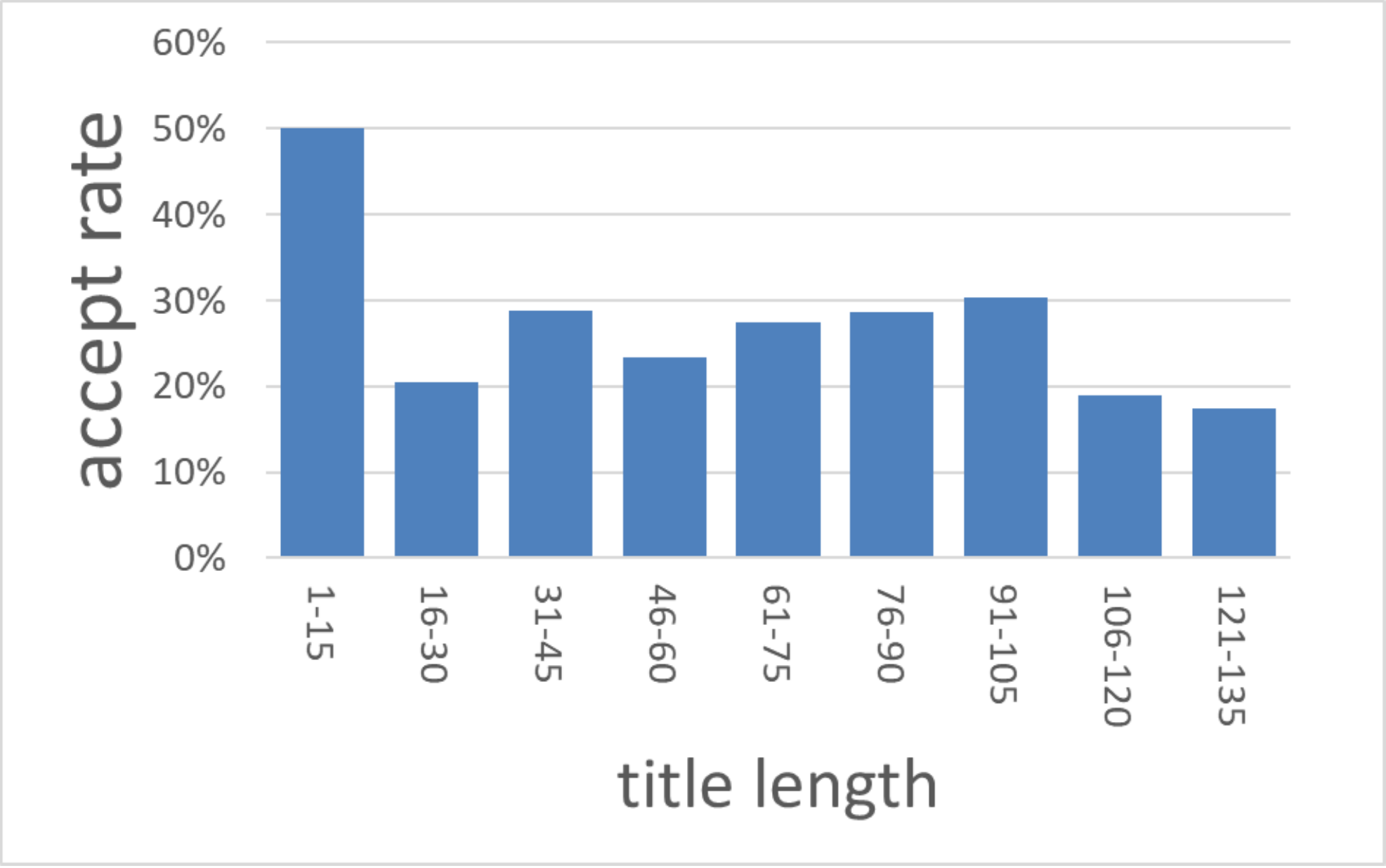}
    \label{fig:reviewer:2018}}
    }
\end{minipage}
}
\caption{Acceptance rate vs. length of paper title (i.e., number of characters). The papers are grouped according to the length of title, and the average acceptance rate of each group is depicted. The papers with a long title are likely to be accepted. }
\end{figure}

\section{The Divergence between Different Confidence Level Reviews}

We show more detailed results on the divergence between different confidence level reviews.


\small
\newcommand{\tabincell}[2]{\begin{tabular}{@{}#1@{}}#2\end{tabular}}
\makeatletter\def\@captype{table}\makeatother
\begin{minipage}{.5\textwidth}
\centering
\renewcommand\arraystretch{0.8}
\begin{tabular}{c|c|c|c|c|c} \hline 
      & conf1 & conf2 & conf3 & conf4 & conf5 \\ \hline
      num & 74 & 455 & 2330 & 4612 & 1600 & \\ 
      fra(\%)   &0.80 &5.01 &25.67 &50.81 &17.71 & \\ 
      conf1 &\tabincell{l}{\textbf{2.45}} &\tabincell{l}{6.78}&\tabincell{l}{10.63} &\tabincell{l}{11.75} &\tabincell{l}{5.83}& \\ 
      conf2 &\tabincell{l}{6.78} &\tabincell{l}{14.63} &\tabincell{l}{25.90} &\tabincell{l}{34.23} &\tabincell{l}{20.69} &\\ 
      conf3 &\tabincell{l}{10.63}&\tabincell{l}{25.90}&\tabincell{l}{34.23}&\tabincell{l}{75.02}&\tabincell{l}{45.40} &\\
      conf4 &\tabincell{l}{11.75}&\tabincell{l}{34.23}&\tabincell{l}{75.02}&\tabincell{l}{\textbf{109.05}}&\tabincell{l}{69.92} &\\
      conf5 &\tabincell{l}{5.83}&\tabincell{l}{20.69}&\tabincell{l}{45.40}&\tabincell{l}{69.92}&\tabincell{l}{49.72} \\
      \hline 
\end{tabular} 
\caption{Euclidean distance between different confidence level reviews (ICLR 2017-2019)}
\end{minipage}
\begin{minipage}{.5\textwidth}
\centering
\renewcommand\arraystretch{0.8}
\begin{tabular}{c|c|c|c|c} \hline 
       & conf1 & conf2 & conf3 & conf4  \\ \hline
      num &1104 &2554 &2659 & 1449& \\ 
      fra(\%)   &14.22 &32.89 &34.24 &18.66 & \\ 
      conf1 &\tabincell{l}{54.62}  &\tabincell{l}{69.81} &\tabincell{l}{62.04} &\tabincell{l}{\textbf{48.64}}&  \\ 
      conf2 &\tabincell{l}{69.81} &\tabincell{l}{\textbf{106.22}} &\tabincell{l}{101.55} &\tabincell{l}{76.71}&  \\ 
      conf3 &\tabincell{l}{62.04}&\tabincell{l}{101.90}&\tabincell{l}{101.55}&\tabincell{l}{79.95}&\\
      conf4 &\tabincell{l}{\textbf{48.64}}&\tabincell{l}{76.71}&\tabincell{l}{79.95}&\tabincell{l}{59.03}\\
      \hline 
\end{tabular} 
\caption{Euclidean distance between different confidence level reviews (ICLR 2020)}
\end{minipage}
\normalsize

\small
\newcommand{\tabincell}[2]{\begin{tabular}{@{}#1@{}}#2\end{tabular}}
\makeatletter\def\@captype{table}\makeatother
\begin{minipage}{.5\textwidth}
\centering
\renewcommand\arraystretch{0.8}
\begin{tabular}{c|c|c|c|c} \hline 
       & avg var & paper num & review num   \\ \hline
      (1) &\tabincell{l}{no value}  &\tabincell{l}{0} &\tabincell{l}{0} &  \\ 
      (2) &\tabincell{l}{0.39} &\tabincell{l}{4} &\tabincell{l}{12} &  \\ 
      (3) &\tabincell{l}{0.70}&\tabincell{l}{84}&\tabincell{l}{252}&\\
      (4) &\tabincell{l}{0.80}&\tabincell{l}{408}&\tabincell{l}{1224}&\\
       (5) &\tabincell{l}{1.24}&\tabincell{l}{34}&\tabincell{l}{102}&\\
      (1,2) &\tabincell{l}{\textbf{0.11}}  &\tabincell{l}{2} &\tabincell{l}{6} &  \\ 
      (1,3) &\tabincell{l}{0.44} &\tabincell{l}{12} &\tabincell{l}{36} &  \\ 
      (1,4) &\tabincell{l}{0.56}&\tabincell{l}{12}&\tabincell{l}{36}&\\
      (1,5) &\tabincell{l}{0.89}&\tabincell{l}{2}&\tabincell{l}{6}&\\
      (2,3) &\tabincell{l}{0.66}&\tabincell{l}{42}&\tabincell{l}{126}&\\
      (2,4) &\tabincell{l}{0.71}  &\tabincell{l}{103} &\tabincell{l}{309} &  \\ 
      (2,5) &\tabincell{l}{1.25} &\tabincell{l}{14} &\tabincell{l}{42} &  \\ 
      (3,4) &\tabincell{l}{0.76}&\tabincell{l}{817}&\tabincell{l}{2451}&\\
      (3,5) &\tabincell{l}{0.94}  &\tabincell{l}{128} &\tabincell{l}{384} &  \\ 
      (4,5) &\tabincell{l}{0.91} &\tabincell{l}{573} &\tabincell{l}{1719} &  \\ 
      (1,2,3) &\tabincell{l}{0.78}&\tabincell{l}{6}&\tabincell{l}{18}&\\
      (1,2,4) &\tabincell{l}{\textbf{1.30}}&\tabincell{l}{6}&\tabincell{l}{18}&\\
      (1,2,5) &\tabincell{l}{0.22}  &\tabincell{l}{1} &\tabincell{l}{3} &  \\ 
      (1,3,4) &\tabincell{l}{0.94} &\tabincell{l}{16} &\tabincell{l}{48} &  \\ 
      (1,3,5) &\tabincell{l}{0.89}  &\tabincell{l}{5} &\tabincell{l}{15} &  \\ 
      (1,4,5) &\tabincell{l}{0.83} &\tabincell{l}{7} &\tabincell{l}{21} &  \\ 
      (2,3,4) &\tabincell{l}{0.90} &\tabincell{l}{115} &\tabincell{l}{345} &  \\ 
      (2,3,5) &\tabincell{l}{0.76}  &\tabincell{l}{30} &\tabincell{l}{90} &  \\ 
      (2,4,5) &\tabincell{l}{0.89} &\tabincell{l}{43} &\tabincell{l}{129} &  \\ 
      (3,4,5) &\tabincell{l}{0.91} &\tabincell{l}{340} &\tabincell{l}{1020} &  \\ 
      (all) &\tabincell{l}{0.84}&\tabincell{l}{ }&\tabincell{l}{ }&\\
      \hline 
\end{tabular}

\caption{AVG review score variance of different combinations of different confidence level reviews (ICLR 2017-2019)}
\end{minipage}
\hsapce{ }
\begin{minipage}{.5\textwidth}
\centering
\renewcommand\arraystretch{0.8}

\begin{tabular}{c|c|c|c|c} \hline 
       & avg var & paper num & review num   \\ \hline
      (1) &\tabincell{l}{1.78}  &\tabincell{l}{23} &\tabincell{l}{69} &  \\ 
      (2) &\tabincell{l}{2.19} &\tabincell{l}{101} &\tabincell{l}{303} &  \\ 
      (3) &\tabincell{l}{1.83}&\tabincell{l}{118}&\tabincell{l}{354}&\\
      (4) &\tabincell{l}{\textbf{1.41}}&\tabincell{l}{30}&\tabincell{l}{90}&\\
      (1,2) &\tabincell{l}{\textbf{2.30}}  &\tabincell{l}{168} &\tabincell{l}{504} &  \\ 
      (1,3) &\tabincell{l}{1.81} &\tabincell{l}{144} &\tabincell{l}{432} &  \\ 
      (1,4) &\tabincell{l}{1.67}&\tabincell{l}{47}&\tabincell{l}{141}&\\
      (2,3) &\tabincell{l}{1.82}&\tabincell{l}{527}&\tabincell{l}{1581}&\\
      (2,4) &\tabincell{l}{1.95}  &\tabincell{l}{193} &\tabincell{l}{579} &  \\ 
      (3,4) &\tabincell{l}{2.05} &\tabincell{l}{258} &\tabincell{l}{774} &  \\ 
      (1,2,3) &\tabincell{l}{1.81}&\tabincell{l}{176}&\tabincell{l}{528}&\\
      (1,2,4) &\tabincell{l}{1.75}&\tabincell{l}{100}&\tabincell{l}{300}&\\
      (1,3,4) &\tabincell{l}{1.86}  &\tabincell{l}{93} &\tabincell{l}{279} &  \\ 
      (2,3,4) &\tabincell{l}{2.03} &\tabincell{l}{310} &\tabincell{l}{930} &  \\ 

      \hline 
\end{tabular}

\caption{AVG review score variance of different combinations of different confidence level reviews (ICLR 2020)}
\end{minipage}
\normalsize

\small
\newcommand{\tabincell}[2]{\begin{tabular}{@{}#1@{}}#2\end{tabular}}
\makeatletter\def\@captype{table}\makeatother
\begin{minipage}{.5\textwidth}
\centering
\renewcommand\arraystretch{0.8}
\begin{tabular}{c|c|c|c|c} \hline 
       & MJS & paper num & review num   \\ \hline
      (1,2) &\tabincell{l}{0.10}  &\tabincell{l}{18} &\tabincell{l}{38} &  \\ 
      (1,3) &\tabincell{l}{0.07} &\tabincell{l}{44} &\tabincell{l}{103} &  \\ 
      (1,4) &\tabincell{l}{0.08}&\tabincell{l}{48}&\tabincell{l}{115}&\\
      (1,5) &\tabincell{l}{\textbf{0.04}}&\tabincell{l}{16}&\tabincell{l}{35}&\\
      (2,3) &\tabincell{l}{0.08}&\tabincell{l}{215}&\tabincell{l}{492}&\\
      (2,4) &\tabincell{l}{0.11}  &\tabincell{l}{294} &\tabincell{l}{726} &  \\ 
      (2,5) &\tabincell{l}{0.14} &\tabincell{l}{101} &\tabincell{l}{224} &  \\ 
      (3,4) &\tabincell{l}{0.08}&\tabincell{l}{1374}&\tabincell{l}{3700}&\\
      (3,5) &\tabincell{l}{0.10}  &\tabincell{l}{537} &\tabincell{l}{1230} &  \\ 
      (4,5) &\tabincell{l}{0.08} &\tabincell{l}{1027} &\tabincell{l}{2717} &  \\ 
      (1,2,3) &\tabincell{l}{0.09}&\tabincell{l}{8}&\tabincell{l}{25}&\\
      (1,2,4) &\tabincell{l}{\textbf{0.19}}&\tabincell{l}{8}&\tabincell{l}{25}&\\
      (1,2,5) &\tabincell{l}{0.08}  &\tabincell{l}{2} &\tabincell{l}{7} &  \\ 
      (1,3,4) &\tabincell{l}{0.13} &\tabincell{l}{20} &\tabincell{l}{64} &  \\ 
      (1,3,5) &\tabincell{l}{0.07}  &\tabincell{l}{6} &\tabincell{l}{19} &  \\ 
      (1,4,5) &\tabincell{l}{0.13} &\tabincell{l}{7} &\tabincell{l}{21} &  \\ 
      (2,3,4) &\tabincell{l}{0.13} &\tabincell{l}{130} &\tabincell{l}{411} &  \\ 
      (2,3,5) &\tabincell{l}{0.12}  &\tabincell{l}{38} &\tabincell{l}{120} &  \\ 
      (2,4,5) &\tabincell{l}{0.16} &\tabincell{l}{54} &\tabincell{l}{180} &  \\ 
      (3,4,5) &\tabincell{l}{0.13} &\tabincell{l}{366} &\tabincell{l}{1134} &  \\ 
      \hline 
\end{tabular} 

\caption{MJS-divergence of different combinations of different confidence level reviews (ICLR 2017-2019)}
\end{minipage}
\hsapce{ }
\begin{minipage}{.5\textwidth}
\centering
\renewcommand\arraystretch{0.8}
\begin{tabular}{c|c|c|c|c} \hline 
       & MJS & paper num & review num   \\ \hline
       (1,2) &\tabincell{l}{0.28}  &\tabincell{l}{535} &\tabincell{l}{1300} &  \\ 
      (1,3) &\tabincell{l}{0.27} &\tabincell{l}{469} &\tabincell{l}{1110} &  \\ 
      (1,4) &\tabincell{l}{0.30}&\tabincell{l}{298}&\tabincell{l}{675}&\\
      (2,3) &\tabincell{l}{\textbf{0.26}}&\tabincell{l}{1139}&\tabincell{l}{2908}&\\
      (2,4) &\tabincell{l}{0.27}  &\tabincell{l}{693} &\tabincell{l}{1634} &  \\ 
      (3,4) &\tabincell{l}{0.28} &\tabincell{l}{744} &\tabincell{l}{1787} &  \\ 
      (1,2,3) &\tabincell{l}{0.34}&\tabincell{l}{213}&\tabincell{l}{660}&\\
      (1,2,4) &\tabincell{l}{0.34}&\tabincell{l}{138}&\tabincell{l}{434}&\\
      (1,3,4) &\tabincell{l}{0.38}  &\tabincell{l}{122} &\tabincell{l}{380} &  \\ 
      (2,3,4) &\tabincell{l}{\textbf{0.42}} &\tabincell{l}{364} &\tabincell{l}{1132} &  \\ 
      \hline 
\end{tabular}

\caption{MJS-divergence of different combinations of different confidence level reviews (ICLR 2020)}
\end{minipage}
\normalsize

%
%
%

\subsection{Covariance and Pearson Coefficient Between Different Confidence-Leveled Reviews}

\small
\newcommand{\tabincell}[2]{\begin{tabular}{@{}#1@{}}#2\end{tabular}}
\makeatletter\def\@captype{table}\makeatother
\begin{minipage}{.5\textwidth}
\centering
\renewcommand\arraystretch{0.6}
\begin{tabular}{c|c|c|c|c|c} \hline 
      & conf1 & conf2 & conf3 & conf4 & conf5 \\ \hline
      num & 74 & 455 & 2330 & 4612 & 1600 & \\ 
      fra(\%)   &0.80 &5.01 &25.67 &50.81 &17.71 & \\ 
      conf1 &\tabincell{l}{\textbf{0.42} \\ \textbf{0.45}} &\tabincell{l}{0.82 \\ 0.45}&\tabincell{l}{0.46 \\0.33} &\tabincell{l}{0.95 \\ 0.48} &\tabincell{l}{1.22 \\0.59}& \\ 
      conf2 &\tabincell{l}{0.82 \\ 0.45} &\tabincell{l}{0.77 \\ 0.46} &\tabincell{l}{0.68 \\ 0.36} &\tabincell{l}{0.77 \\ 0.38} &\tabincell{l}{0.75 \\ 0.36} &\\ 
      conf3 &\tabincell{l}{0.46 \\ 0.33}&\tabincell{l}{0.68 \\ 0.36}&\tabincell{l}{0.79 \\ 0.44}&\tabincell{l}{0.86 \\ 0.43}&\tabincell{l}{1.13 \\ 0.47} &\\
      conf4 &\tabincell{l}{0.95 \\ 0.48}&\tabincell{l}{0.77 \\ 0.38}&\tabincell{l}{0.86 \\ 0.43}&\tabincell{l}{1.10 \\ 0.47}&\tabincell{l}{1.41 \\ 0.51} &\\
      conf5 &\tabincell{l}{1.22 \\ 0.59}&\tabincell{l}{0.75 \\ 0.36}&\tabincell{l}{1.13 \\ 0.47}&\tabincell{l}{1.41 \\ 0.51}&\tabincell{l}{\textbf{1.67} \\ \textbf{0.51}} \\
      \hline 
\end{tabular} 
\caption{Covariance (up) and Pearson coefficient (down) between different confidence level reviews (ICLR 2017-2019).}
\end{minipage}
\hsapce{ }
\begin{minipage}{.5\textwidth}
\centering
\renewcommand\arraystretch{0.8}
\begin{tabular}{c|c|c|c|c} \hline 
       & conf1 & conf2 & conf3 & conf4  \\ \hline
      num &1104 &2554 &2659 & 1449& \\ 
      fra(\%)   &14.22 &32.89 &34.24 &18.66 & \\ 
      conf1 &\tabincell{l}{\textbf{1.03} \\ \textbf{0.26}}  &\tabincell{l}{1.41 \\ 0.32} &\tabincell{l}{1.90 \\ 0.40} &\tabincell{l}{1.80 \\ 0.38}&  \\ 
      conf2 &\tabincell{l}{1.41 \\ 0.32} &\tabincell{l}{1.50 \\ 0.32} &\tabincell{l}{1.86 \\ 0.40} &\tabincell{l}{2.05 \\ 0.41}&  \\ 
      conf3 &\tabincell{l}{1.90 \\ 0.40}&\tabincell{l}{1.86 \\ 0.40}&\tabincell{l}{\textbf{2.23} \\ \textbf{0.46}}&\tabincell{l}{1.93 \\ 0.39}&\\
      conf4 &\tabincell{l}{1.80 \\ 0.38}&\tabincell{l}{2.05 \\ 0.41}&\tabincell{l}{1.90 \\ 0.39}&\tabincell{l}{1.87 \\ 0.41}\\
      \hline 
\end{tabular} 
\caption{Covariance (up) and Pearson coefficient (down) between different confidence level reviews (ICLR 2020).}
\end{minipage}

\section{The Impact of Low Confidence Reviews to Acceptance Rate}

Suppose that we trust the opinions of reviewers, and that reviewers of papers are distributed uniformly among papers of different levels. We divide reviewers into two categories: professional and non-professional. A paper is reviewed by at least three reviewers. Therefore, it can be divided into three cases. First, one situation is when all reviewer levels are non-professional. Second, all reviewer levels are professional. Finally, professional and non-professional reviewers jointly review a paper. The scoring variance can indicate that there is a difference between different confidence level reviews, but sometimes does not. For example, the two review scores are 2 and 4 and the two review scores are 4 and 6. The variance between them is  same. The difference between non-professionals and professionals cannot be explained, and if variance is used, the paper's score cannot be related to the reviewer's professionalism. Therefore, we use positive and negative differences when analyzing different levels of reviewers.

Positive and negative difference: The calculation method of positive and negative difference is as follows. For example, an article is assigned three reviewers, a professional reviewer is divided into 4, and two non-professional reviewers are divided into 6 and 5. Then the positive and negative difference is (4) / 1- (6 + 5) / 2.

The reason for using positive and negative differences is that the average score of a paper can measure whether a paper is accepted or not, which is obvious. However, sometimes the reason for the high average may be high professional scores, low non-professional scores, low professional scores, high non-professional scores, or high professional and non-professional reviewers. At this time, you need to use an indicator to see if the professional reviewer's scoring plays a greater role, that is, if the average reviewer is equal, whether a high professional reviewer's scoring is more conducive to the paper Admission. In the case of large scoring differences, whether professional reviewers' scoring is more influential. This is why the design difference between positive and negative is used to judge the scoring differences between professional and non-professional reviewers. A positive and negative difference indicates that professionals are more optimistic, and a negative and negative difference indicates that professionals are not optimistic, and the same is true.

It is important to note that when all reviewers of a paper are professionals or non-professionals, the positive and negative difference is zero. Because the difference between positive and negative represents the differences between professional and non-professionals in giving thesis. When the reviewers are all professionals or non-professionals, this is no longer a condition for calculating the positive and negative difference, so the positive and negative difference is set to 0 at this time.

\begin{figure}
     \centering 
     \includegraphics[width=4in]{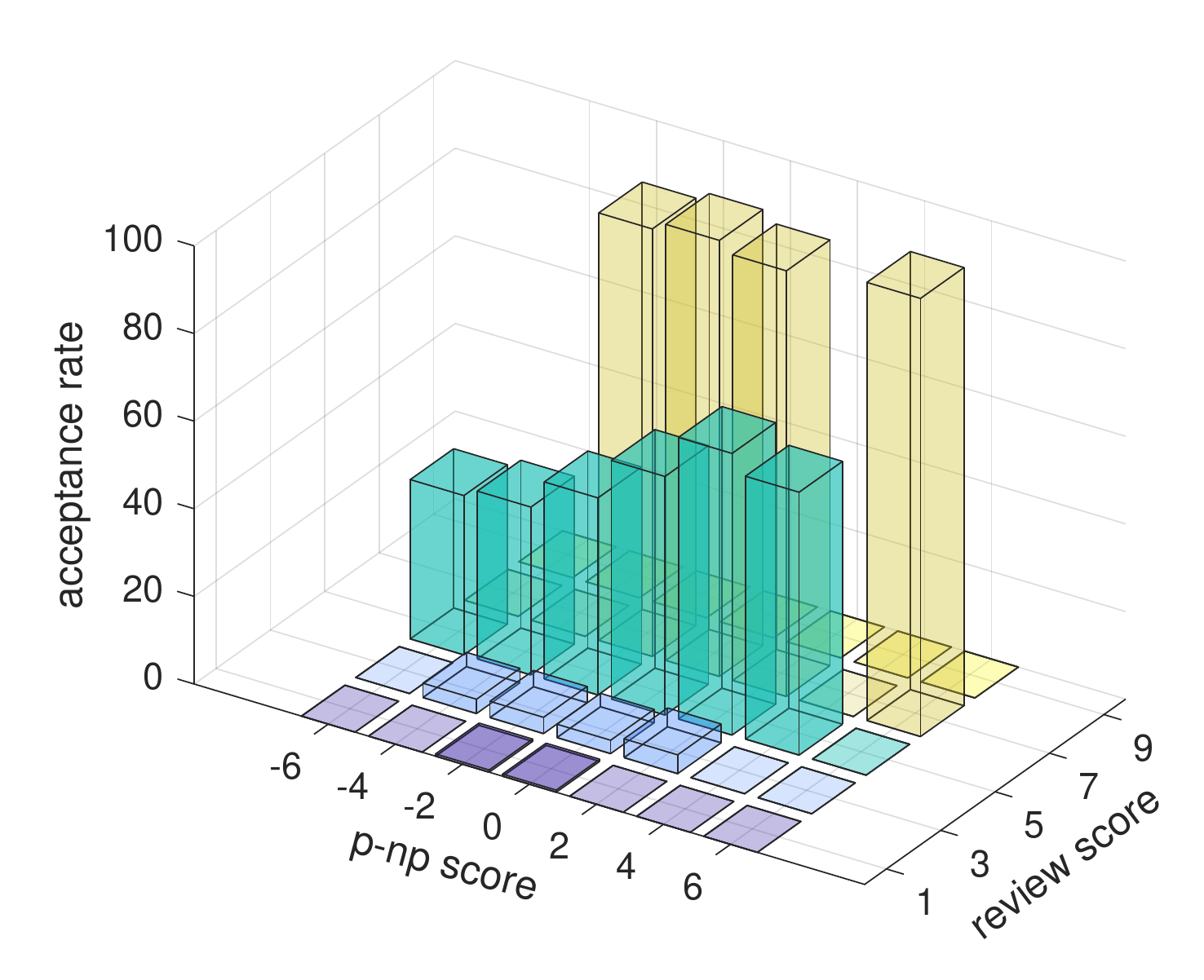}
     \caption{The impact of low confidence reviews to the acceptance rate} 
     \label{fig:pnp} 
 \end{figure}

\section{Text Sentiment Analysis Details}

\newcommand{\tabincell}[2]{\begin{tabular}{@{}#1@{}}#2\end{tabular}}
\begin{tabular}{|c|c|} \hline 
     id  & sentence  \\ \hline
     
      1 &\tabincell{l}{It defines the samples whose average probability on assigned label in recent q iterations is largest\\ among all labels as memorized samples, in the sense of the network memorize these samples.}    \\ \hline 
      2 &\tabincell{l}{Then authors proposed two stage method which firstly early-stops at minimum validation error (or\\ memorized rate), and then trains on maximal safe set that gathers memorized samples. }  \\ \hline 
      3 &\tabincell{l}{The experiments compared several state-of-art approaches and showed that the proposed method\\ benefits from early-stopping and safe set. }\\\hline 
      4 &\tabincell{l}{Authors also showed that the prestopping idea can also be used to improve other approaches.}\\\hline 
      5 &\tabincell{l}{Pros: The proposed method achieves better performance than state-of-art methods. Authors\\ have good experiments which evaluate on multiple datasets and algorithms. }    \\ \hline 
      6 &\tabincell{l}{Authors also investigate the relation between model complexity and performance of co-teaching+ }  \\ \hline 
      7 &\tabincell{l}{Cons: Many recent papers indicate the “error-prone period”, authors should include related works\\ about early-stopping on label noise training.}\\\hline 
      8 &\tabincell{l}{
      Although the method achieves good performance, since the idea is a bit straightforward especially\\ after exploring above papers, I am slightly worried about novelty of the ideas.}\\
      \hline 
\end{tabular}

\begin{center}
    \begin{tabular}{|c|c|c|c|} \hline 

\centering
      id & Exist aspects & Sentiment analysis &Auto tag  \\ \hline\hline 
     
      1 &\tabincell{l}{no}  &\tabincell{l}{} &\tabincell{l}{} \\ \hline
      2 &\tabincell{l}{no} &\tabincell{l}{} &\tabincell{l}{}  \\ \hline
      3 &\tabincell{l}{yes}&\tabincell{l}{1}&\tabincell{l}{} \\ \hline
      4 &\tabincell{l}{yes}&\tabincell{l}{1}&\tabincell{l}{} \\  \hline
      5 &\tabincell{l}{yes}  &\tabincell{l}{2} &\tabincell{l}{2}   \\ \hline
      6 &\tabincell{l}{no} &\tabincell{l}{2} &\tabincell{l}{2}  \\ \hline
      7 &\tabincell{l}{yes}&\tabincell{l}{0}&\tabincell{l}{0} \\ \hline
      8 &\tabincell{l}{no}&\tabincell{l}{0}&\tabincell{l}{0} \\ \hline
      
\end{tabular} 
\end{center}
This is an example taken randomly from the 2020 review. Break this review into periods. It is divided into 8 sentences. First match these sentences with words in 5 dimensions. The match was successful for yes. No match is no. For the automatic labeling method, one method is to match angles and then perform emotional word matching. Another method is to use emotional words first. For example, pros and cons in this paragraph. Think of a positive expression between pros and cons. After cons is a negative expression. These two methods can reduce the workload of manual labeling. In the second column, predictions are made for sentences where the human annotation and angle match. The results showed as expected.

%
Here we provide a typical review sample that contains ``pros and cons'':

pros:

1. the proposed method shows the ability to learn the nested distributions with the help of hierarchical structure information. it is a reasonable way to model the general multimodal i2i. I think the authors work in the correct direction. 

2. to model the partial order relation in the hierarchy, the authors borrow the thresholded divergence technique from the natural language processing field (athiwaratkun \& wilson (2018)) and use the kl divergence.

cons: 

1. some figures are hard to understand without looking at the text. for example, in figure 1, the caption does not explain the figure well. what does each image, the order, and the different sizes mean? as to figure 3, the words “top left image”, “right purple arrows” are a bit confusing. 

2. the “coarse to fine conditional translation” section describes the conditional translation in the shallow layers. i suggest mentioning it in previous sections for easy understanding. 

3. as to the t-sne visualization in figure 9, different methods seem to use different n-d to 2-d mapping functions. this may lead to an unfair comparison. suggestions: 1. the authors use the pre-trained classification network vgg for feature extraction and then train dedicated translators based on these features. i wonder if the authors also tried finetuning vgg on the two domains or training an auto-encoder on the two domains. the domain-specific knowledge may help to improve the results and alleviate the limitations presented in the paper, e.g. background of the object is not preserved, missing small instances or parts of the object due to invertible vgg-19.

We perform sentiment analysis for each aspect using a pre-trained text model ELECTRA \cite{Clark2020ELECTRA}. The hyper-parameters of ELECTRA are listed as follows.
\begin{table}[]
    \centering
    \begin{tabular}{c|c|c|c|c|c} \hline
         & novelty & motivation & experiment & related work & presentation \\ \hline
      epoch &7 &7 &3 & 10&10 \\ 
      batchsize  &64 &64 &64 &64&64 \\ 
      maxlen  &256 &256 &256 &256&256 \\ 
      \hline 
    \end{tabular}
    \caption{Hyper-parameters of ELECTRA}
    \label{tab:my_label}
\end{table}

\bibliographystyle{splncs04}
\bibliography{ref}

%% file: intro.tex
\section{Introduction}
\label{sec:intro}

Peer review is a widely adopted quality control mechanism in which the value of scientific paper is assessed by several reviewers with a similar level of competence. The primary role of the review process is to decide which papers to publish and to filter information, which is particularly true for a top conference that aspires to attract a broad readership to its papers. The novelty, significance, and technical flaws are identified by reviewers, which can help PC chair make the final decision. 

Anonymous peer review (no matter single-blind or double-blind), despite the criticisms often leveled against it, is used by the great majority of computer science conferences, where the reviewers do not identify themselves to the authors. It is understandable that some authors are uncomfortable with a system in which their identities are known to the reviewers while the latter remain anonymous. Authors may feel themselves defenseless against what they see as the arbitrary behavior of reviewers who cannot be held accountable by the authors for unfair comments. On the other hand, apparently, there would be even more problems if letting authors know their reviewers' identities. Reviewers would give more biased scores for fear of retaliation from the more powerful colleagues. Given this contradiction, opening up the reviews to public seems to be a good solution. The openness of reviews will force reviewers to think more carefully about the scientific issues and to write more thoughtful reviews, since PC chairs know the identities of reviewers and bad reviews would affect their reputations.


OpenReview\footnote{https://openreview.net/} is such a platform that aims to promote openness in peer review process. The paper, (meta) reviews, rebuttals, and final decisions are all released to public. Colleagues who do not serve as reviewers can judge the paper's contribution as well as judge the fairness of the reviews by themselves. Reviewers will have more pressure under public scrutiny and force themselves to give much fairer reviews. On the other hand, previous works on peer-review analysis  \cite{10.1145/2979672,DBLP:journals/corr/abs-1804-09635,DBLP:conf/sigir/WangW18,DBLP:conf/emnlp/LiZYW19,DBLP:conf/acl/GhosalVEB19,DBLP:conf/nips/StelmakhSS19} are often limited due to the lack of rejected paper instances and their corresponding reviews. Given these public reviews (for both accepted papers and rejected ones), studies towards multiple interesting questions related to peer-review are made available.

Given these public reviews, there are multiple interesting questions raised that could help us understand the effectiveness of the public-accessible double-blind peer review process: a) As known, AI conferences have extremely heavy review burden in 2020 due to the explosive number of submissions \cite{twitter}. These AI conferences have to hire more non-experts to involve in the double-blind review process. How is the impact of these non-experts on the review process (Sec. \ref{sec:review:nonpro})? b) Reviewers often evaluate a paper from multiple aspects, such as motivation, novelty, presentation, and experimental design. Which aspect has a decisive role in the review score (Sec. \ref{sec:review:weakpoint})? c) The OpenReview platform provides not only the submission details (e.g., title, keywords, and abstract) of accepted papers but also that of rejected submissions, which allows us to perform a more fine-grained cluster analysis. Given the fine-grained hierarchical clustering results, is there significant difference in the acceptance rate of different research fields (Sec. \ref{sec:review:cluster})? d) A posterior quantitative method for evaluating papers is to track their citation counts. A high citation count often indicates a more important, groundbreaking, or inspired work. OpenReview releases not only the submission details of accepted papers but also that of rejected submissions. The rejected submissions might be put on arXiv.org or published in other venues to still attract citations. This offers us opportunities to analyze the correlation between review scores and citation numbers. Is there a strong correlation between review score and citation number for a submission (Sec. \ref{sec:cite:correlation})?  e) Submissions might be posted on arXiv.org before the accept/reject notification, which might be the rejected ones from other conferences. They are special because they could be improved according to the rejected reviews and their authors are not anonymous. Are these submissions shown higher acceptance rate (Sec. \ref{sec:review:arxiv})? 


In this paper, we collect 5,527 (accepted and rejected) submissions and their 16,853 reviews from ICLR 2017-2020 venues\footnote{International Conference on Learning Representations. https://iclr.cc/} on the OpenReview platform as our main corpus. By acquiring deep insights into these data, we have several interesting findings and aim to answer the above raised questions quantitatively. Our submitted supplementary file also includes more data analysis results. We expect to introduce more discussions on the effectiveness of peer-review process and hope that treatment will be obtained to improve the peer-review process.




%% file: dataset.tex
\section{Dataset}
\label{sec:dataset}

ICLR has used OpenReview to launch double-blind review process for 8 years (2013-2020). Similar to other major AI conferences, ICLR adopts a reviewing workflow containing double-blind review, rebuttal, and final decision process. After paper assignment, typically three reviewers evaluate a paper independently. After the rebuttal, reviewers can access the authors' responses and other peer reviews, and accordingly modify their reviews. The program chairs then write the meta-review for each paper to make the final accept/reject decision according to the three anonymous reviews. Each official review mainly contains a review score (integer between 1 and 10), a reviewer confidence level (integer between 1 and 5), and the detailed review comments. The official reviews and meta-reviews are all open to the public on the OpenReview platform. Public colleagues can also post their reviews on OpenReview. We will present the collected dataset of submissions and reviews from OpenReview, these submissions' citation data from Google Scholar, and their non-peer-reviewed versions from arXiv.org\footnote{These datasets and the source code for the analysis experiment are available at https://github.com/Seafoodair/Openreview/}. 

\begin{table}[t]
	\vspace{-0.1in}
	\caption{Statistics of ICLR reviews dataset}
	\vspace{-0.05in}
	\label{tab:data}
	\centering
	\small
	{
	\begin{tabular}{p{0.6in} p{0.8in}  p{0.8in}  p{0.8in}  p{0.8in}  p{0.8in}}
		\hline
		{\textbf{year}} &
		{\textbf{\#papers}} &
		{\textbf{\#authors}}&
		{\textbf{accept rate}} &
		{\textbf{\#reviews}} &
		{\textbf{review len.}} \\
		\hline\hline
		{\textbf{2017}} & 489 & 1,417 & 50.1\% & 1,495 & 295.11 \\
		\hline
		{\textbf{2018}} & 939 & 2,882 & 49.0\% & 2,849 & 372.07 \\
		\hline
		{\textbf{2019}} & 1,541 & 4,332 & 32.5\% & 4,733 & 403.22 \\
		\hline
		{\textbf{2020}} & 2,558 & 7,765 & 26.5\% & 7,766 & 407.08 \\
		\hline
		{\textbf{total}} & 5,527 & 16,396 & 39.5\%  & 16,843 & 369.37 \\
		\hline
	\end{tabular}
	}
	\vspace{-0.2in}
	\normalsize
\end{table}

\Paragraph{Submissions and Reviews} We have collected 5,527 submissions and 16,853 official reviews from ICLR 2017-2020 venues on the OpenReview platform. We only use the review data since 2017 because the submissions before 2017 is too few. Though a double-blind review process is exploited, the authors' identities of the rejected submissions are also released after decision notification. Thus, we can also access the identity information for each rejected submission, which is critical in most of our analysis. Some statistics of the reviews data are listed in Table \ref{tab:data}, in which \textbf{review len.} indicates the average length of reviews. 

\Paragraph{Citations} In order to investigate the correlation between review scores and citation numbers, we also collect the citation information from Google Scholar for all the 1,183 accepted papers from 2017 to 2019. Since the rejected submissions might be put on arXiv.org or published in other venues, they might also attract citations. We also collect the citation information for 2,054 rejected submissions that have been published elsewhere (210 for 2017, 324 for 2018, 493 for 2019, and totally 1027 rejected papers). All the citation numbers are gathered up to 20 Jan. 2020. We do not collect citation information of ICLR 2020 papers because they have not yet accumulated enough citations yet.

\Paragraph{arXiv Submissions} In order to investigate whether the submissions that have been posted on arXiv.org before notification have a higher acceptance rate, we also crawl the arXiv versions of ICLR 2017-2020 submissions if they exist. We record the details of an arXiv preprint if its title matches an ICLR submission title. Note that, their contents might be slightly different. We totally find 1,158 matched arXiv papers and 948 among them were posted before notification (178/150 for 2017, 103/79 for 2018, 420/303 for 2019, and 457/416 for 2020) up to 18 Feb 2020.

%% file: proreview.tex
\section{Results Learned from Open Reviews}

\subsection{How is the Impact of Non-Expert Reviewers?}
\label{sec:review:nonpro}


\begin{table}[t]
\vspace{-0.1in}
    \caption{Statistics of different confidence level reviews}
    \vspace{-0.05in}
    \label{tab:conf}
    \centering
    \small
    \begin{tabular}{c | c|c|c|c|c| c|c|c|c}
    \hline
 \multirow{2}{*}{} & \multicolumn{5}{c|}{2017-2019} & \multicolumn{4}{c}{2020} \\
 \cline{2-10}
     & \textbf{level1} & \textbf{level2} & \textbf{level3} & \textbf{level4} & \textbf{level5} & \textbf{level1} & \textbf{level2} & \textbf{level3} & \textbf{level4}\\
\hline
\textbf{\#reviews} & 74 & 455 & 2,330 & 4,612 & 1,600 & 1,104 & 2,554 & 2,659 & 1,449\\
\hline
\textbf{fraction} & 0.80\% & 5.01\% & 25.67\% & 50.81\% & 17.71\% & 14.22\% & 32.89\% & 34.24\% & 18.66\%\\
\hline
\end{tabular}
\vspace{-0.2in}
\normalsize
\end{table}

Due to the extensively increasing amount of submissions, ICLR 2020 hired much more reviewer volunteers. There were complaints about the quality of reviews (47\% of the reviewers have not published in the related areas \cite{twitter}). Similar scenarios have been observed in other AI conferences, such as NIPS, CVPR, and AAAI. Many authors complain that their submissions are not well evaluated because the assigned ``non-expert'' reviewers lack of enough technical background and cannot understand their main contributions. How is the impact of these ``non-experts'' on the review process? In this subsection, we aim to answer the question through quantitative data analysis. 
\begin{figure}
 \vspace{-0.4in}
 	\centerline{
 	\subfloat[2017]{\includegraphics[width=1.25in]{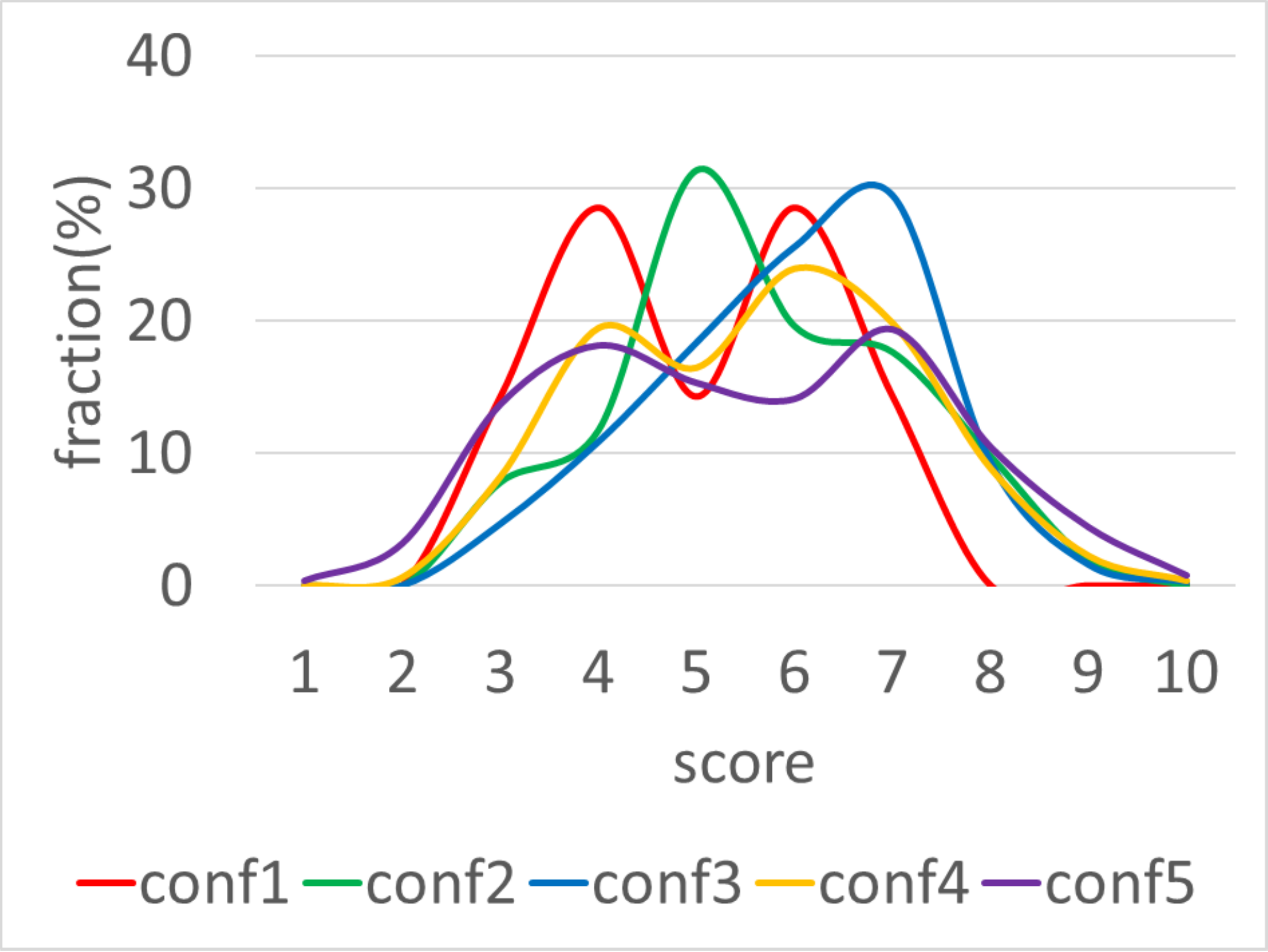}
     \label{fig:reviewer:2017}}
     \subfloat[2018]{\includegraphics[width=1.25in]{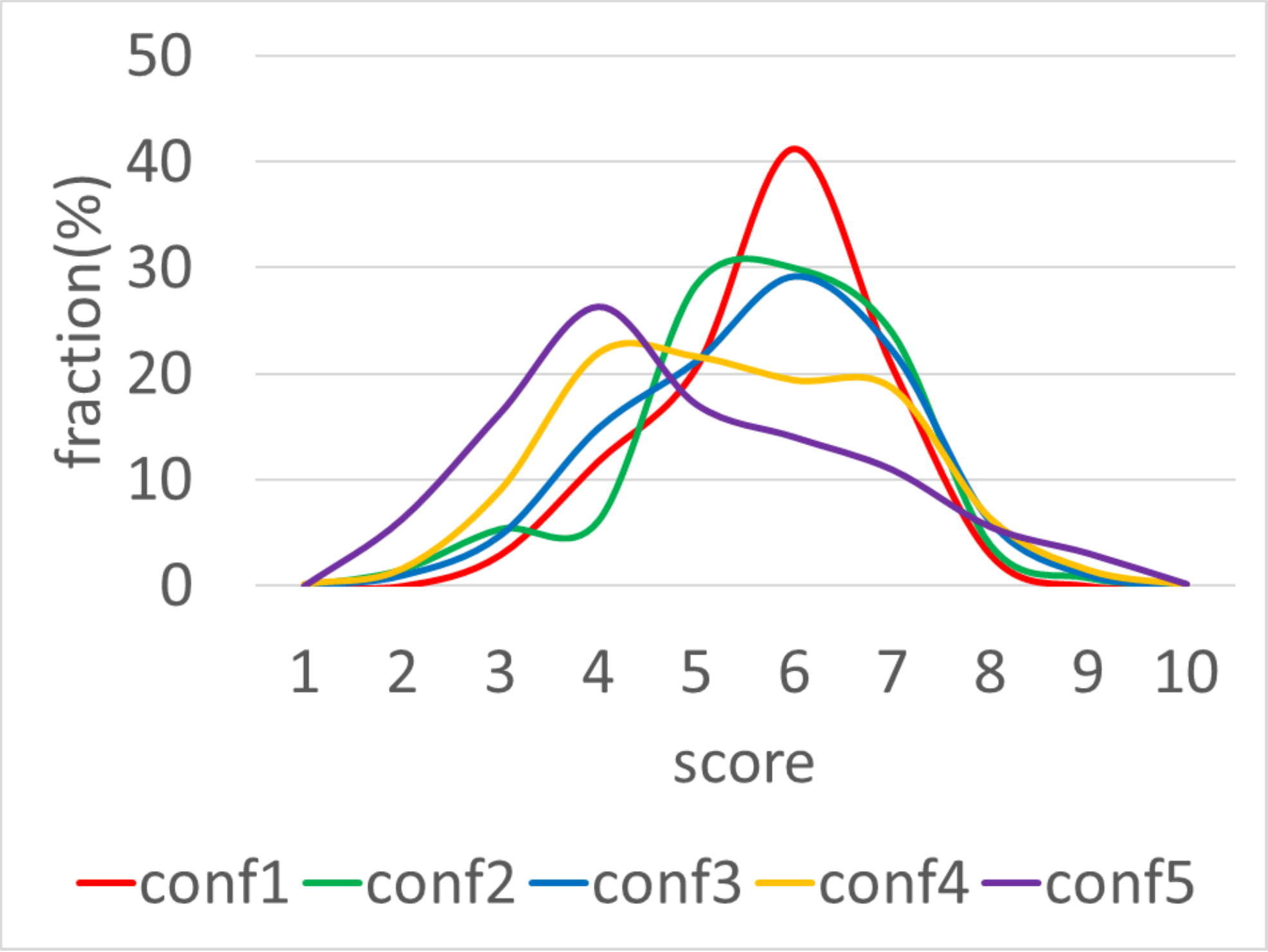}
     \label{fig:reviewer:2018}}
     \subfloat[2019]{\includegraphics[width=1.25in]{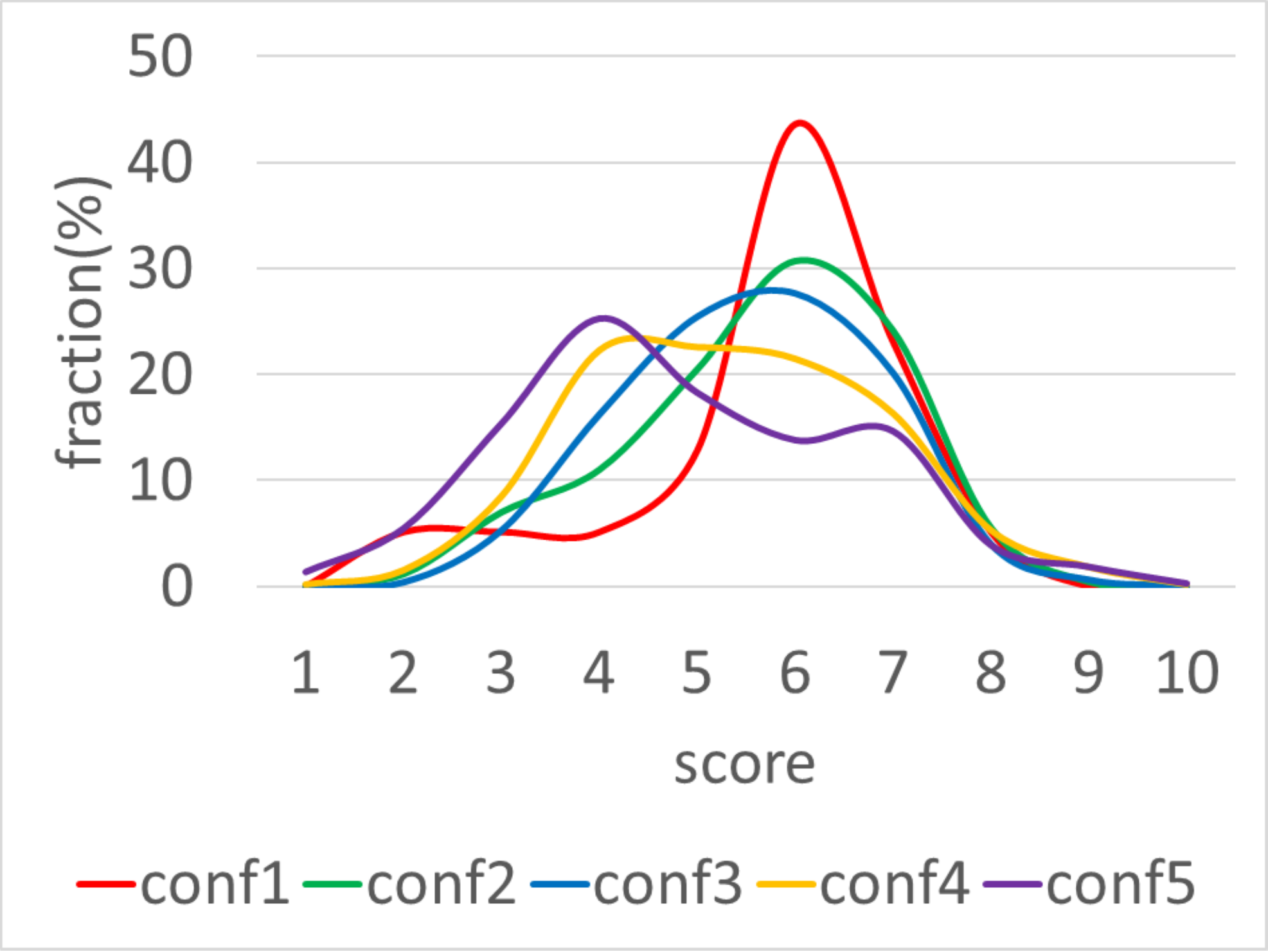}
    \label{fig:reviewer:2019}}
     \subfloat[2020]{\includegraphics[width=1.25in]{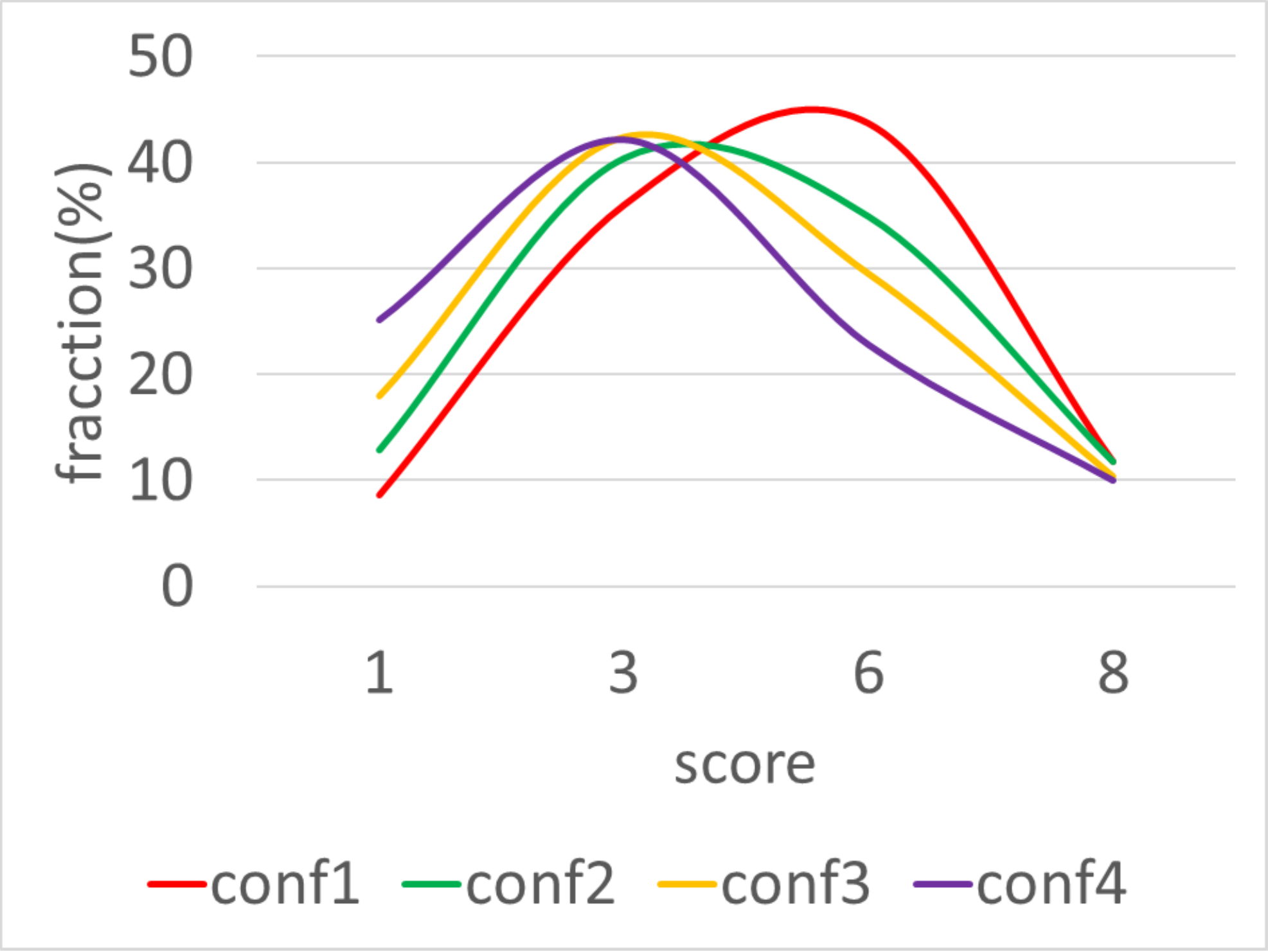}
     \label{fig:reviewer:2020}}
     }
    \vspace{-0.05in}
	\caption{The review score distributions of different confidence level (conf) reviews}
 	\label{fig:reviewer}
 \vspace{-0.3in}
\end{figure}

\Paragraph{Review Score Distribution} For ICLR 2017-2019, reviewer gives a review score (integer) from 1 to 10, and is asked to select a confidence level (integer) between 1 and 5. For ICLR 2020, reviewer gives a rating score in \{1, 3, 6, 8\} and should select an experience assessment score (similar to confidence score) between 1 and 4. We divide the reviews into multiple subsets according to their confidence levels. Fig. \ref{fig:reviewer} shows the smoothed review score distributions for each subset of reviews. For 2018-2020, we consistently observe that the scores of reviews with confidence level 1 and 2 are likely to be higher than those reviews with confidence level 4 and 5. The trend of ICLR 2017 is not clear because it contains too few samples to be statistically significant (e.g., only 7 level-1 reviews). In 2017-2019, the lowest confidence level reviews has an average review score 5.675, while the highest confidence level reviews has an average review score 4.954. In 2020, the numbers for the lowest and highest confidence level reviews are 4.726 and 3.678, respectively. Our results show that the`low-confidence reviewers (e.g., level 1 and 2) tend to be more tolerant because they may be not confident about their decision, while the high-confidence reviewers (e.g., level 4 and 5) tend to be more tough and rigorous because they may be confident in the identified weakness.

\begin{figure}
\vspace{-0.4in}
	\centerline{
	\subfloat[2017]{\includegraphics[width=1.2in]{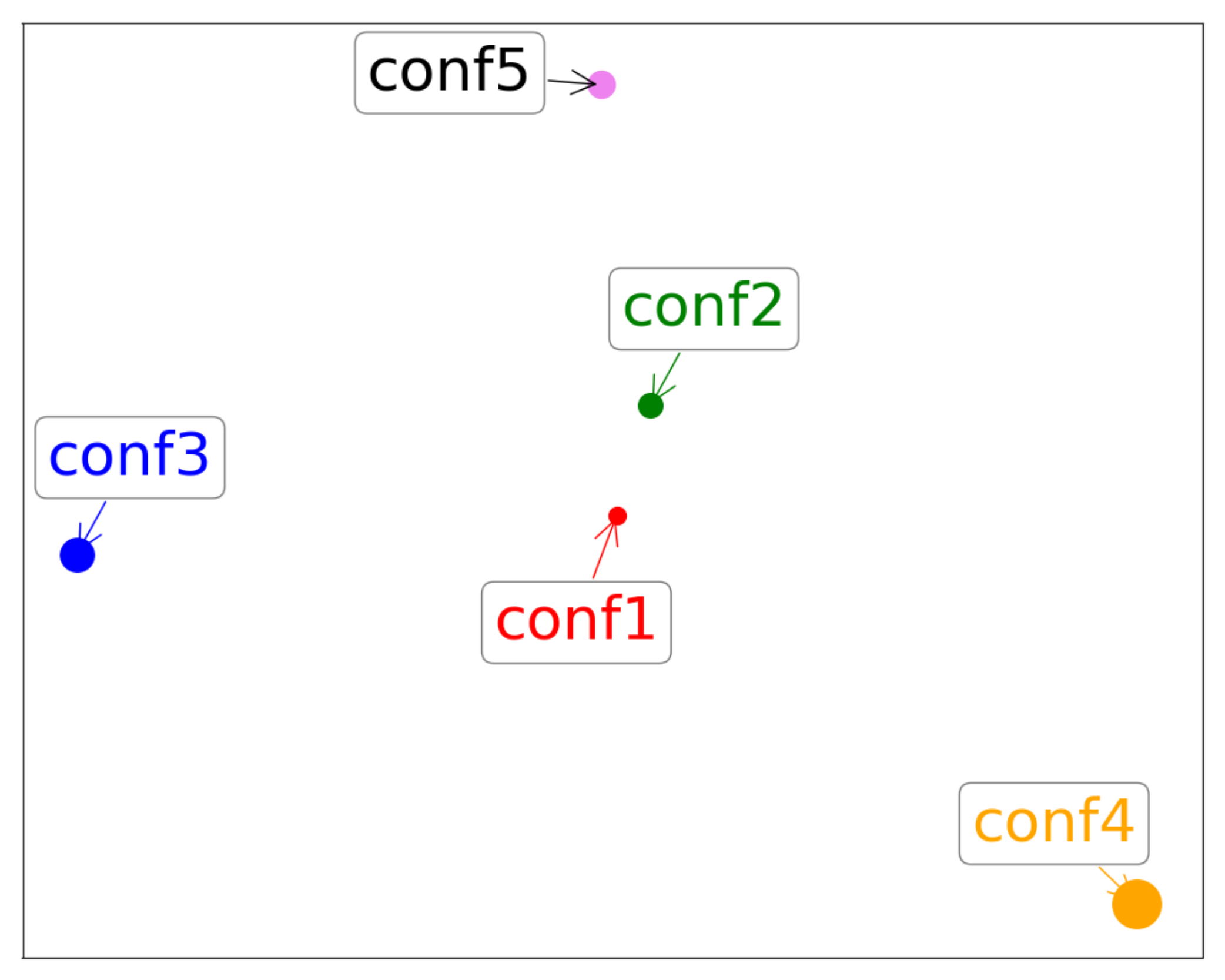}
    \label{fig:reviewer:2017}}
    \subfloat[2018]{\includegraphics[width=1.2in]{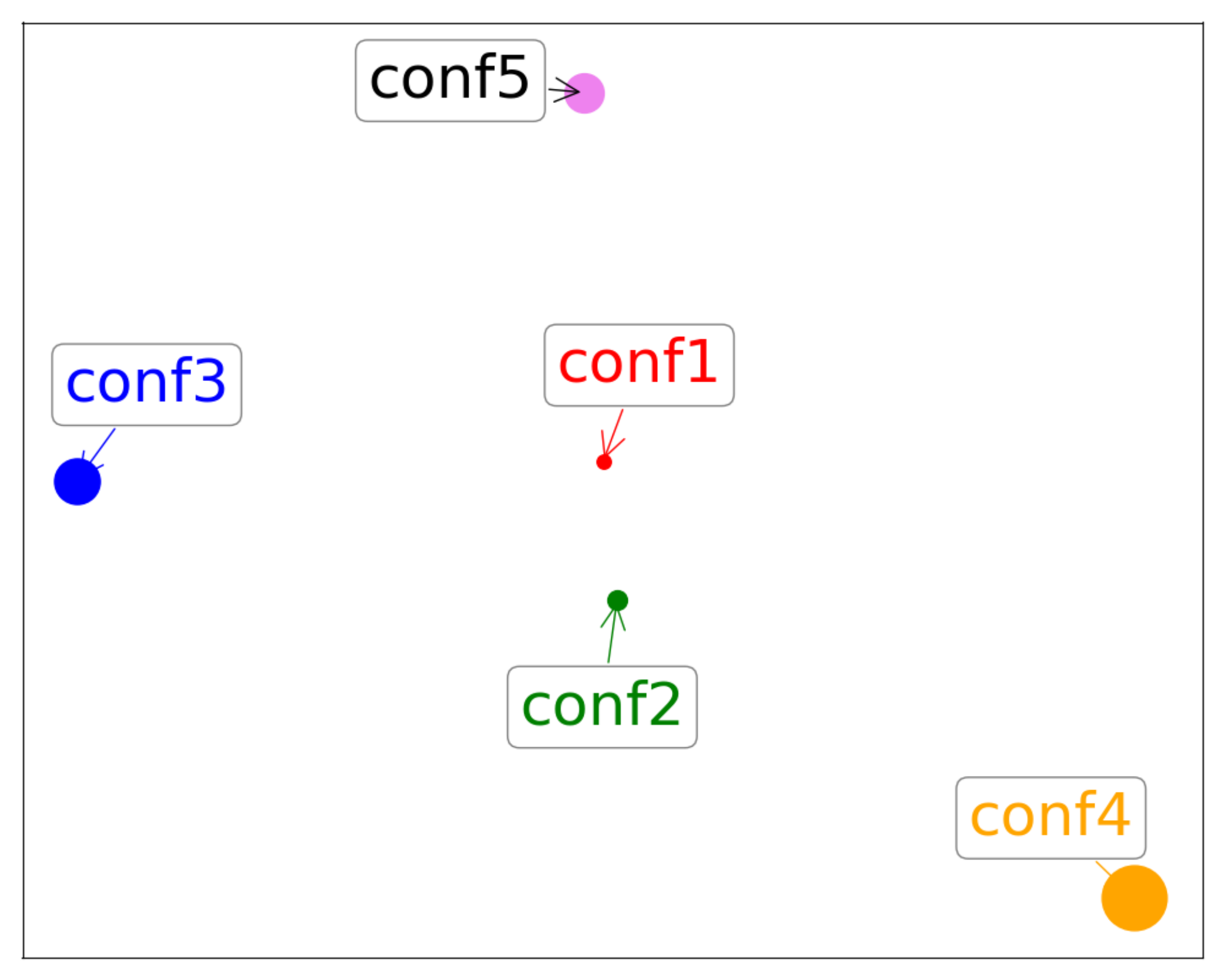}
    \label{fig:reviewer:2018}}
    \subfloat[2019]{\includegraphics[width=1.2in]{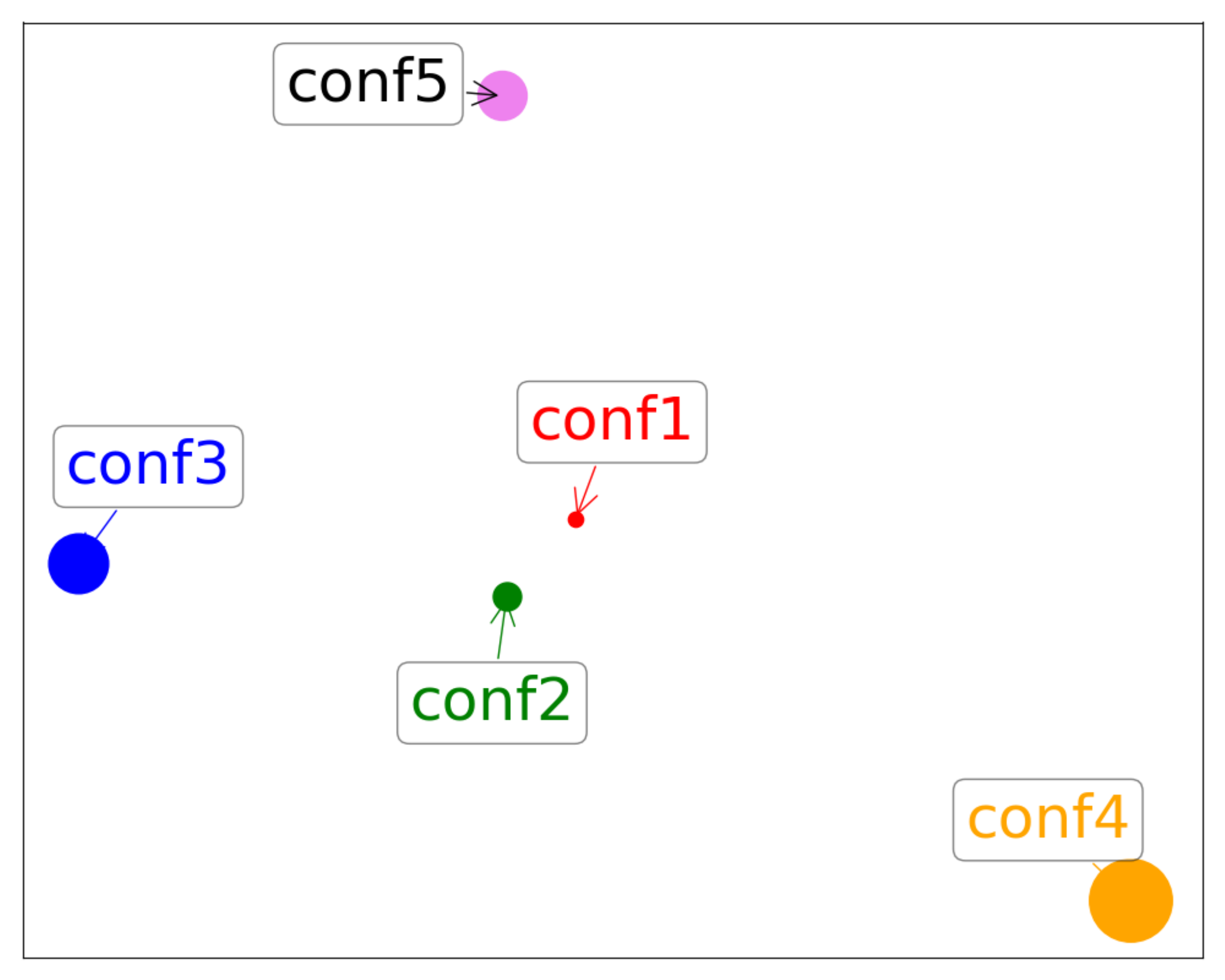}
    \label{fig:reviewer:2019}}
    \subfloat[2020]{\includegraphics[width=1.2in]{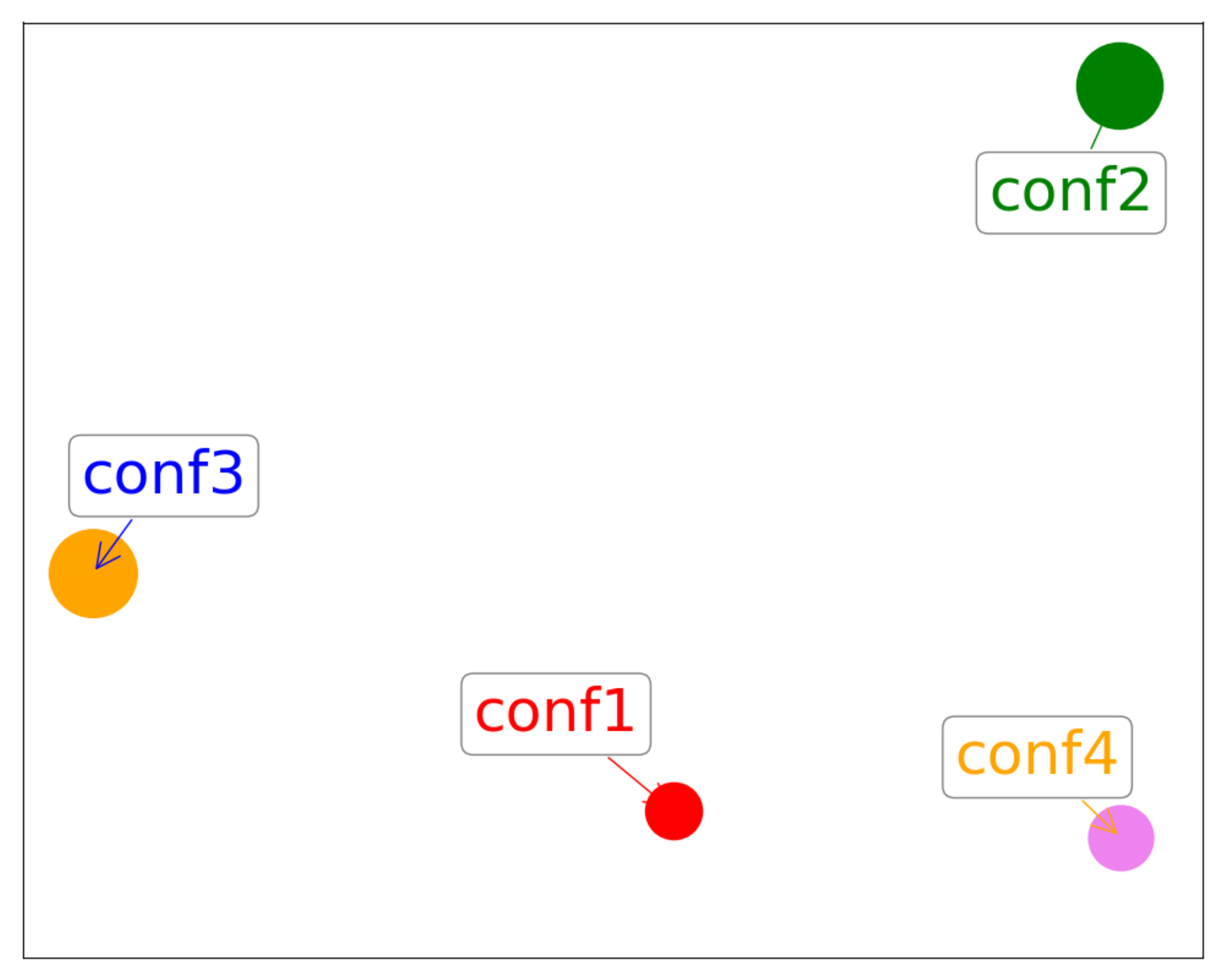}
    \label{fig:reviewer:2020}}
    }
    \vspace{-0.05in}
	\caption{The visualized layout of groups of reviews with different confidence scores. Each point indicates a group of reviews with a specific confidence level (abbrv. conf). The size of point indicates the relative number of reviews in that group. The distance between two points indicates the divergence of review scores between two groups.}
	\label{fig:visuallayout}
\vspace{-0.3in}
\end{figure}

\Paragraph{Divergence Reflected by Euclidean Distance} On the other hand, peoples are worrying that non-expert reviewers are not competent to give a fair evaluation of a submission (e.g., fail to identify key contributions or fail to identify flaws) and will ruin the reputation of top conferences \cite{twitter}. Particularly, these non-expert reviewers may have different opinions with the expert reviewers regarding the same paper. Actually, opinion divergence commonly exists between reviewers in the peer-review process. Each paper is typically assigned to 3 reviewers. These 3 reviewers may have significantly different review scores. In order to illustrate the difference between the reviews with different confidence scores, we first compute the euclidean distance $DIS(l_i,l_j)$ between between group $l_i$ and group $l_j$ as follows. Let $R_{l_i,l_j}$ be the set of paper IDs, where each paper concurrently has both confidence-$l_i$ review(s) and confidence-$l_j$ review(s). Let $\overline{s_p^i}$ be paper $p$'s average review score from $l_i$-confidence reviews. Then, the distance between the group of confidence-$l_i$ reviews and that of confidence-$l_j$ reviews is:
\begin{equation}
\label{eq:euclidean}
DIS(l_i,l_j)=\sqrt{\sum_{p\in R_{l_i,l_j}}\Big(\overline{s_p^i}-\overline{s_p^j}\Big)^2}.
\end{equation}
After computing the distance betwenn each pair of groups, we can construct a distance matrix. According to the distance matrix, we use t-SNE \cite{maaten2008visualizing} to plot the visualized layout of different groups of reviews of each year in Fig. \ref{fig:visuallayout}. For ICLR 2017-2019, we can see similar layout, where conf1 reviews are close to conf2 reviews in the central part and the groups of conf3, conf4, and conf5 locate around. The group of conf4 reviews is far apart from the most professional reviews (conf5). In ICLR 2020, there are 4 confidence levels. Surprisingly, we observe that the most professional reviews (conf4) and least professional reviews (conf1) are closest to each other. Conf2 reviews and conf3 reviews are both far apart from conf4 reviews.

\begin{figure}
\vspace{-0.3in}
	\centerline{
	\subfloat[2017-2019]{\includegraphics[width=2.2in]{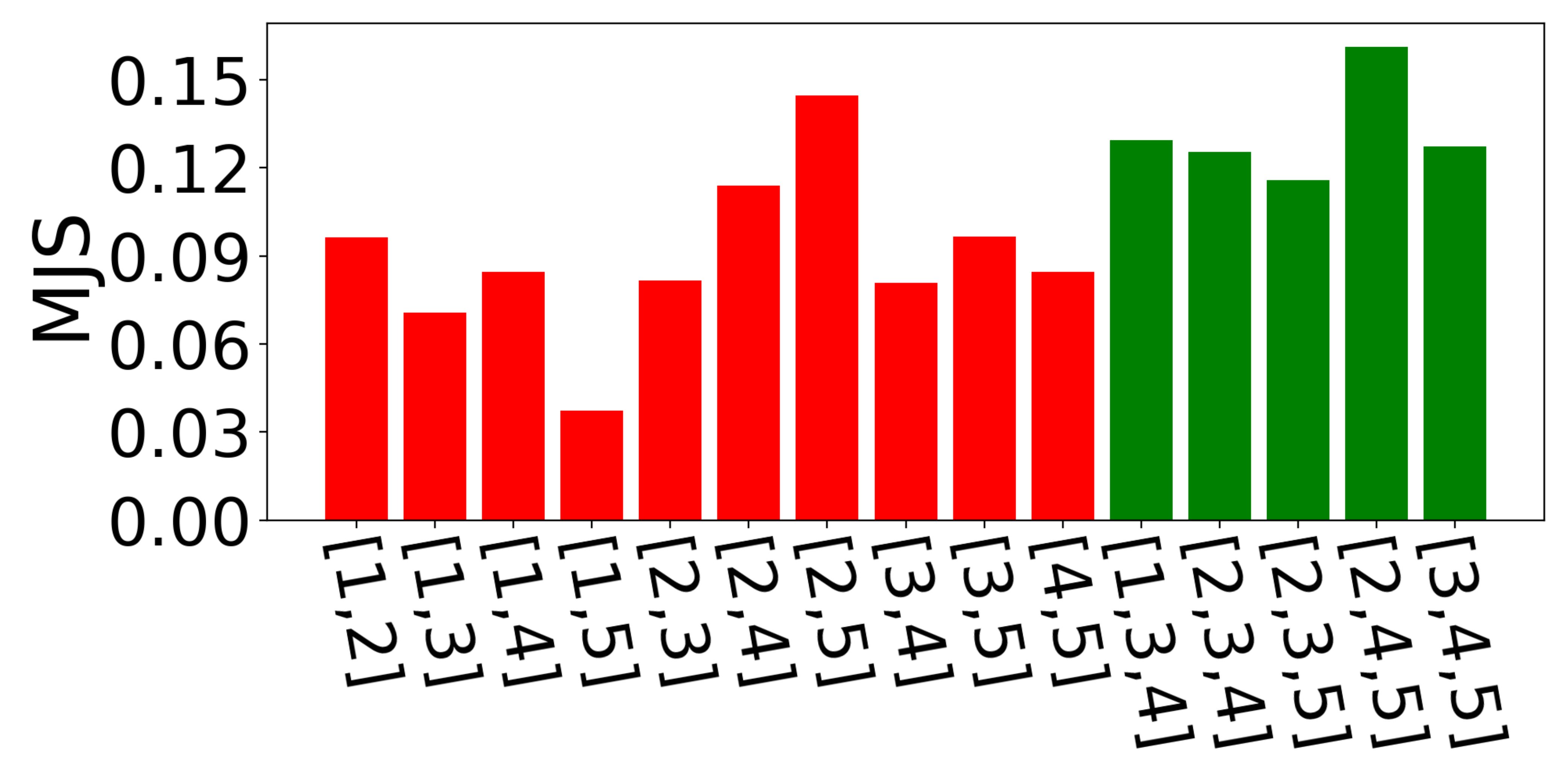}
    \label{fig:reviewer:2017-2019}}
    \subfloat[2020]{\includegraphics[width=2.2in]{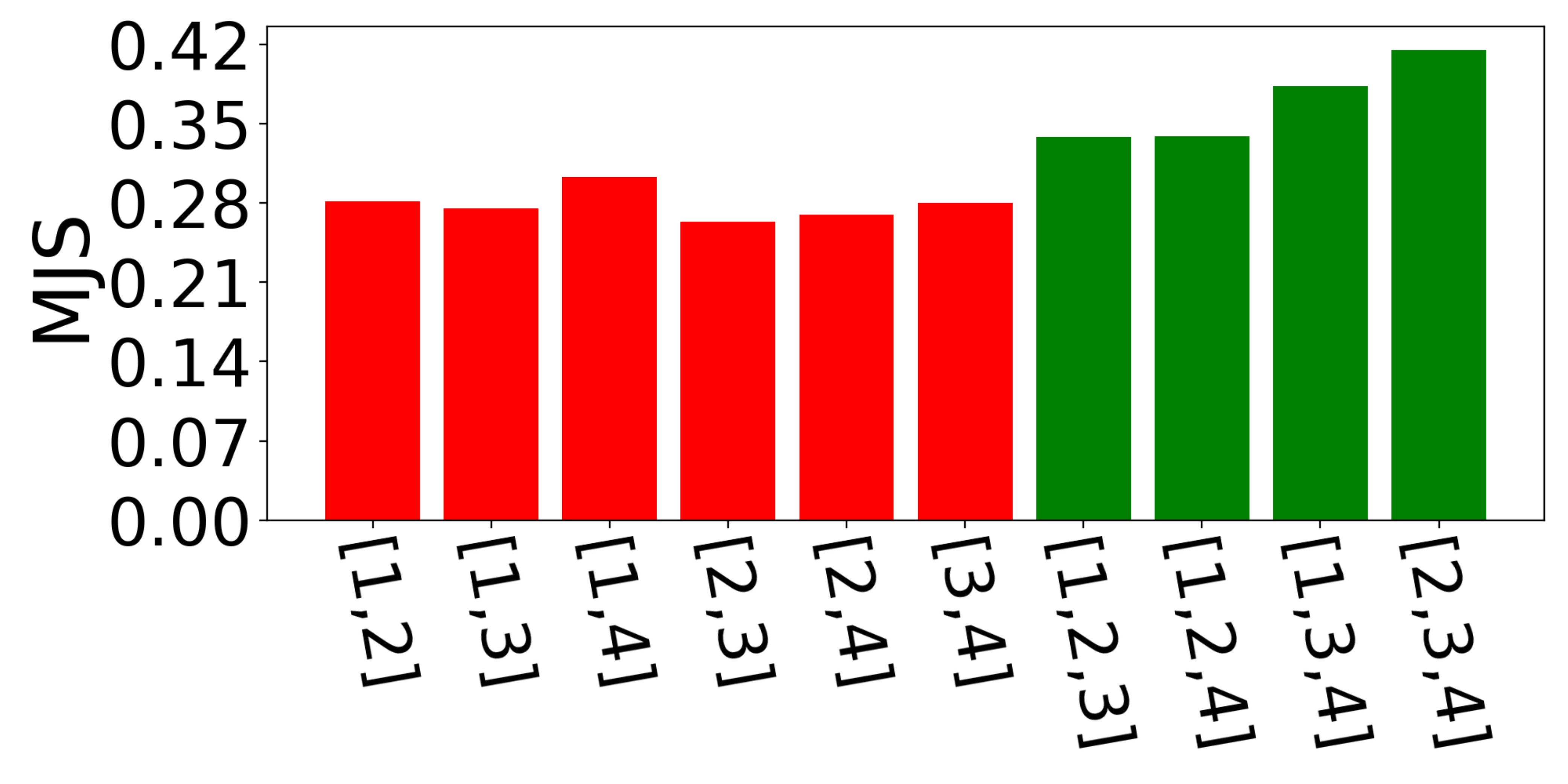}
    \label{fig:reviewer:2020}}
    }
    \vspace{-0.05in}
	\caption{MJS divergence of different combinations of different confidence level reviews}
	\label{fig:mjs}
\vspace{-0.3in}
\end{figure}

\Paragraph{Divergence Reflected by Jensen-Shannon Divergence} By using euclidean distance, we can only measure the divergence of two sets of different level reviews. Inspired by Jensen-Shannon Divergence for multiple distributions (MJS) \cite{10.5555/1763653.1763679}, we design the MJS metric to measure the divergence between multiple sets of reviews. The MJS of $m$ sets ($m\geq 2$) of different confidence reviews is defined as follows:
\begin{equation}
\label{eq:mjs}
MJS(l_1,\ldots,l_m)=\frac{1}{m}\sum_{i\in\{l_1,\ldots,l_m\}}\Bigg(\frac{1}{|R_{l_1,\ldots,l_m}|}\sum_{p\in R_{l_1,\ldots,l_m}}\overline{s_p^i}\cdot log\Big(\frac{\overline{s_p^i}}{\overline{s_p^{[1,m]}}}\Big)\Bigg),
\end{equation}
where $R_{l_1,\ldots,l_m}$ is the set of paper IDs, where each paper concurrently has reviews with confidence levels $l_1,\ldots,l_m$, $|\cdot|$ returns the size of a set, $\overline{s_p^i}$ is paper $p$'s average review score of $l_i$-confidence reviews, and $\overline{s_p^{[1,m]}}$ is paper $p$'s average review score of reviews with confidence levels $l_1,\ldots,l_m$. The bigger the $MJS$ is, the significant the opinion divergence is. A nice property of MJS metric is that it is symmetric, e.g., $MJS(i,j)=MJS(j,i)$ and $MJS(i,j,k)=MJS(k,j,i)$. We measure the MJS divergence of different combinations of confidence levels and show the results in Fig. \ref{fig:mjs}. Note that, the results of combinations that contain less than 10 reviews are not shown since they are too few to be statistically significant. In 2017-2019 the MJS divergence between conf1 reviews and conf5 reviews is the smallest. In 2020, it shows bigger divergence than 2017-2019 on different combinations but relatively similar divergence results among different combinations. In addition, a combination of three different confidence levels is likely to result in bigger divergence than a combination of two confidence levels.

\smallskip\noindent{\bf How is the Impact of Non-Expert Reviewers?} All these facts demonstrate that the opinion divergence is not greater after introducing more non-expert reviewers. We also observe that the opinion divergence between non-expert reviewers and other reviewers is often relatively small. The reason behind might be that the non-expert reviewers often have a more neutral opinion rather than clear yes or no. They are more cautious to give positive or negative recommendations.

%% file: weakpoint.tex
\subsection{Which Aspects Play Important Roles in Review Score?}
\label{sec:review:weakpoint}

Reviewers often evaluate a paper from various aspects. There are five most important aspects, i.e., novelty, motivation, experimental results, completeness of related workers, and presentation quality. Some conferences provide a peer-review questionnaire that requires reviewer to evaluate a paper from various aspects and give a score with respect to each aspect. Unfortunately, ICLR does not ask reviewers to answer such a questionnaire. Then a question arises accordingly. Which aspects play more important roles in determining the review score? We aim to answer this question by analyzing the sentiment of each aspect.

\Paragraph{Corpus Creation} For each review, we first extract the related sentences that describe different aspects by matching a set of predefined keywords. The keywords  ``novel, novelty, originality, and idea'' are used to identify a sentence that describes novelty of the paper, ``motivation, motivate, and motivated'' are used to identify a sentence related to motivation, ``experiments, empirically, empirical, experimental, evaluation, results, data, dataset, and data set'' are used to identify a sentence related to experiment results, ``related work, survey, review, previous work, literature, cite, and citation'' are used to identify a sentence related to the completeness of related work, and ``presentation, writing, written, structure, organization, structured, and explained'' are used to identify a sentence related to presentation quality. We have collected a corpus containing 95,208 sentences which are divided into five subsets corresponding to the five aspects. Specifically, we have 11,916 sentences related to ``novelty'', 5,107 sentences related to ``motivation'', 62,446 sentences related to ``experimental results'', 8,710 sentences related to ``completeness of related work'', and 7,029 sentences related to ``presentation quality''. 

\Paragraph{Automatic Annotation} 
In order to train a sentiment analysis model, we need to first annotate enough number of sentences with sentiment label (i.e., positive, negative, and neutral). However, this workload of manual annotation is huge due to the large size of review corpus. Fortunately, we find a possibility of automatic annotation after analyzing the reviews. A large number of reviewers write their positive reviews and negative reviews separately by using the keywords such as ``strengths/weaknesses'', ``pros/cons'', ``strong points/weak points'', ``positive aspects/negative aspects'', and so on. We segment the review text and identify the positive/negative sentences by looking up these keywords. The boundaries are identified when meeting an opposite sentiment word for the first time. By intersecting the set of positive/negative sentences with the set of aspect-specific sentences, we obtain a relatively large set of sentiment-annotated corpus for each aspect. Particularly, we have 2,893 sentiment-annotated sentences for ``novelty'', 1,057 for ``motivation'', 8,956 for ``experimental results'', 1,402 for ``completeness of related work'', 1,644 for ``presentation quality'', and 15,952 in total. We also manually annotate 6,095 sentences including 1,227 corrected automatically annotated sentences since some neutral sentences are incorrectly annotated with positive or negative sentiment. Finally, we have 20,820 labeled sentences\footnote{All of the annotated data including manually annotated ones are publicly available at at https://github.com/Seafoodair/Openreview/.}, i.e., 21.87\% of the total number of sentences (95,208) in corpus. Note that, there might be more than one sentences describing one aspect but having different sentiments. In such a case, we label the sentiment by a majority vote.

\Paragraph{Sentiment Analysis} Given these five datasets including the labeled data, we perform sentiment analysis for each aspect using a pretrained text model ELECTRA \cite{Clark2020ELECTRA} which was recently proposed in ICLR 2020 with state-of-the-art performance. The detailed hyper-parameter settings of ELECTRA are described in our support materials. We split the annotated dataset of each aspect into training/validation/test sets (8:1:1), and use 10-fold cross validation to train five sentiment prediction models for the five aspects. We obtain five accuracy results 93.96\%, 88.46\%, 94.99\%, 85.12\%, and 93.38\% for novelty, motivation, experimental results, completeness of related workers, and presentation quality, respectively. We then use the whole annotated dataset of each aspect to train the corresponding sentiment analysis model and use this model to predict the sentiment of the other unlabeled sentences of each aspect. Finally, for each review, we can obtain the sentiment score of each aspect. Note that, some individual aspects might not be mentioned in a review, which are labeled with neutral.

\begin{figure}[t]
\vspace{-0.2in}
\centering
\includegraphics[width=4.8in]{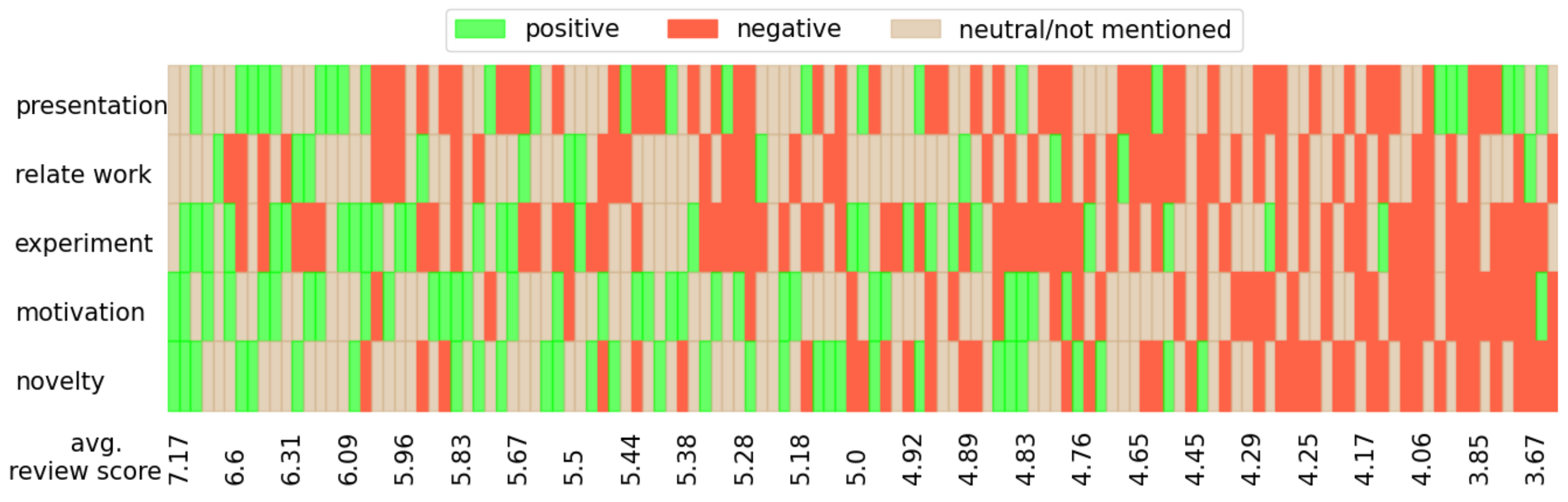}
\vspace{-0.3in}
\caption{The sentiment of each aspect vs. the review score. Each column represents a group of reviews with the same combination of aspect sentiments. These groups are sorted in the descending order of the average review score of a group of reviews.}
\label{fig:sentiment}
\vspace{-0.3in}
\end{figure}


\Paragraph{Sentiment of Each Aspect vs. Review Score} Given the sentiment analysis results of all aspects of each review and the review score, we perform the correlation analysis. We group the reviews with the same combination of aspect sentiments and compute the average review score of each group. The groups that receive less than 3 reviews are not considered since they have too few samples to be statistically significant. We visualize the result as shown in Fig. \ref{fig:sentiment}. We can see that the higher review score often comes with more positive aspects from a macro perspective, which is under expectation. We observe that most of the reviews with score higher than 6 do NOT have negative comments on novelty, motivation, and presentation, but may allow some flaws in related work and experiment. The reviewers that have overall positive to the paper are likely to pose
improvement suggestions on related work and experiment to make the paper perfect. The presentation quality and experiment seem to be mentioned more frequently than the other aspects, and the positive sentiment on presentation
is distributed more evenly from high-score reviews to low-score reviews. This implies that presentation does not play important role in making the decision. It is also interesting that there is no review in which all aspects are positive or negative. It is unlikely that a paper is perfect in all aspects or has no merit. Reviewers are also likely to be more rigorous in ` papers and be more tolerant with poor papers.

\Paragraph{Causality Analysis}
In order to explore which aspect determines the final review score, we perform causal inference following \cite{gelman_hill_2006}. Besides the above five aspects, we also include the factor of reviewer confidence. The process of causal analysis includes four steps: modeling, intervention, evaluation, and inference. In the modeling process, we use multivariate linear regression method\cite{allison1999multiple} to perform regression task on the ICLR reviews dataset, where the six evaluated parameters are the sentiment scores of the five review aspects and a reviewer confidence score, and the regression label is the review score. Each parameter is standardized to [-1,1]. To avoid randomness of model training, we launch 1000 times of training and obtain the average MSE (Mean Square Error) 0.24. The intervention process removes each factor $x$ one by one and performs multiple times of model evaluation to obtain multiple average MSE results, each corresponding to an $x$-absence model. In the absence of overfitting, the MSE value of any $x$-absence model should be larger than 0.24. The MSE value of the $x$-absence model implies the causality. A larger MSE value of an $x$-absence model implies that the factor $x$ is more dominant in determining the final score, and vice versa. In the inference process, we compare the MSE values to infer the causality. The average MSE values of the reviewer confidence, novelty, motivation, experiment, related work, and presentation are 0.84, 0.77, 0.34, 0.86, 0.33, and 0.34, respectively. We observe that the factors of reviewer confidence, novelty, and experiment change the MSE greatly, so they are more dominant in determining the final score.

%% file: cluster.tex
\subsection{Which Research Field has Higher/Lower Acceptance Rate?}
\label{sec:review:cluster}

AI conferences consider a broad range of subject areas. Authors are often asked to pick the most relevant areas that match their submissions. Area chair could exist who makes decisions for the submissions of a certain research area. Different areas may receive different number of submissions and also may have different acceptance rates. Program chairs sometimes announce the number of submissions and the acceptance rate of each area in the opening event of a conference, which could somehow indicate the popularity of each area. But, the classification by areas is coarse. A more fine-grained classification that provides more specific information is desired. Thanks to the more detailed submission information provided by OpenReview, we utilize the title, abstract, and keywords of each submission to provide a more fine-grained clustering result and gather the statistics of acceptance rate of each cluster of submissions.

\begin{figure}
\vspace{-0.15in}
\centering
  \includegraphics[width=4.4in]{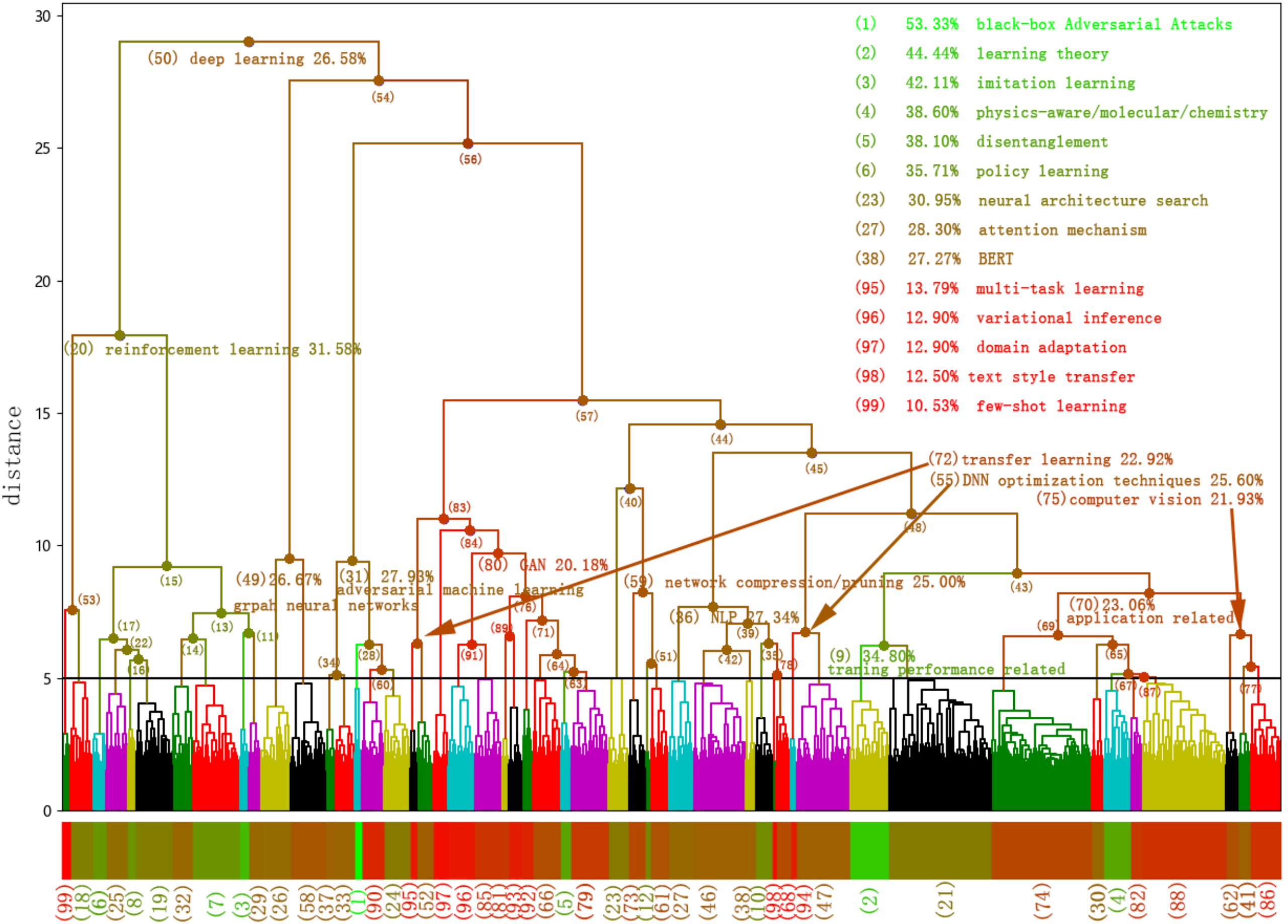}
  \vspace{-0.2in}
\caption{Visualized hierarchical clustering result of ICLR 2020 submissions. Each leaf node represents a submission. Cosine distance 5 is selected as the threshold to control the granularity of leaf-level clusters. There are 99 clusters in total, including both fine-grained clusters and coarse-grained clusters. Clusters are numbered in the order of their acceptance rate. The color of keywords indicates the acceptance rate of that cluster. Light green means a high acceptance rate, while light red means a low acceptance rate. The keywords of some typical clusters are labeled.}
\label{fig:cluster}
\vspace{-0.3in}
\end{figure}

We first concatenate the title, abstract, keywords of each ICLR 2020 submission and preprocess them by removing stop words, tokenizing, stemming list, etc. We leverage an AI terminology dictionary \cite{aiterm} during the tokenizing process to make sure that an AI terminology containing multiple words is not split. We then formulate term-document matrix (i.e., AI term-submission matrix) by applying TF-IDF and calculate cosine distance matrix. The size of the term-document TF-IDF matrix for ICLR 2020 is 12436 x 2558, and the size of the cosine distance matrix is 2558 x 2558. We then apply the Ward clustering algorithm \cite{doi:10.1080/01621459.1963.10500845} on the matrix to obtain submission clusters. Ward clustering is an agglomerative hierarchical clustering method, meaning that at each stage, the pair of clusters with minimum between-cluster distance are merged. We use silhouette coefficient to finalize the number of clusters and plot a dendrogram to visualize the hierarchical clustering result as shown in Fig. \ref{fig:cluster}.

From Fig. \ref{fig:cluster}, we observe three aspects of insights. \textbf{(a) Overall Structure of Deep Learning Research}. We observe the correlation between research topics. For example, the submissions in the left part belong to reinforcement learning field (20), which is far apart from all the other research topics (because it is the last merged cluster and its distance to the other clusters is more than 27). Another independent research field is Graph Neural Networks (GNNs) (49), as a promising field, becomes really hot in only 2-3 years, which distinguishes itself from others by focusing on graph structure. Adversarial Machine Learning (31) is also an independent research field that attempts to fool models through malicious input and different from others. The next independent subject is Generative Adversarial Networks (GANs) (80). But GANs is not completely independent since we found that many submissions on NLP (36) and CV (75) are mixed with GANs as well. We also observe that Transfer Learning (72) is close to GANs, since some works have applied transfer learning to GANs. Most of the submissions in the right part are applications related (e.g., vision, audio, NLP, biology, chemistry, and robotics). They are mixed with DNN optimization techniques since many optimizations are proposed to improve DNN on a specific application field. \textbf{(b) Popularity Difference between Clusters}. We observe that multiple areas attract large amount of submission. For example, Reinforcement Learning (20), GNNs (49), GANs (80), NLP (36), and Computer Vision (75) have attracted more than 50\% of the submissions, which are really hot topics in today's deep learning research. \textbf{(c) Acceptance Rate Difference between Clusters}. There exists significant difference on acceptance rate between clusters, say ranging from 53.33\% to 10.53\%. The cluster of submissions on ``Black-Box Adversarial Attacks'' has the highest acceptance rate (53.33\%), which is a subject belongs to ``Adversarial Machine Learning'' area. The top-6 highest acceptance rate topics are listed in the figure. The cluster of submissions on ``Few-Shot Learning'' has the lowest acceptance rate (10.53\%), which is a subject belongs to ``Reinforcement Learning'' area. The top-5 lowest acceptance rate topics are listed in the figure. We also list some typical topics in the figure. For example, the cluster on ``Graph Neural Networks (49)'' has an acceptance rate of 26.67\%. The cluster on ``BERT (38)'' has an acceptance rate of 27.27\%. The cluster on ``GANs (80)'' has an acceptance rate of 20.18\%. The cluster on ``Reinforcement Learning (20)'' has an acceptance rate of 31.58\%.

%% file: citation.tex
\subsection{Review Score vs. Citation Number}

The citation number quantitatively indicates a paper's impact. In this subsection, we show several interesting results on the correlation between review scores and citation numbers.

\label{sec:cite:correlation}



\begin{figure}[t]
 \vspace{-0.3in}
	\centerline{
	\subfloat[2017]{\includegraphics[width=1.7in]{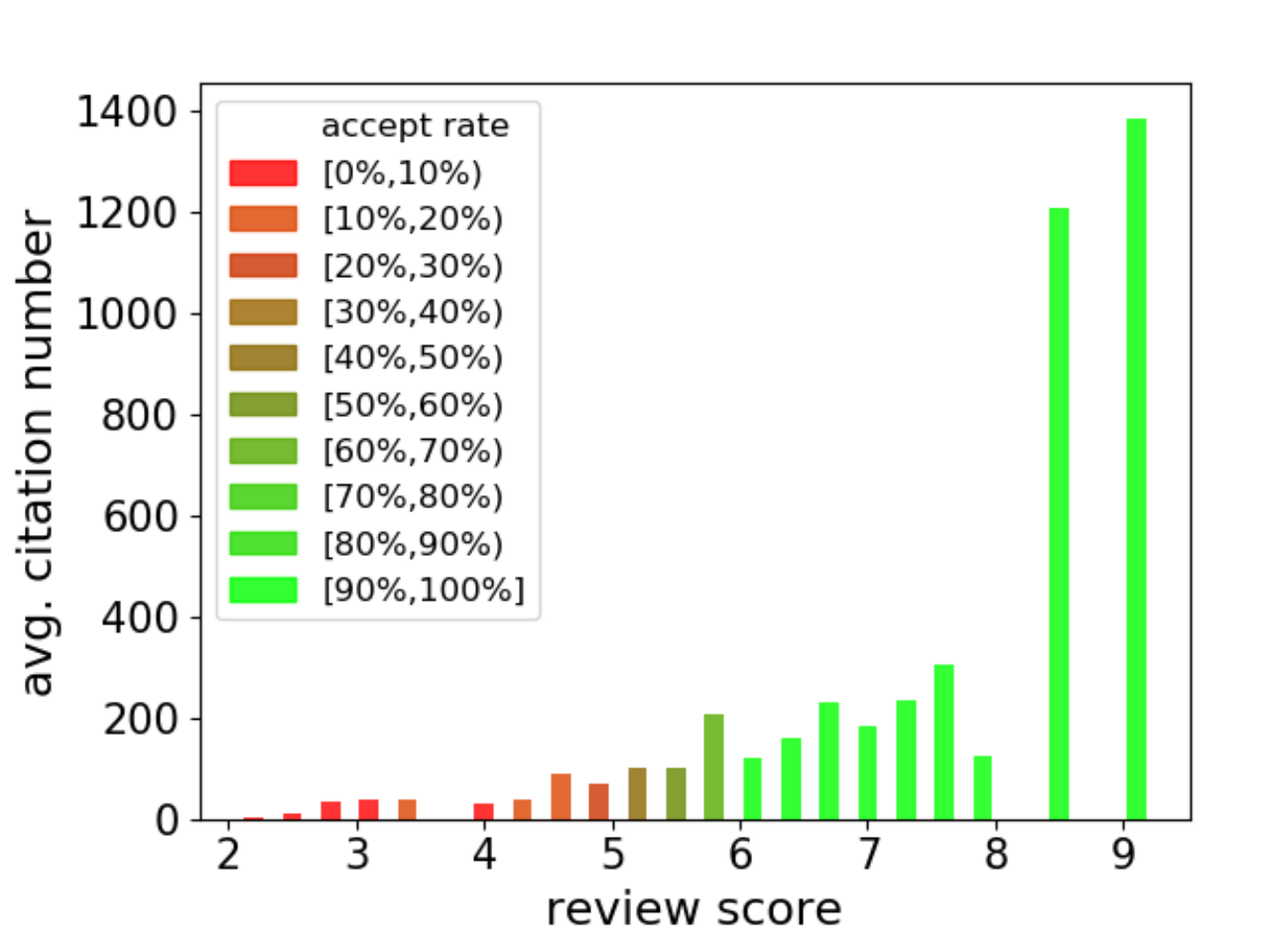}
    \label{fig:avgcite:2017}}
    \subfloat[2018]{\includegraphics[width=1.7in]{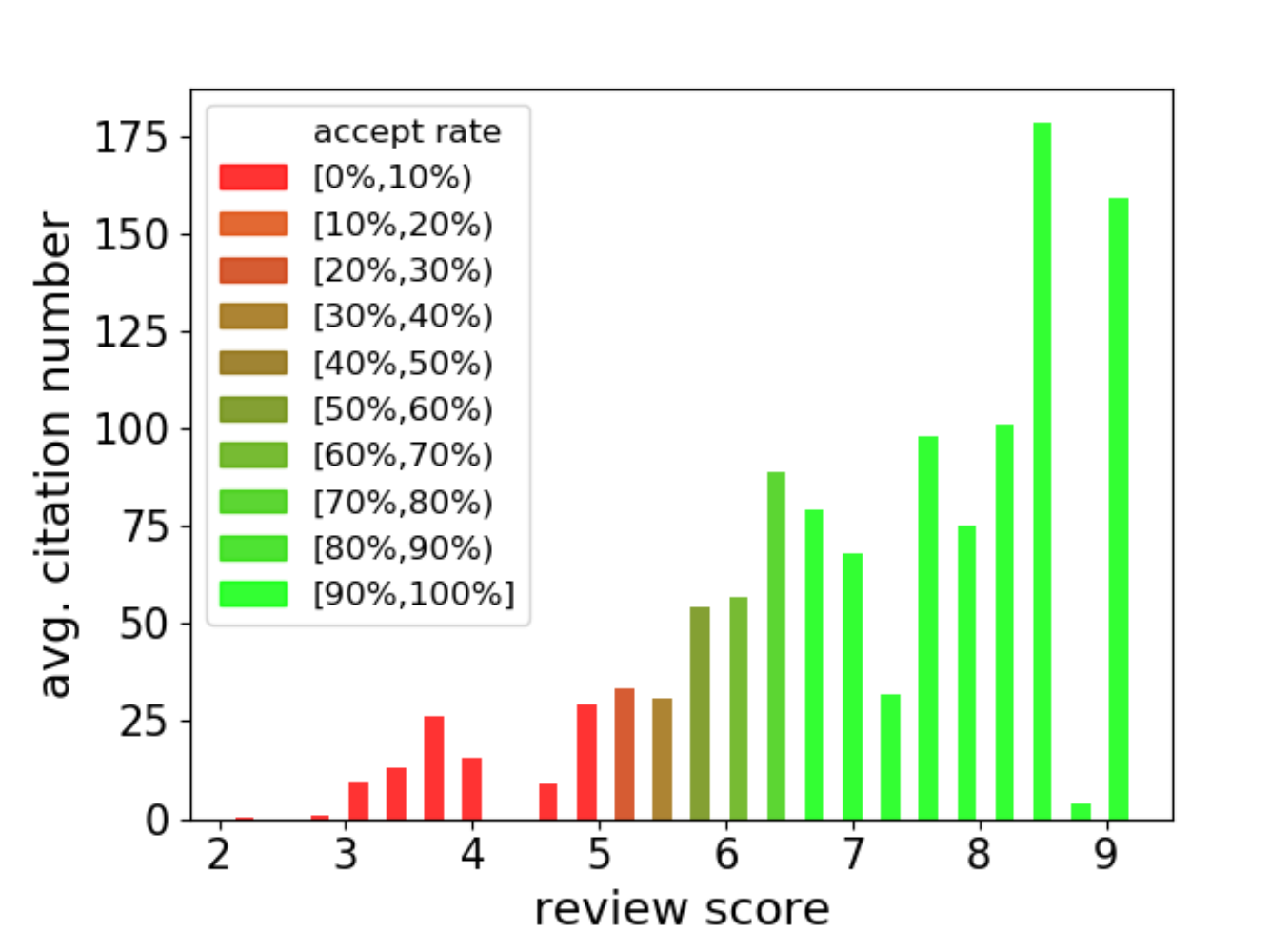}
    \label{fig:avgcite:2018}}
    \subfloat[2019]{\includegraphics[width=1.7in]{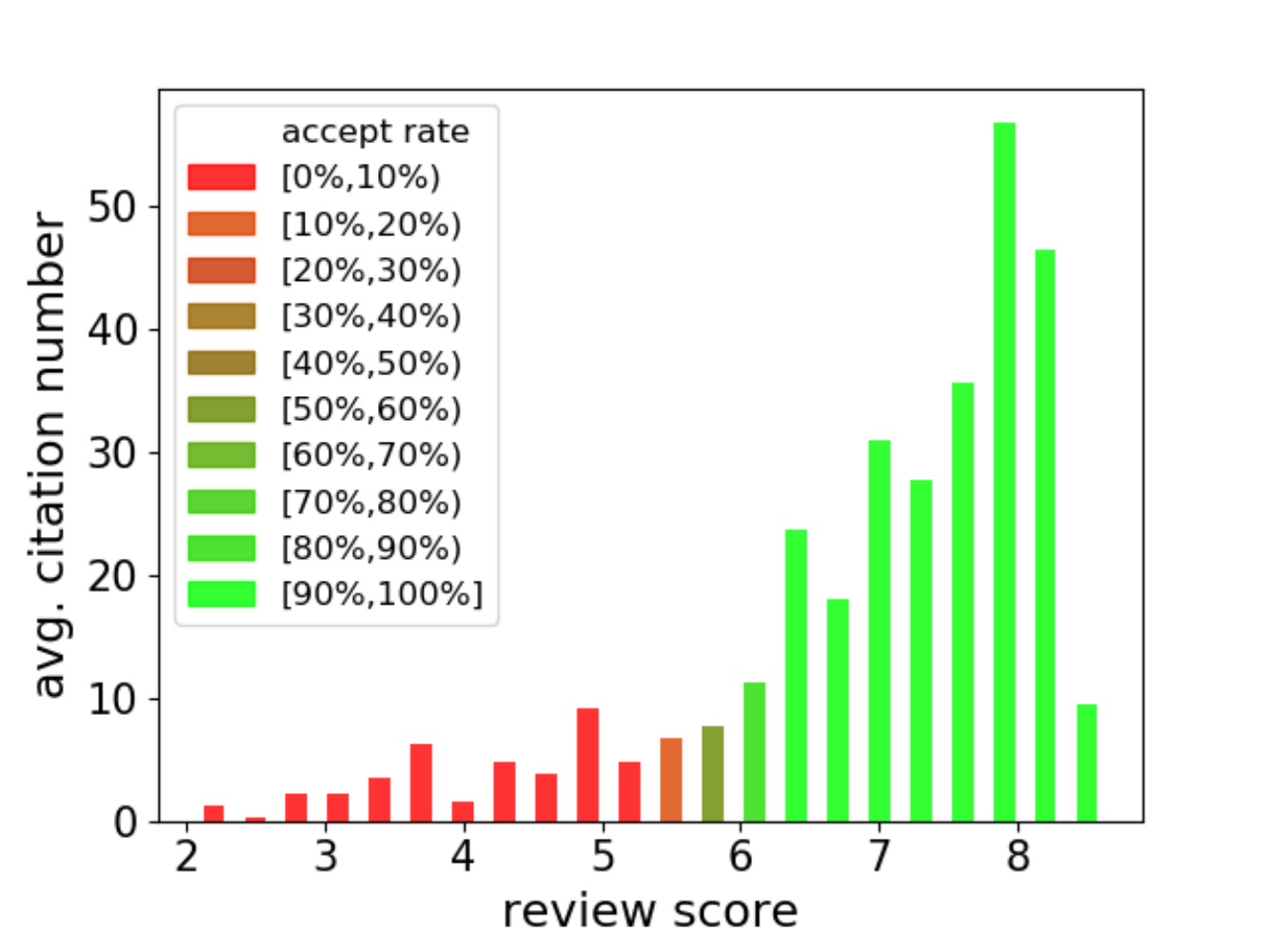}
    \label{fig:avgcite:2019}}
    }
     \vspace{-0.05in}
	\caption{The histogram of citation numbers against 0.3-intervals of average review score}
	\label{fig:avgcite}
\end{figure}

\smallskip\noindent{\bf Is There a Strong Correlation between Review Score and Citation Number?} OpenReview releases not only the submission details and reviews of the accepted papers but also that of the rejected submissions. These rejected submissions might be put on arXiv.org or published in other venues and still make an impact. We collect the citation number information of both accepted papers and rejected papers and study the correlation between their review scores and their citation numbers. We plot the histogram of average citation numbers of ICLR 2017-2019 submissions as shown in Fig. \ref{fig:avgcite}. The papers are divided into multiple subsets according to their review scores. Each bin of the histogram corresponds to a subset of papers with similar review scores (with an interval of 0.3). Then the average citation number of each subset is calculated. The color of bin indicates the acceptance rate of the corresponding subset of papers. From the figure, we can observe that the papers with higher review score are likely to have higher citation numbers, which is under expectation.


\begin{figure}[t]
 \vspace{-0.3in}
	\centerline{
	\subfloat[2017]{\includegraphics[width=1.7in]{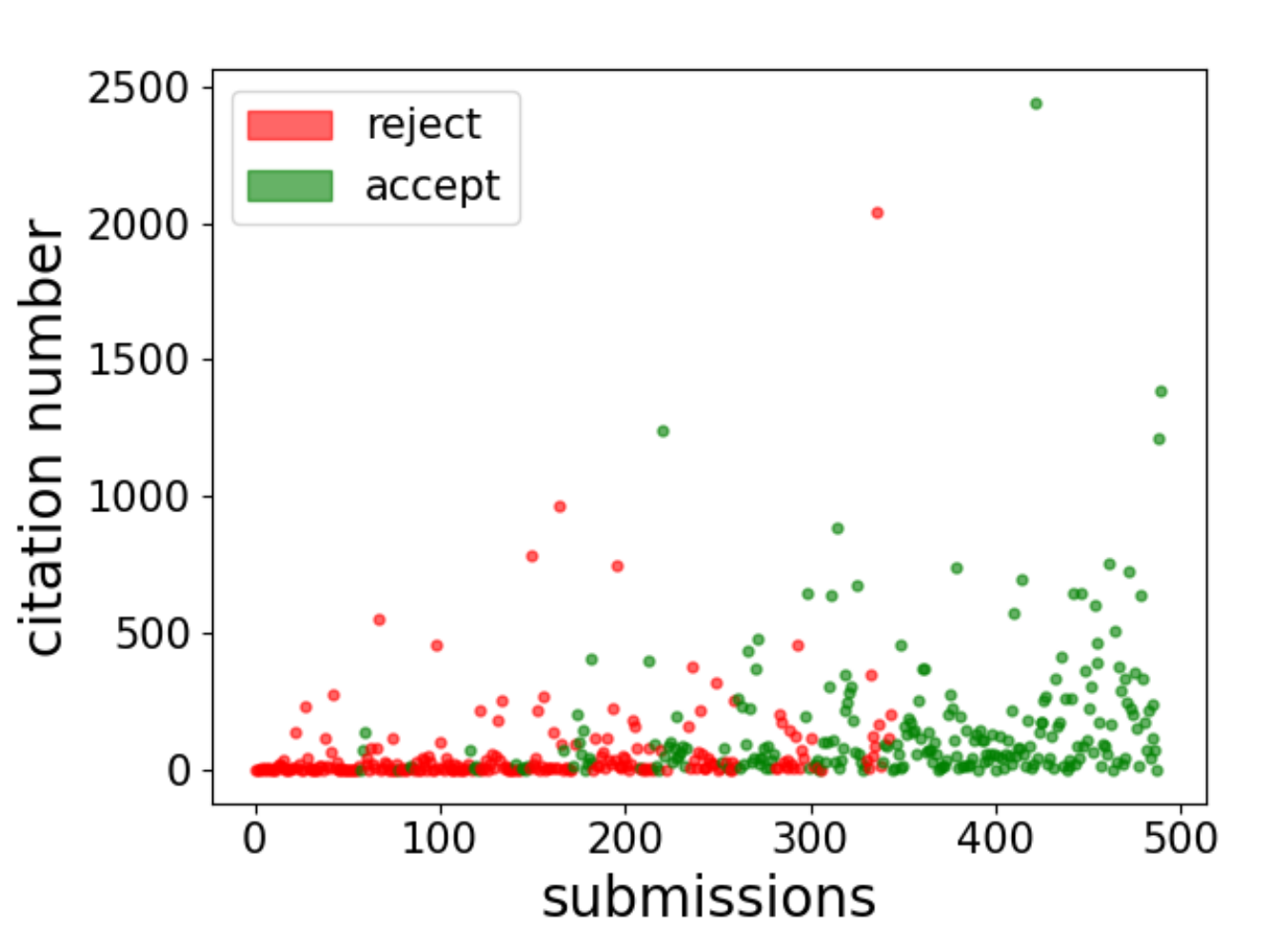}
    \label{fig:avgcite:2017}}
    \subfloat[2018]{\includegraphics[width=1.7in]{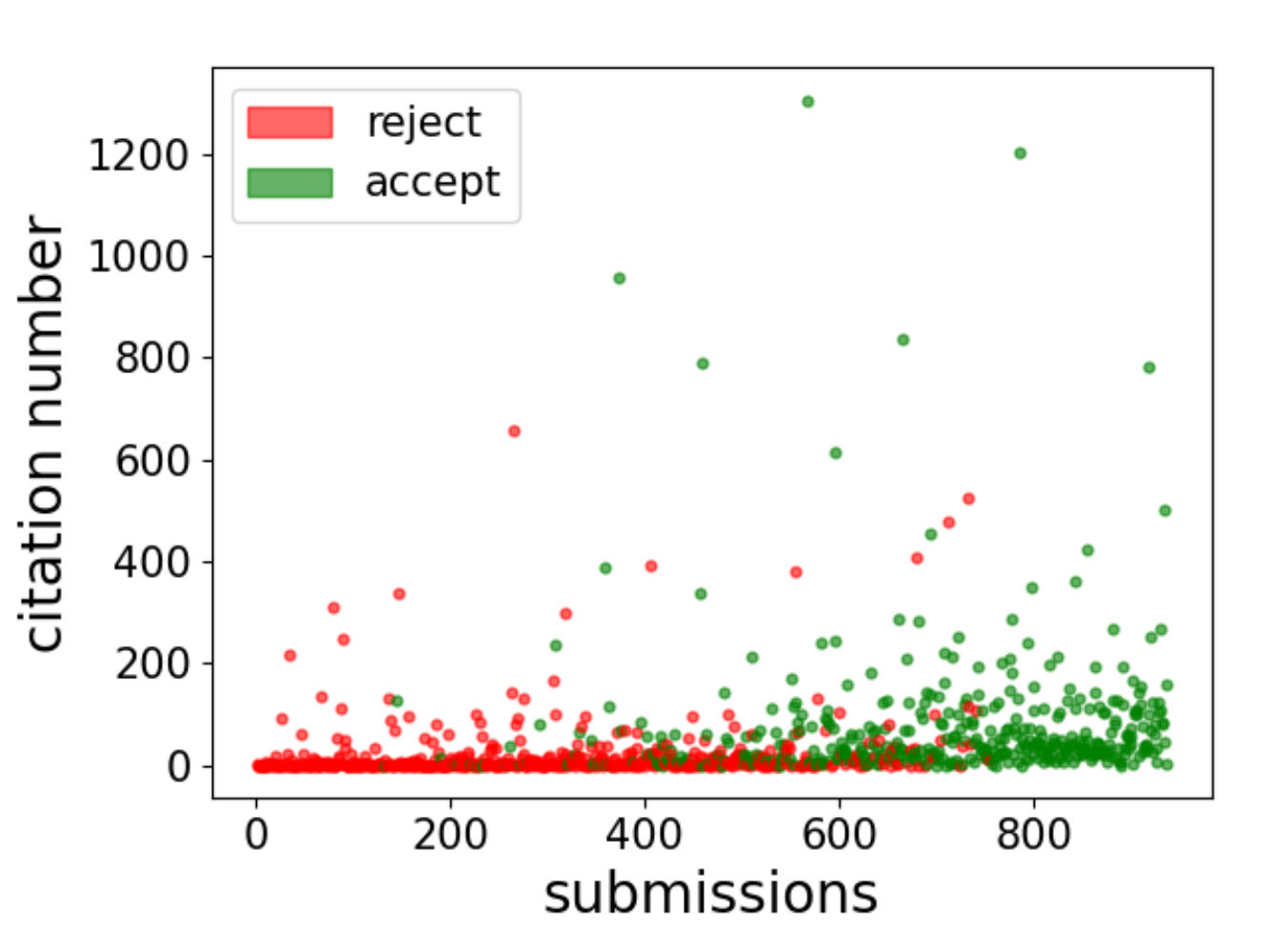}
    \label{fig:avgcite:2018}}
    \subfloat[2019]{\includegraphics[width=1.7in]{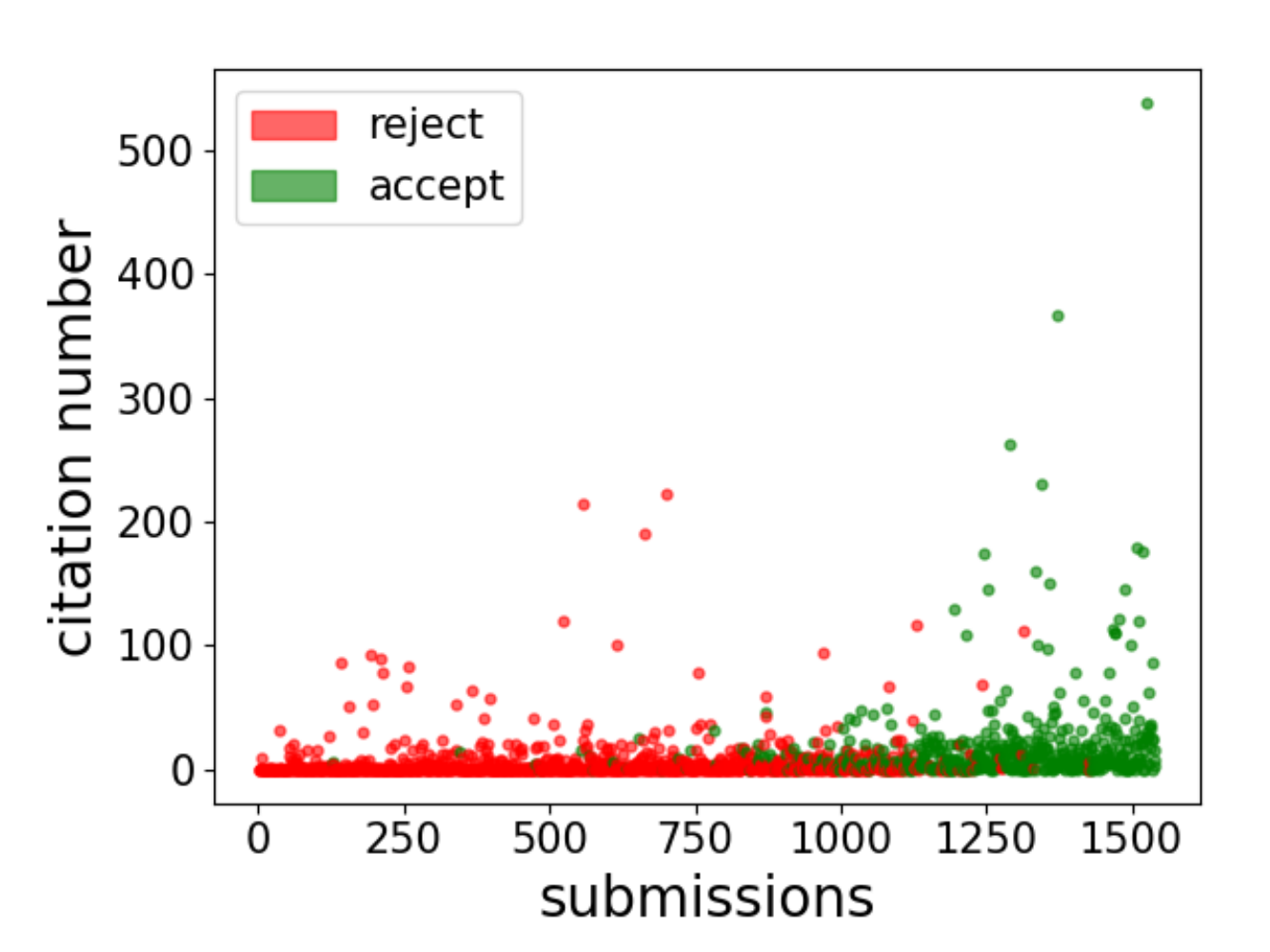}
    \label{fig:avgcite:2019}}
    }
     \vspace{-0.1in}
	\caption{The distribution of citation numbers of individual papers, where the papers on the x-axis are sorted in the ascending order of their review scores}
	\label{fig:cite}
\end{figure}

We further investigate the citation numbers of individual papers as shown in Fig. \ref{fig:cite}. Each point represents a paper. Green color indicates an accepted paper and red color indicates a rejected one. The papers are sorted on the x-axis in the ascending order of their review scores. The distribution of citation numbers is messy. We can see that many rejected papers gain a large number of citations (i.e, red points in the top-left part), which is a bit surprised. Generally speaking, the accepted papers will attract more attentions since they are officially published in ICLR. However, the rejected papers may be accepted later at other venues and still attract attentions. In addition, a few papers with high review score are rejected (i.e., red points on the right side). We observe that the reject decision does not impact their citation numbers. Though rejected, the papers with higher review score are still likely to have higher citation numbers.

\begin{figure}[t]
\vspace{-0.1in}
\centering
  \includegraphics[width=4.3in]{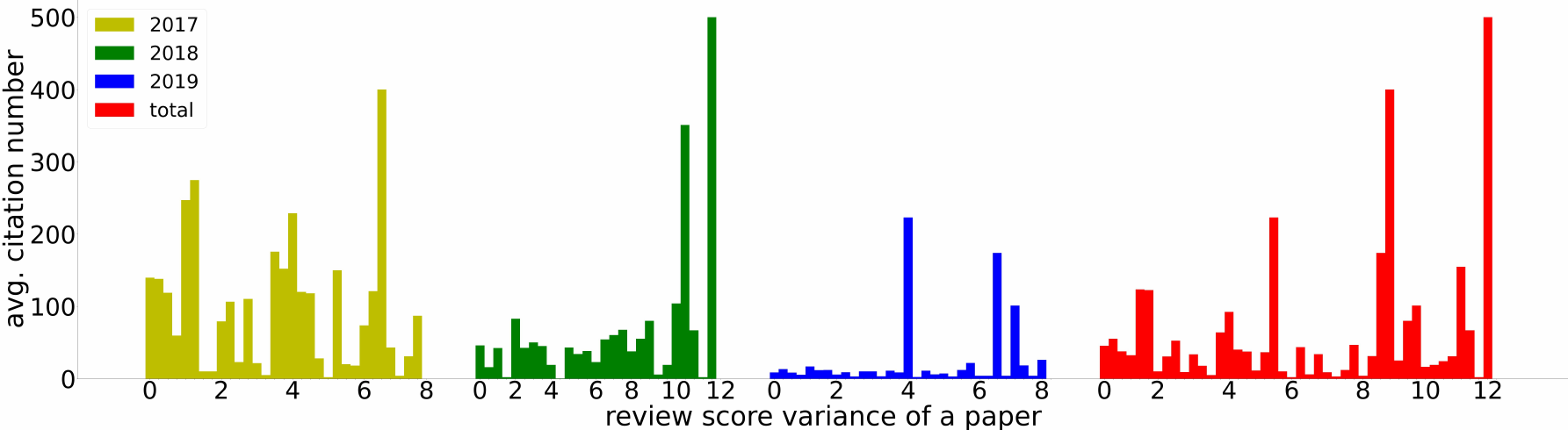}
  \vspace{-0.1in}
\caption{Review score variance of a paper vs. average citation number}
\label{fig:pnp}
\vspace{-0.2in}
\end{figure}

\noindent{\bf Do Great Papers Gain More Diverse Review Scores?} People always have diverse opinions on the breakthrough works. Reviewer A thinks it novel and is happy to give higher scores, but reviewer B may think it too crazy or unrealistic and reject it. There might be a big debate between reviewers. But it is usually hard to reach consensus. We investigate the relationship between the variance of review scores of a submission and its citation number. We group papers according to their review score variances and calculate the average citation number of each group. Fig. \ref{fig:pnp} shows the statistical results of the submissions of ICLR 2017-2019. We observe that the papers that have large number of citations are indeed more likely to gain diverse review scores. Note that a paper that has diverse review scores (big review score variance) does not necessarily have high review scores.

%% file: arxiv.tex
\subsection{Do Submissions Posted on arXiv have Higher Acceptance Rate?}
\label{sec:review:arxiv}

We found 1,083 submissions that have been posted on arXiv before accept/reject notification\footnote{We compare paper creation date on arXiv with ICLR official notification date.}, which account for about 19.59\% of the total submissions. The arXiv versions are not anonymous, which bring unfairness to the double-blind review process. We refer to the submissions that have been posed on arXiv before notification as ``arXived submissions''. We investigate the acceptance rates of the arXived and non-arXived submissions. The acceptance rates of the arXived submissions in 2017, 2018, 2019, and 2020 are 59.33\%, 62.39\%, 45.36\%, and 30.48\%, respectively. The acceptance rates of the non-arXived submissions in 2017, 2018, 2019, and 2020 are 45.88\%, 41.23\%, 26.37\%, and 17.22\%, respectively. We observe that the arXived submissions have significantly higher acceptance rate than the non-arXived submissions (49.39\% vs. 32.68\% on average). 

\begin{figure}
\vspace{-0.4in}
	\centerline{
	\subfloat[2017]{\includegraphics[width=1.25in]{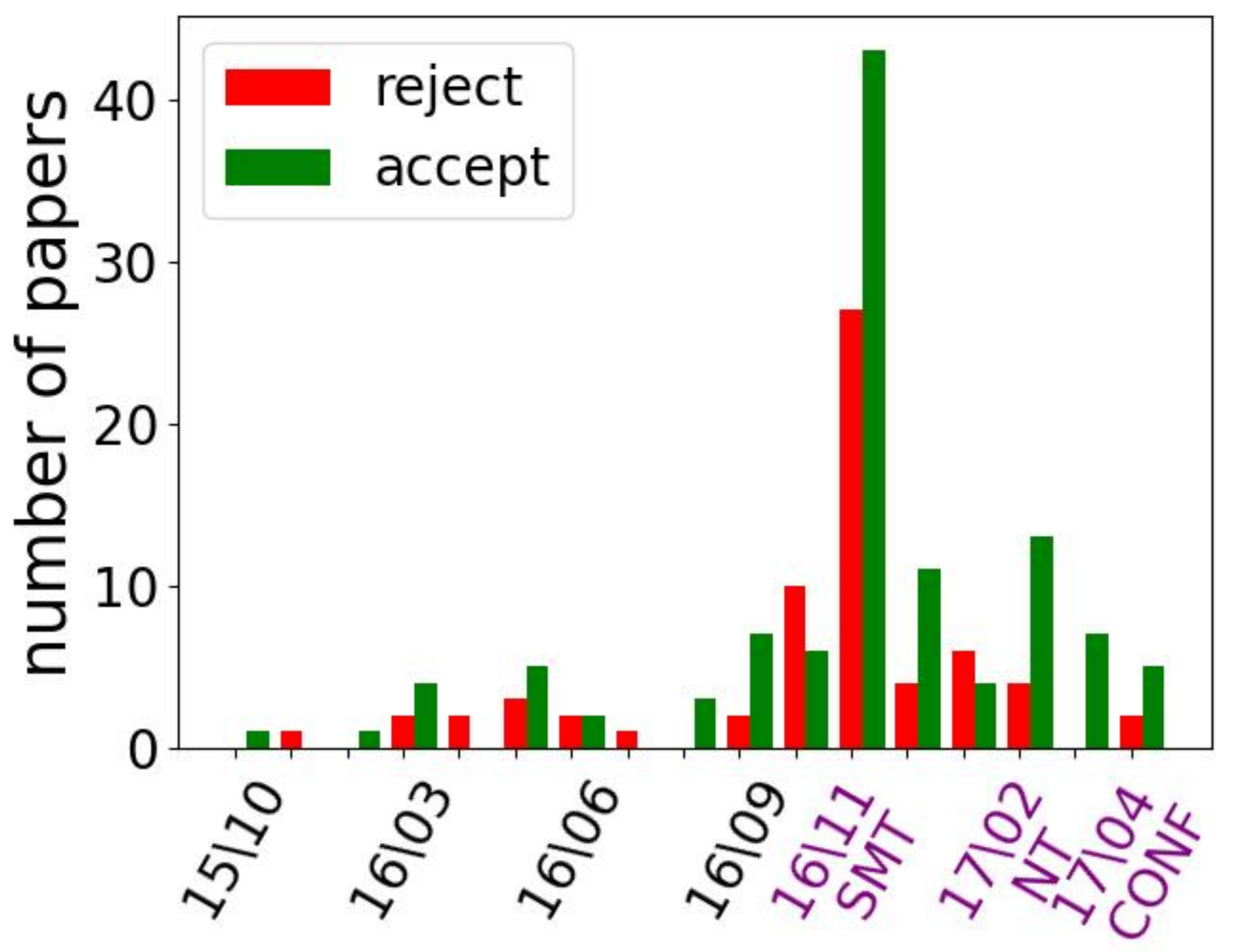}
    \label{fig:reviewer:2017}}
    \subfloat[2018]{\includegraphics[width=1.25in]{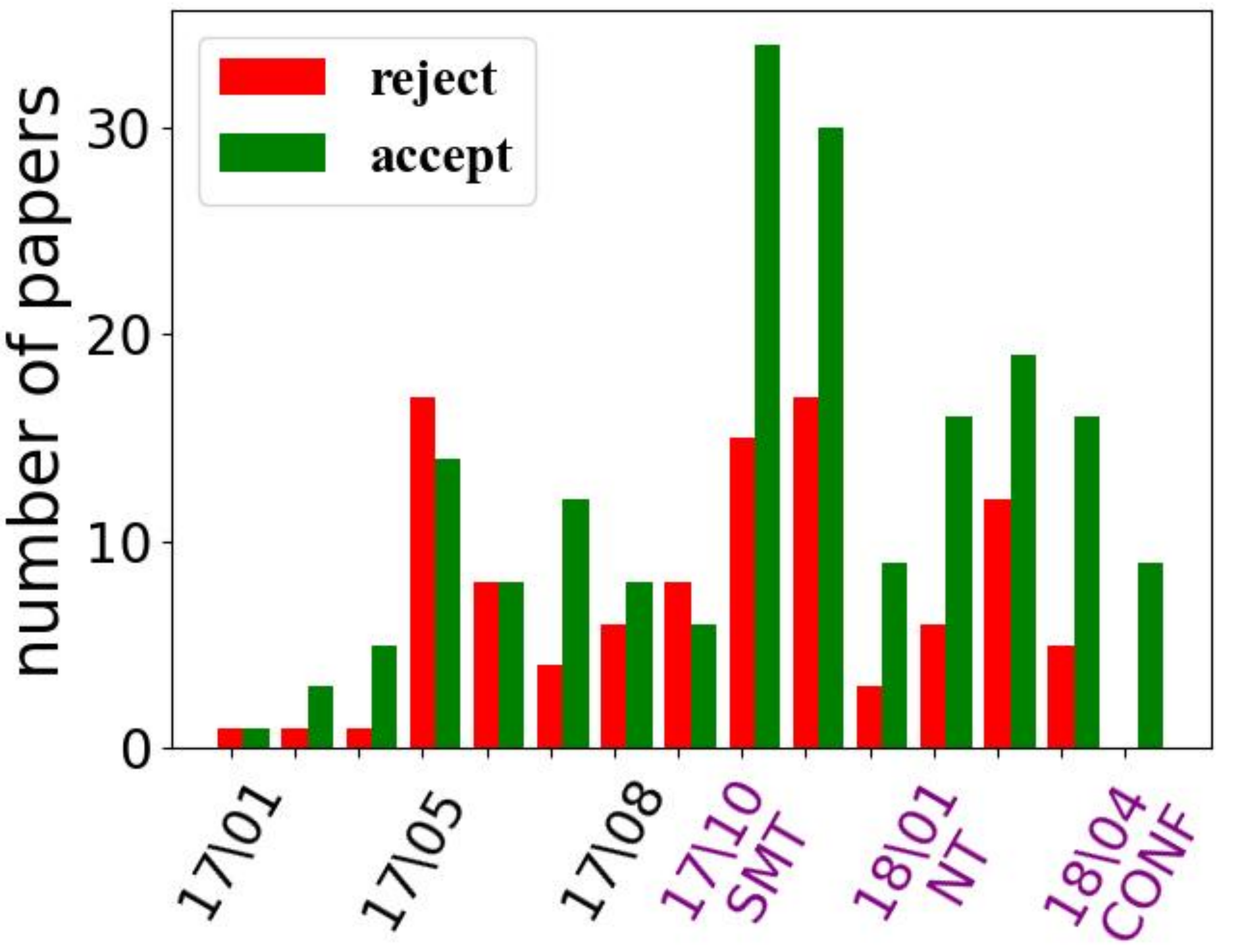}
    \label{fig:reviewer:2018}}
    \subfloat[2019]{\includegraphics[width=1.25in]{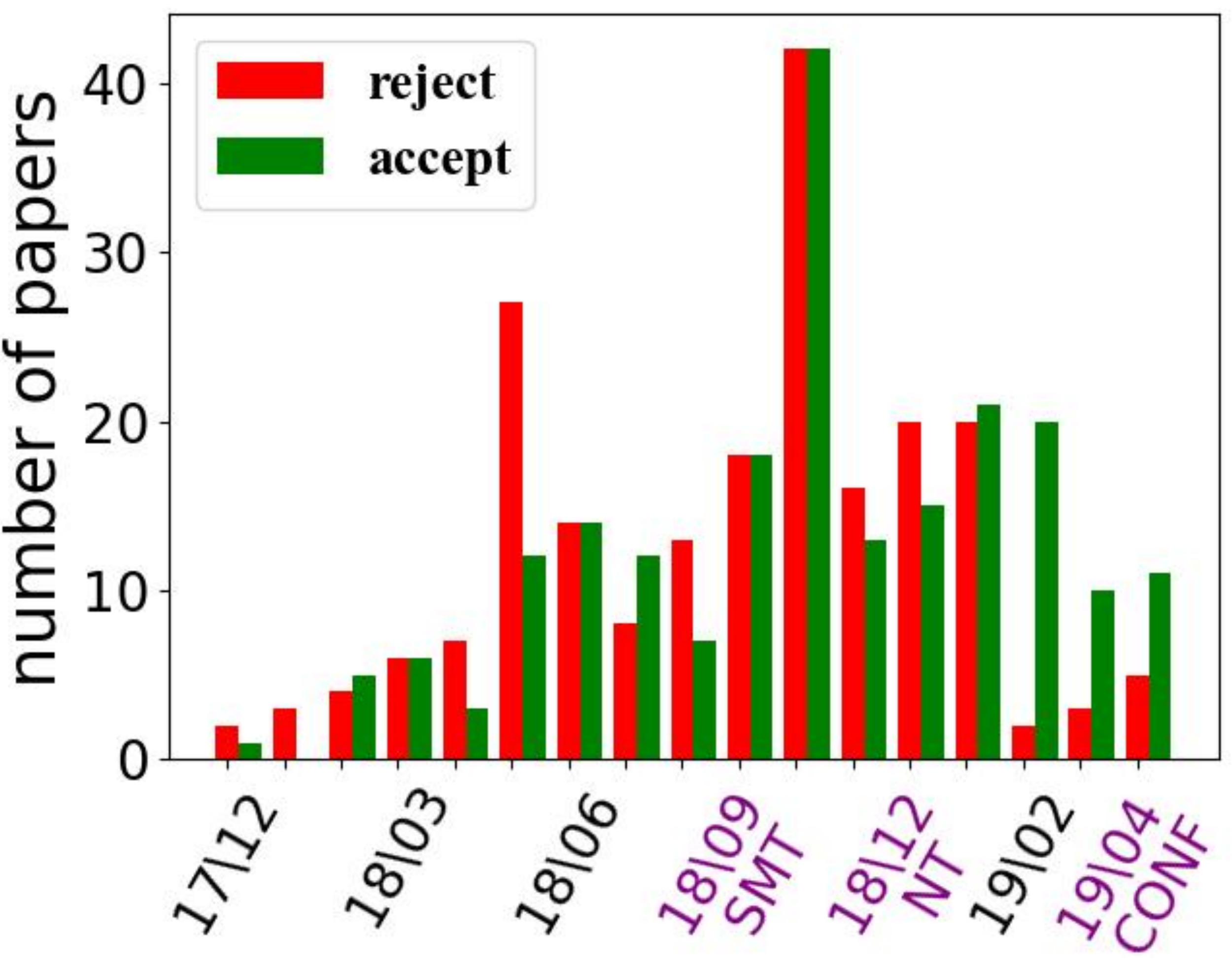}
    \label{fig:reviewer:2019}}
    \subfloat[2020]{\includegraphics[width=1.25in]{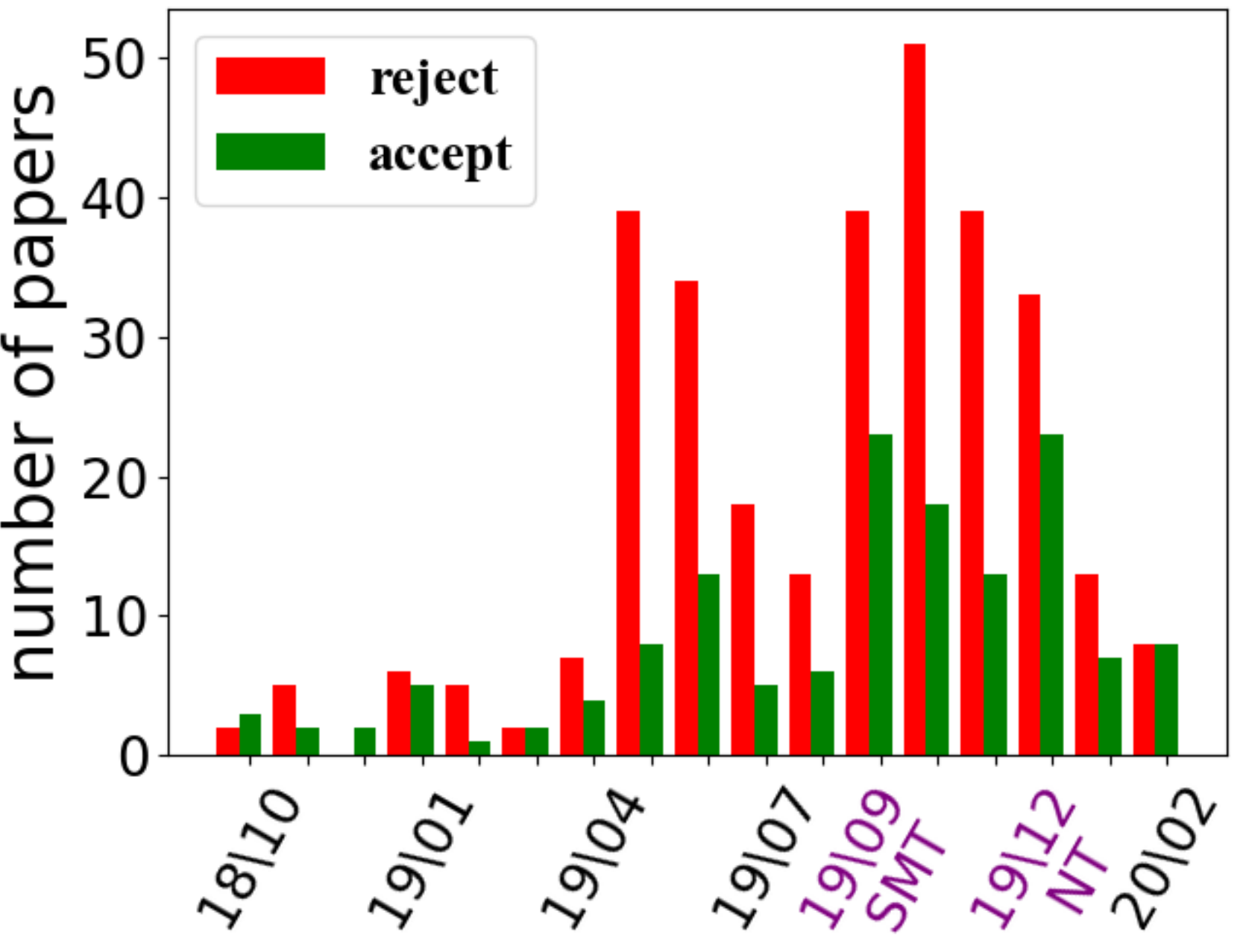}
    \label{fig:reviewer:2020}}
    }
    \vspace{-0.1in}
	\caption{The number of accepted/rejected arXived submissions posted on arXiv by month. SMT means the submission deadline. NT means the notification date. CONF means the conference date.}
	\label{fig:arxiv}
	\vspace{-0.2in}
\end{figure}

We think the reason should be not only anonymity but also that the arXived  ICLR submissions have higher quality. These arXived submissions might attract more feedbacks from colleagues, according to which the authors can improve their manuscripts. The arXived submissions might also be the rejected ones from other conferences and might have been improved according to the rejection reviews. We also observe that some arXived submissions are posted on arXiv one year before the submission deadline. Fig. \ref{fig:arxiv} shows the number of arXived submissions posted on arXiv by month, including both accepted ones and rejected ones. We can see that the papers posted on arXiv are more and more when approaching the submission deadline. There are also a large number of papers posted on arXiv between the submission date and the notification date. From the aspect of acceptance rate, we observe that the earlier the papers are posted on arXiv, the more likely they are accepted. In addition, the papers posted on arXiv after notification date have a higher acceptance rate. The reason might be that the authors cannot wait to share their research results after their papers are accepted.





%% file: related.tex
\section{Related Work}

There exist many interesting works related to peer-review analysis. We list several related works as follows.

Ivan Stelmakh's Blog \cite{ML-CMU} shares a lot of interesting findings: First, reviewers give lower scores once they are told that a paper is a resubmission. Second, there is no evidence of herding in the discussion phase of peer review. Third, A combination of the selection and mentoring mechanisms results in reviews of at least comparable and on some metrics even higher-rated quality as compared to the conventional pool of reviews.
Nihar B.Shah et al. \cite{shah2018design} analyzed the influence of reviewer and AC bid, reviewer assignment, different types of reviewers, rebuttals and discussions, distribution across subject areas in detail. Homanga Bharadhwaj et al. \cite{bharadhwaj2020anonymization} provide an analysis on whether there is a positive impact if his/hers paper is upload on arXiv before the submission deadline. They suggest that the paper arXived will have a higher acceptance rate. David Tran et al. \cite{2010.05137} analyzed ICLR conferences and quantified reproducibility/randomness in review scores and acceptance decisions, and examined whether scores correlate with paper impact. Their results suggest that there exists strong institutional bias
in accept/reject decisions, even after controlling for paper quality. They analyzed the influence of scores among gender, institution, scholar reputation in detail. Emaad Manzoor et al. \cite{manzoor2020uncovering} proposed a framework to nonparametrically estimate biases expressed in text. The authors leveraged the framework to accurately detect these biases from the review text without having access to the review ratings. Birukou et al. \cite{birukou2011alternatives} analyzed ten CS conferences and found low correlation between review scores and the impact of papers in terms of future number of citations. Gao et al. \cite{DBLP:conf/naacl/0023EKGM19} predict after-rebuttal (i.e., final) scores from initial reviews and author responses. Their results suggest that a reviewer’s final score is largely determined by her initial score and the distance to the other reviewers’ initial scores. Li et al. \cite{DBLP:conf/emnlp/LiZYW19} utilize peer review data for the citation count prediction task with a neural prediction model.

In this paper, we investigate ICLR 2017-2020's submissions and reviews data on OpenReview and show more different interesting results, e.g., the effect of low confidence reviews, the sentiment analysis of review text on different aspects, the hierarchical relationships of different research fields, etc, which have not been studied before.

\section{Conclusion}
We perform deep analysis on the dataset including review texts collected from OpenReivew, the paper citation information collected from GoogleScholar, and the non-peer-reviewed papers from arXiv.org. All of these collected data are publicly available on Github, which will help other researchers identify novel research opportunities in this dataset. More importantly, we investigate the answers to several interesting questions regarding the peer-review process. We aim to provide hints to answer these questions quantitatively based on our analysis results. We believe that our results can potentially help writing a paper, reviewing it, and deciding about its acceptance. 
